# Mass supply from Io to Jupiter's magnetosphere

*Authors: L. Roth[1,*], A. Blöcker[2,1], K. de Kleer[3], D. Goldstein[4], E. Lellouch[5], J. Saur[6], C. Schmidt[7], D.F. Strobel[8], C. Tao[9], F. Tsuchiya[10], V. Dols[11], H. Huybrighs[12,13,14], A. Mura[15], J. R. Szalay[16], S. V. Badman[17], I. de Pater[18], A.-C. Dott[6], M. Kagitani[10], L. Klaiber[19], R. Koga[20], A. McEwen[21], Z. Milby[3], K.D. Retherford[22,23], S. Schlegel[6], N. Thomas[19], W.L. Tseng[24], A. Vorburger[20]*

[1] *KTH Royal Institute of Technology, Space and Plasma Physics, Stockholm, Sweden*
[2] *Department of Earth and Environmental Sciences, Ludwig Maximilian University of Munich, Munich, Germany*
[3] *Division of Geological and Planetary Sciences, California Institute of Technology, Pasadena, CA 91125 USA*
[4]*Dept. Aerospace Engineering and Engineering Mechanics, The University of Texas at Austin, Austin, TX USA*
[5] *Laboratoire d'Etudes Spatiales et d'Instrumentation en Astrophysique (LESIA), Observatoire de Paris, Meudon, France*
[6] *Institute of Geophysics and Meteorology, University of Cologne, Cologne, Germany*
[7] *Center for Space Physics, Boston University, Boston, MA, USA*
[8] *Departments of Earth & Planetary Science and Physics & Astronomy, Johns Hopkins University, Baltimore, MD 21218, USA*
[9] *National Institute of Information and Communications Technology, Koganei, Japan*
[10] *Graduate School of Science, Tohoku University, Sendai, Japan*
[11] *Institute for Space Astrophysics and Planetology, National Institute for Astrophysics, Italy*
[12] *School of Cosmic Physics, DIAS Dunsink Observatory, Dublin Institute for Advanced Studies, Dublin 15, Ireland*
[13] *Space and Planetary Science Center, Khalifa University, Abu Dhabi, UAE*
[14] *Department of Mathematics, Khalifa University, Abu Dhabi, UAE*
[15] *Istituto Nazionale di Astrofisica - Istituto di Astrofisica e Planetologia Spaziali (INAF-IAPS), Roma, Italy*
[16] *Department of Astrophysical Sciences, Princeton University, Princeton, NJ, USA*
[17] *Department of Physics, Lancaster University, Lancaster, LA1 4YB, UK*
[18] *Department of Astronomy and Department of Earth & Planetary Science, University of California, Berkeley, CA 94720, USA*
[19] *Physics Institute, University of Bern, 3012 Bern, Switzerland*
[20] *School of Data Science, Nagoya University, Nagoya, Aichi 464-8601, Japan*
[21] *Department of Planetary Sciences, University of Arizona, Tucson, AZ, USA*
[22] *Southwest Research Institute, San Antonio, TX, USA*
[22] *University of Texas at San Antonio, San Antonio, Texas, USA*
[24] *Department of Earth Sciences, National Taiwan Normal University, Taiwan*
[*] *Corresponding author, email: lorenzr@kth.se*

**Conflict of interest statement**

The authors have no relevant financial or non-financial interests to disclose.

# Abstract

Since the Voyager mission flybys in 1979, we have known the moon Io to be both volcanically active and the main source of plasma in the vast magnetosphere of Jupiter. Material lost from Io forms neutral clouds, the Io plasma torus and ultimately the extended plasma sheet. This material is supplied from Io's upper atmosphere and atmospheric loss is likely driven by plasma-interaction effects with possible contributions from thermal escape and photochemistry-driven escape. Direct volcanic escape is negligible. The supply of material to maintain the plasma torus has been estimated from various methods at roughly one ton per second.

Most of the time the magnetospheric plasma environment of Io is stable on timescales from days to months. Similarly, Io's atmosphere was found to have a stable average density on the dayside, although it exhibits lateral (longitudinal and latitudinal) and temporal (both diurnal and seasonal) variations. There is potential positive feedback in the Io torus supply: collisions of torus plasma with atmospheric neutrals are probably a significant loss process, which increases with torus density. The stability of the torus environment may be maintained by limiting mechanisms of either torus supply from Io or the loss from the torus by centrifugal interchange in the middle magnetosphere.

Various observations suggest that occasionally (roughly 1 to 2 detections per decade) the plasma torus undergoes major transient changes over a period of several weeks, apparently overcoming possible stabilizing mechanisms. Such events (as well as more frequent minor changes) are commonly explained by some kind of change in volcanic activity that triggers a chain of reactions which modify the plasma torus state via a net change in supply of new mass. However, it remains unknown what kind of volcanic event (if any) can trigger events in torus and magnetosphere, whether Io's atmosphere undergoes a general change before or during such events, and what processes could enable such a change in the otherwise stable torus. Alternative explanations, which are not invoking volcanic activity, have not been put forward.

We review the current knowledge on Io's volcanic activity, atmosphere, and the magnetospheric neutral and plasma environment and their roles in mass transfer from Io to the plasma torus and magnetosphere. We provide an overview of the recorded events of transient changes in the torus, address several contradictions and inconsistencies, and point out gaps in our current understanding. Lastly, we provide a list of relevant terms and their definitions.

# 1. Introduction

## 1.1 Objective of this review

Io, the most volcanically active body in our solar system, plays a key role in the magnetospheric system of Jupiter, our largest planet with the strongest planetary magnetic field. With a radius of 1821 km (=1 $R_{Io}$), Io is embedded in the Io plasma torus and orbits Jupiter at a distance of 421700 km (or 5.9 Jupiter radii, $R_J$, 1 $R_J$= 71492 km) with a period of 42 hours and 28 minutes. The influence of Io on the huge magnetospheric system is manifold, but the supply of material to the magnetosphere most significantly affects the dynamics of the whole system.

This is a review of the current understanding of mass transfer from Jupiter's moon Io to the Io plasma torus, the magnetospheric plasma sheet and to regions beyond the magnetosphere. Our goal is to clarify the connections in the Io-Jupiter system and the limitations on the exchange of mass between Io's surface and atmosphere and the magnetospheric environment.

The understanding of Io's role in the Jupiter system has changed quickly and significantly between 1970 and today. Partly because of the rapid development, some misconceptions exist today. The primary example of such misconceptions is that an eruption at a volcanic site on Io is capable of directly and immediately triggering changes in the magnetosphere of Jupiter. This is not known to be the case, and most eruptions likely do not affect the magnetosphere at all. Yet, there are many different phenomena that are possibly interconnected in the system: the interior and surface on Io, Io's atmosphere, neutral gas in the magnetosphere, plasma and dust in the magnetosphere and beyond, as well as even dust dynamics in the magnetosphere and auroral processes in Jupiter's upper atmosphere. This means also that most readers will have expertise in some of these aspects but likely not know much about other aspects. Key for making the right connections and drawing correct conclusions is yet to know enough about all involved processes and the observations and measurements thereof.

## 1.2 Structure of this review

We structured this review such that readers can select and jump to specific sections that are most relevant for their purposes or interests. This means Sections 2-5 do not strictly build on another and can be read separately. We will briefly introduce each section and its purpose in the following.

Section 2 provides an overview on how the understanding of Io's role has developed and presents a selection of key publications that have essentially coined the current comprehension. This section provides insights into how different the perception was at different times given the available information. This might be particularly helpful to put other publications from different decades into perspective or for learning where different views today originated from.

Section 3 reviews all the relevant parts of the system separately; namely Io's volcanic activity (3.1), the bound atmosphere (3.2), escape from the atmosphere (3.3), the electrodynamic interaction with the plasma environment (3.4), Io's neutral gas environment and clouds (3.5), the plasma torus and sheet (3.6), Jupiter's aurora (3.7) and the dust in the system (3.8). In contrast to other reviews, the focus in the subsections here is on the relation to the mass transfer from Io to Jupiter's magnetosphere.

Section 4 provides an overview on the connections in the system as we understand it today as well as on the transient events in the different parts. In Section 4.1 and 4.2, we discuss the understanding of the stable conditions, based on the current knowledge on the different parts presented in Section 3. In Section 4.3 we present an overview on transient events that were reported in different parts of the magnetosphere and that are commonly interpreted to be triggered by Io. After that, in Section 4.4, we highlight caveats and gaps in our understanding concerning the connections in the system.

Section 5 discusses prospects for future observations from both telescopes and planetary missions, as well as modeling efforts that may help to improve our understanding of the supply of mass from Io, the environment and its short-term variability.

This review focuses on aspects related to the topic of Io as a source for the plasma torus. For a comprehensive review of all aspects around Io, we refer the reader to the recently published book "Io: A New View of Jupiter's Moon" (eds. Lopes, de Kleer, and Keane, 2023).

# 2. Development of the understanding of Io's role in the system

## 2.1 Io as the main source of mass for the magnetosphere

In the 1970s, the two Pioneer and the two Voyager spacecraft enabled many important discoveries, but ground-based observations also played a key role. In the 1960's and early 70's, Io was considered an airless body (e.g., with electrically conductive surface material to explain electromagnetic coupling to Jupiter; Goldreich and Lynden-Bell, 1969) in a comparably low density ($<10^2$ $cm^{-3}$) hydrogen-dominated magnetospheric environment, populated by ion outflow from the upper atmosphere of Jupiter (e.g., Goertz, 1973). A series of discoveries changed this view: radio occultations by Pioneer 10 revealed an ionosphere at Io, suggesting the presence of a substantial atmosphere (Kliore et al., 1974). Optical emissions from sodium and potassium were detected using ground-based telescopes with signal peaks near Io (Brown & Chaffee 1974, Trafton et al. 1974, Trafton 1975). Two years later, sulfur ion optical emissions from Jupiter's magnetosphere were detected (Kupo et al., 1976). Shortly thereafter these sulfur ions were suggested to be sourced from Io based on modeling of the electron-excited sulfur ion emissions (Brown, 1976).

The next leap forward came from the Voyager 1 and later Voyager 2 flybys. In December 1978, prior to arrival at Jupiter, the Voyager Ultraviolet Spectrometers (UVS) detected ultraviolet (UV) emissions that revealed the five major ions of sulfur and oxygen, with a luminosity 50 times brighter than one would have inferred from the Brown (1976) planetary nebula analysis of sulfur ions in orbit at Io. Dust plumes from active volcanoes were seen for the first time at Io in high-phase-angle images (Morabito et al., 1979), infrared observations identified $SO_2$ gas over Loki (Pearl et al., 1979), in-situ plasma measurements revealed five heavy ion species of sulfur and oxygen in the environment (Sullivan and Bagenal, 1979), and UV emissions from these ions allowed for the first time the identification of a torus-shaped plasma nebula (Broadfoot et al., 1979). These findings were all consistent with a dense plasma environment linked to Io's volcanic activity with peak ion densities $>10^3$ $cm^{-3}$ near the orbit of Io.

Several estimations for the mass provided by Io to the magnetosphere were put forward based on the new results: Broadfoot et al. (1979) used the UV power emitted by the torus to derive a value of $7\times10^{29}$ amu/s (or 1.2 tons/s) of fresh, slow ions supplied to the torus and accelerated to corotation; Hill et al. (1979) estimated an outward transport and thus mass loading of $\sim10^{30}$ amu/s of ions (or 1.7 tons/s) from his analytical model to explain the radial distance where corotation breakdown occurs; and Dessler (1980) found a similar value of ~1 ton/s using modeling for various observed phenomena such as the Jovian aurora. After these findings, the understanding of the system had changed as summarized by Dessler (1980): “We now know from direct, in situ measurements that Io is the plasma source and that plasma input from the Jovian ionosphere and/or the solar wind amount to less than 1% of the total ionic mass.” The value of about 1 tons/s of mass sourced from Io into the magnetosphere was never severely challenged in later studies but instead reached canonical status. It is still considered the average rate at which neutrals are ionized becoming part of the plasma torus, often termed “neutral source rate” (e.g., Smyth, 1992; Delamere and Bagenal, 2003; Hikida et al., 2020; Bagenal and Dols, 2020), but also used for actually different rates of mass transfer in the system (see definitions in Appendix).

## 2.2 Stability of the Io torus system

The available data from the Voyager 1 and 2 flybys and continued ground-based observations, together with newly developed models, allowed more detailed characterization of the distribution, velocity and energy of the neutral and plasma environments. It was found that the loss processes are likely driven by the interaction of Io’s atmosphere and possibly surface with the corotating plasma that overtakes the moon at a relative speed of 57 km/s with a synodic period of 13 h (e.g., Schneider et al. 1989). This suggests a positive feedback because increased loss from Io would enhance the plasma density in the torus, which in turn should enhance the loss rate through increased collisions between torus plasma and the atmosphere. However, the torus was found to be stable, evidenced mostly through neutral sodium cloud observations, which was the only part of the system that could be observed well from Earth at the time (e.g., Thomas, 1993). Therefore, a mechanism is required to limit the potential positive feedback. Schneider al. (1989) discusses different possibilities, including non-linear (exponential) loss of torus material with increasing torus density (Figure 1a), non-linear (e.g., logarithmic) supply to the torus from Io (Figure 1b) or linear dependencies but a steeper slope for the loss (Figure 1c). Several later studies suggest an increase in net radial transport in the torus with increasing torus density, thus supporting the supply-limiting hypothesis (e.g., Yang et al., 1994; Delamere et al., 2004; Hill, 2006; Hikida et al., 2020).

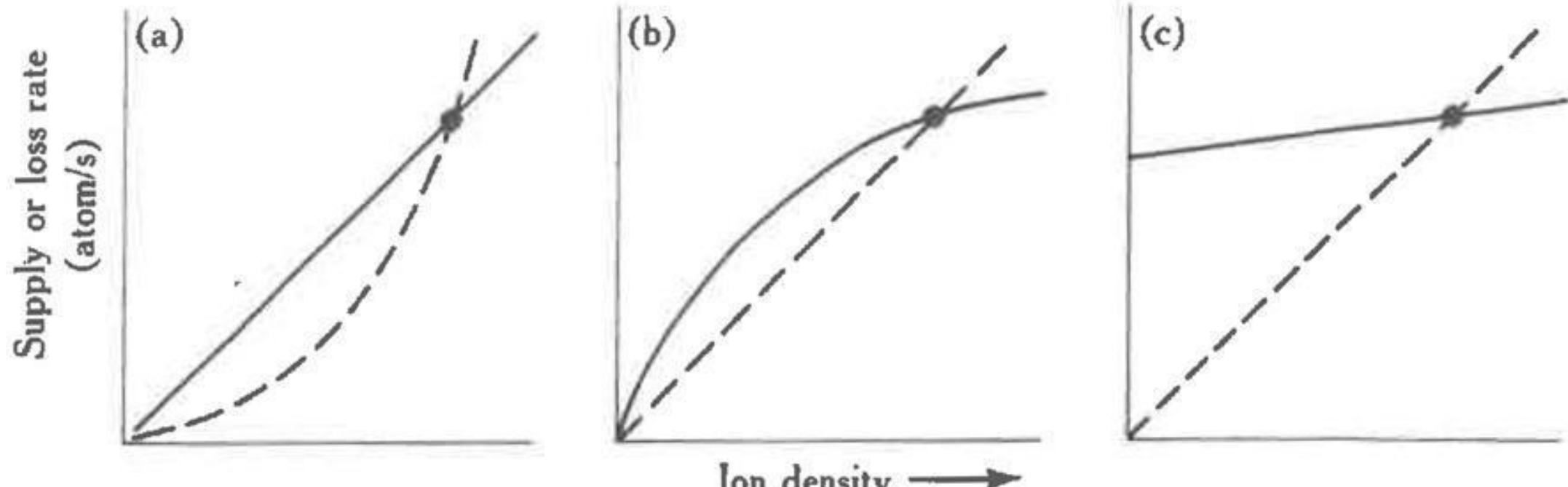

***Figure 1.*** *Different scenarios for a stable plasma torus based on the curves of supply to the torus (solid) and of loss from the torus (dashed) as a function of torus ion density (Schneider et al., 1989). Equilibrium points are reached where the lines cross (black dot). All shown scenarios lead to a stable torus at some plasma (ion) density.*

Overall, the torus and cloud system was considered stable in the early to mid 1990s (Thomas, 1993; Moos et al., 1985). Long-term changes (on scales of decades) were proposed in later studies (e.g., Delamere and Bagenal, 2003, Delamere et al. 2004, Smyth et al. 2011) based on differences between the different epochs of the Pioneer and Voyager measurements (Broadfoot et al., 1979). The seasonal modulation of Io's $SO_2$ column with distance to the Sun was also not discovered until later (Tsang et al., 2013), and no connection to long term modulations in the torus or neutral cloud density has yet been established. By the mid-1990s, there were still no clear hints for changes on shorter time scales. The only part of the system that could be observed (and frequently was) was the sodium cloud, which at the time did not reveal any obvious changes (Schneider et al., 1989; Flynn et al., 1994). Indications of transient events in the torus were only reported later and are discussed in the next section (2.3).

## 2.3 Hypothesized volcanic mass supply events

Strong enhancements in thermal emissions from Io have been observed occasionally since 1978 and were dubbed 'outbursts' (see review by Spencer and Schneider, 1996). Such outbursts are now understood to represent extremely high effusion rates of high-temperature lava, often accompanied by large plumes of gas and dust (Davies, 1996). However, aperiodic or transient major changes in the environment or atmosphere had not been observed prior to 1996. Spencer and Schneider (1996) only speculate in their review: "As we improve our sensitivity to volcanic emissions, atmospheric abundances, and torus densities, we may identify the ways in which volcanoes modulate the Jovian system." Two publications then essentially coined the idea that a volcanic event (like those seen in thermal outbursts) can trigger a transient change in the environment.

In the first of these, Brown and Bouchez (1997) detected a 4-fold increase in emissions from the sodium cloud followed by a 30% increase in sulfur ion emissions (Figure 2, left). The transient change lasted for about 70 days. The sodium was seen as an indicator for a change in mass supply from Io, which was assumed to be triggered by a volcanic outburst. The sulfur emissions reflect the state of the plasma torus, and the temporal curves were interpreted to be consistent with a loss-limiting scenario (Figure 1a).

The second key publication, Mendillo et al. (2004), reported long-term monitoring of the hot spot thermal emissions and the sodium nebula emission and found a putative correlation between the two parts (Figure 2, right), with the Loki volcano presumably playing a key role for the thermal increases. The authors interpret the results (although based on relatively few data with apparently some additional variability) as evidence for volcanic activity controlling the abundance of trace gas sodium. They, however, refrain from drawing conclusions about the effects on the bulk species (S and O) based on the Na observations.

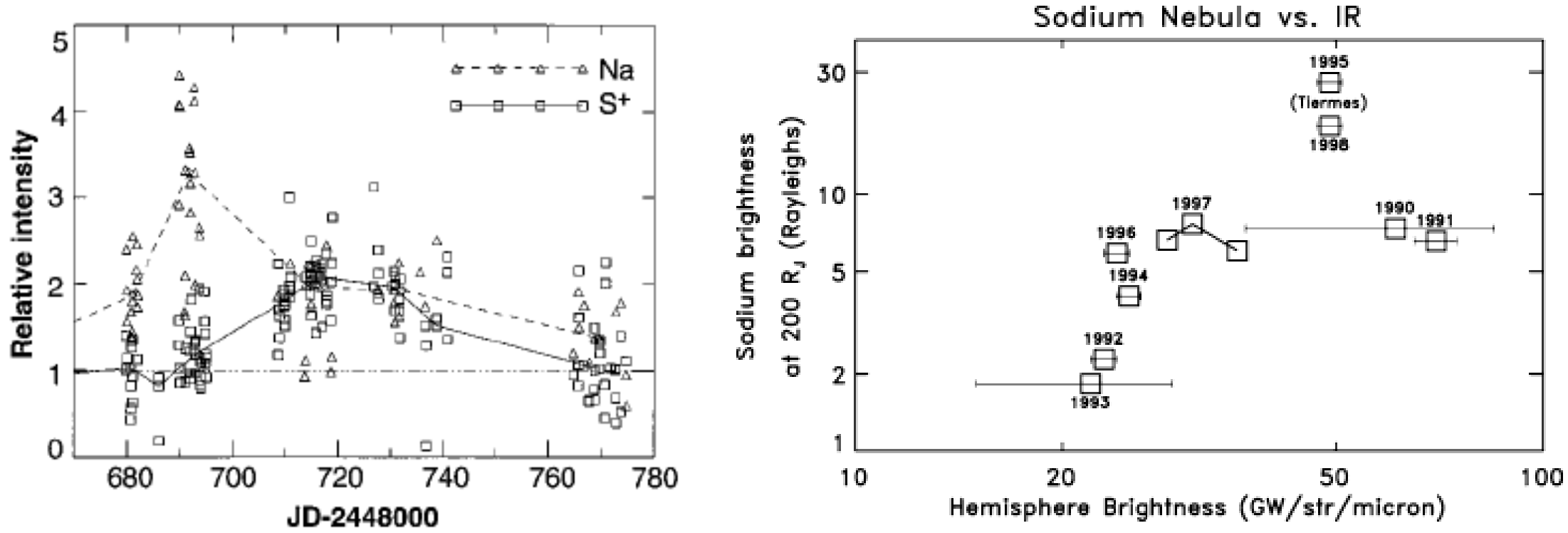

***Figure 2.*** *(Left) Transient enhancement of the sodium neutral cloud and sulfur torus ion emissions from Brown and Bouchez (1997), interpreted as evidence for a change in the torus triggered by a volcanic outburst. (Right) Comparison of the brightness of the wide sodium nebula and emitted thermal power revealing a (putative) correlation (Mendillo et al., 2004).*

After the studies of Brown and Bouchez (1997) and Mendillo et al. (2004), the narrative of a volcano-triggered transient change in the torus and magnetosphere (on time scales of weeks) was coined and many studies built upon these results to interpret their findings of transient events (e.g., Delamere et al., 2004; Yoneda et al., 2009; Bonfond et al., 2012).

A transient change in the torus in 2015, with a similar time scale of weeks, is the most well documented event to date thanks to the nearly continuous ultraviolet observations of the Hisaki satellite (Section 3.6). In addition to the observed transient change in the sulfur and oxygen ion emissions, Hisaki also measured for the first time an increase in the *neutral* oxygen emissions from Io's orbital environment, simultaneously or marginally preceding the changes in *ion* emissions (Koga et al., 2018b; 2019). Enhancement in the sodium emissions (scattered sunlight) showed a common temporal envelope to that in the oxygen cloud (Yoneda et al., 2015). This affirms that the neutral cloud density was indeed elevated, as opposed to a brightening merely caused by increased torus electron impacts with neutral oxygen atoms. Some observations and modeling work had by this time indicated that the supply to the plasma torus comes from ionization of neutrals that had earlier escaped Io's gravity, forming a cloud or complete orbital torus (Durrance et al., 1983; Skinner and Durrance, 1986; Bagenal et al., 1997; Saur et al., 2003). Whether the torus is sourced from these large scale neutral clouds orbiting Jupiter or by the localized ion pick-up at Io itself has been a major outstanding question (e.g., Thomas et al., 2004). The observation of a transient change in the neutral oxygen emission is consistent with the former: changes in the plasma torus are caused by (and preceded by) a change in neutral gas loss from Io to the larger scale neutral clouds (Koga et al., 2019). This 2015 event is discussed in detail in Section 3.6.

The change in total mass supply from Io that was inferred via modeling from the observed torus emissions is on the order of a factor 2 and can be up to 10 (e.g., Delamere et al., 2004; Koga et al., 2019; Hikida et al., 2020). Such changes are, however, difficult to reconcile with the current understanding of the exchange of mass (volatiles) between Io and its orbital environment, as we will discuss in detail in Section 4.4. Furthermore, many of the often assumed correlations and connections between different parts of the system (like hot spots and volcanic plumes, or plume activity and mass supply to the torus) are unclear and not fully understood today.

# 3. Review of the relevant components of the Io-Jupiter system

## 3.1. Volcanic activity: hot spots and plumes

The material that constitutes Io's atmosphere, and that is ultimately lost from it, originates as molecules outgassed via volcanic activity. However, the path that gas takes from the moment it emerges from a volcanic vent to the time and place where it is lost from the upper atmosphere is far from clear. Moreover, while lava effusion should be accompanied by gas emission, and indeed all plumes observed by spacecraft are associated with thermal anomalies when observed with sufficient sensitivity, the association between lava flows and gas emission is complex. Ground-based and Earth-bound remote observations, which are sensitive to the largest lava flows and gas plumes, find that the brightest hot spot thermal activity is frequently not associated with the largest gas plumes, and vice versa (e.g. de Pater et al. 2020a;b; and see Section 3.1.3). In this section, we review Io's hot spot and plume activity, and we discuss what has been observed of the connections between hot spot activity and Io's atmosphere, plumes, and mass loss. Note that "hot spot activity" here refers to the detection of enhanced thermal emission at the surface arising from volcanism; there may also be undetected subsurface thermal anomalies.

Broader reviews of Io's hot spot and plume activity can be found in de Kleer and Rathbun (2023), de Pater et al. (2023), Williams and Howell (2007), and Geissler and Goldstein (2007) among others.

### 3.1.1 Io's hot spot activity

Thermal emission from Io's hot spots was seen by the *Voyager* spacecraft during their flybys of the Jupiter system in 1979 (Pearl and Sinton, 1982) and even before (Witteborn et al., 1979). When it can be localized, the thermal emission is associated with surface features seen in optical imagery, and the thermal behavior combined with the geological context indicate the most plausible style of volcanism at each site (e.g., Davies, 1996). Io hosts over 400 active volcanoes (Radebaugh et al., 2001; Williams et al., 2011) with over 250 of them active recently enough to still be producing thermal emission (Veeder et al., 2015; Cantrall et al. 2018). These occur predominantly in the form of lava lakes (Mura et al., 2024; Lopes et al., 2004; Radebaugh et al. 2004) and effusive lava flows, with occasional dramatic lava fountaining events (Keszthelyi et al., 2001). These volcanic types are analogous to the expressions of effusive volcanism observed on Earth, albeit much larger in scale (Davies et al., 2001; Davies, 2007). Thermal and visible observations suggest that Io's magmas are of high-temperature mafic or ultramafic compositions (McEwen et al., 1998); such low-viscosity lava does not commonly produce explosive eruptions on Earth (although the difference in atmospheric pressure between Earth and Io also affects the explosivity of eruptions). Additionally, the enormous effusion rates of some of Io's eruptions are not observed anywhere on Earth today.

Io's hot spots are spatially distributed over all regions of Io and exhibit a high degree of temporal variability, which can be used to search for connections between thermal volcanic activity and gas input into the atmosphere or even the Jovian environment. Io's hot spots are classified as either persistent or transient. Persistent hot spots exhibit thermal emission consistently and typically do not exhibit large-scale transient events, whereas transient hot spots are not consistently active but do exhibit large-scale transient events (e.g., Lopes-Gautier et al., 1999; de Kleer and de Pater, 2016a). Events of very strong and usually transient thermal emission were dubbed 'outbursts' in the literature (e.g., Veeder et al., 2012). They have an eruptive power that is more than an order of magnitude higher than Io's typical volcanic hot spot. A handful of sites (e.g., Loki Patera, Pillan Patera and Pele) are known to fall into both categories (both persistent and transient hot spot activity); observations are limited so this may in fact be more common.

If bright, transient thermal events occur because of pressure build-up in a subsurface magma system leading to eruption, then transient volcanoes may be expected to produce more gas than persistent volcanoes. At more persistently active sites, magma may be either already degassed or may produce plumes by simply sublimating $SO_2$ at a slow but steady rate as lava fronts advance across the frost patches.

The relation between lava lakes and plumes is not straightforward: intuitively, a passively-cooling lava lake is unlikely to produce a gas plume, but an active lava lake connected to a deep magma reservoir could. In practice, large plumes have indeed been seen associated with lava lakes, for example at Pele (Davies et al., 2001).

If volcanic gases are lost from the atmosphere close to where they are emitted from volcanoes, then certain volcanoes, or even eruptions that take place at certain times of the day, may preferentially contribute to mass loss as the effects of the plasma on the atmosphere varies with surface location and time of day (see Section 3.4).

Between 2013 and 2022, detected bright transient thermal events occurred preferentially on Io's trailing hemisphere (de Kleer et al., 2019; Tate et al., 2023). The (sparser) data prior to 2012 do not show this preference as clearly (Tate et al., 2023), but the distribution of large red plume deposits associated with bright transient events do follow this same distribution (McEwen and Soderblom, 1983).

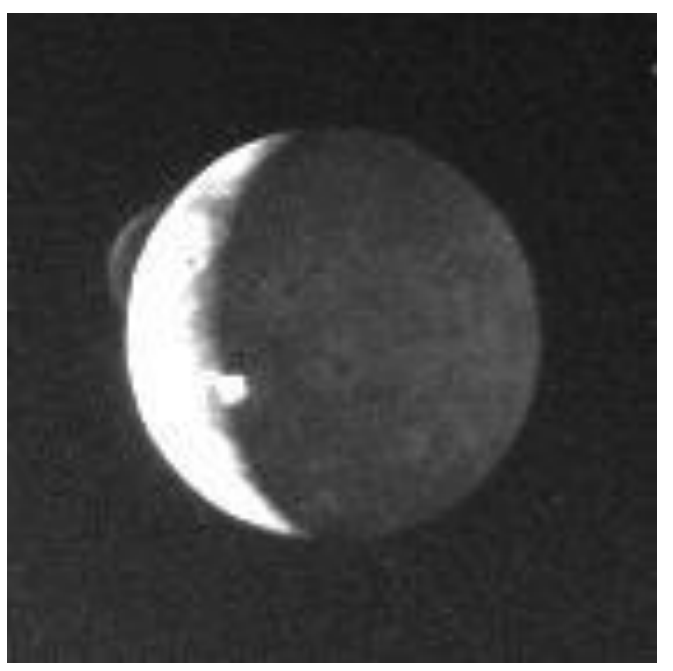

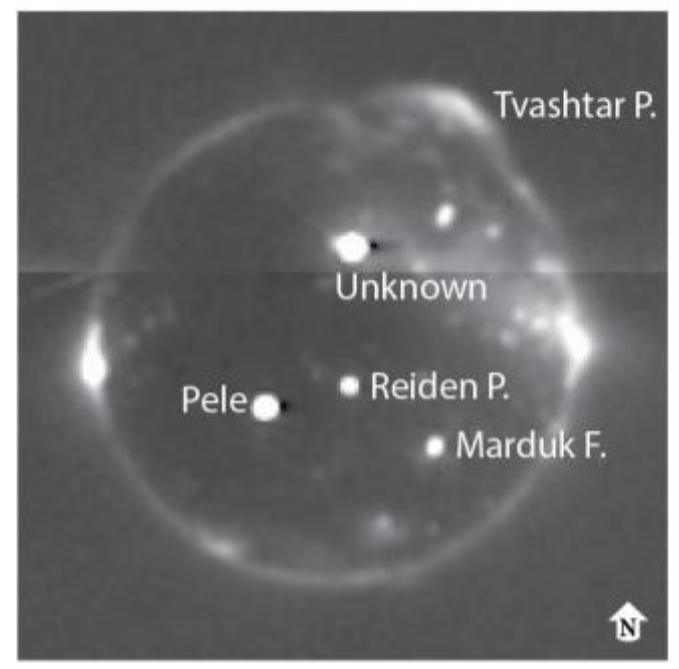

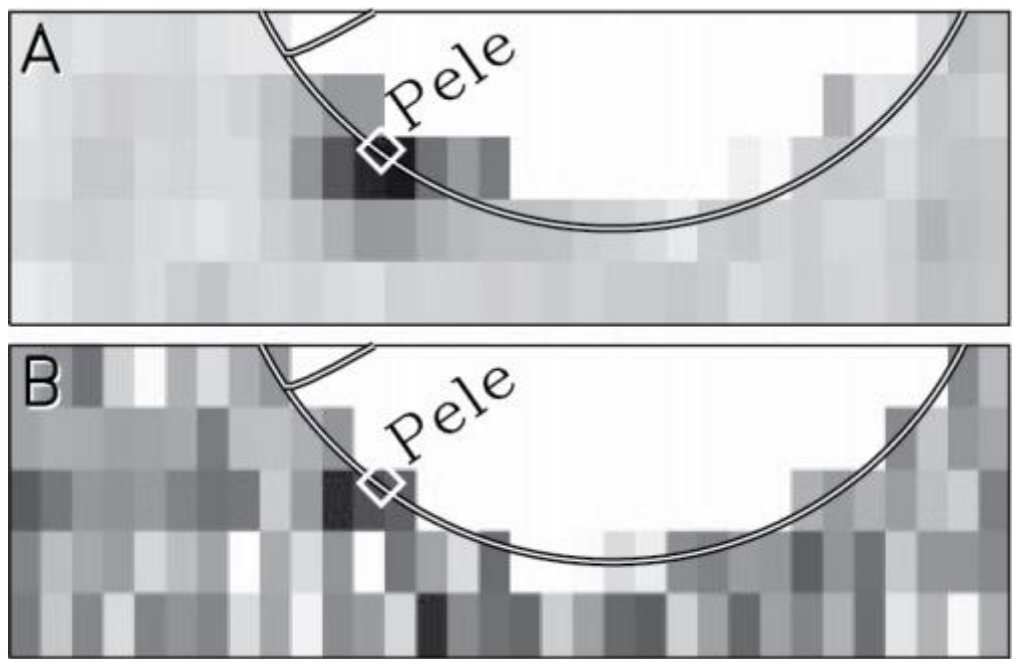

***Figure 3**. Left: Voyager volcanism discovery image through scattering by plume dust: Pele on the sunlight left, Loki at the terminator (NASA PIA00379). Middle: Visible image of Io in eclipse from the New Horizons spacecraft showing emissions from both hot spots (bright and round) and excited gases above volcanic sites like the plume of Tvastar above the north pole and from the global atmosphere as equatorial spots on the left and right. Right: $S_2$ and $SO_2$ plume gas absorption measurements by the Hubble Space Telescope (Spencer et al., 2000).*

### 3.1.2. Dust and gas plumes

Volcanic plumes on Io are mostly observed through visible light scattering by entrained dust particulates with grain sizes estimated to be in the range of tens to hundreds of nanometers (Figure 3 left; Geissler and Goldstein, 2007; Geissler and McMillan, 2008). Plume gases can be identified in eclipse observations, when localized electron-impact excited (auroral-like) emissions are seen above volcanic sites in spacecraft images (Figure 3 middle; Geissler et al., 2004; 2007; Spencer et al., 2007; Roth et al., 2011). In addition, transit spectroscopy of Pele's plume on Io's limb against Jupiter by the Hubble Space Telescope (HST) allowed measurements of $S_2$ and $SO_2$ plume-only gases (Figure 3 right; Spencer et al., 2000; Jessup et al. 2007). Signals from other molecular species, which can only be explained to be produced in active plume outgassing, are also observed (Section 3.1.3). Dust to gas ratios in the range of $10^{-1}$ to $10^{-2}$ were inferred from HST images of dust reflections and gas absorptions in large plumes (Jessup and Spencer, 2012). Detection of scattered light by dust grains, and of emissions or absorptions from plume gas components, are typically not possible within the same type of observation, preventing clear constraints on the gas to dust ratio. The observed plumes mirror the dichotomy seen in the surface thermal emissions (Section 3.1.1), with long-lived small "Prometheus-type" plumes and short-lived large "Pele-type" plumes (McEwen and Soderblom, 1983). Pele's plume now appears to have been long-lived but with a short phase in which it was easily visible.

The Pele-type high energy plumes rise a few hundred kilometers and are surrounded by large reddish deposition rings consisting of $SO_2$ ice/frost, sulfur allotropes and metastable polymorphs of elemental sulfur (Moses and Nash, 1991; Carlson et al., 2007). Within the main red ring, which corresponds to the visible extent of the aloft particulates (or grains), are generally other sprays/deposits probably consisting of larger refractory particulates and $SO_2$ frost (McDoniel et al., 2015). Simulations suggest that Pele-type plumes are predominantly gas (1 to 10% mass-loaded by micron-scale or less particulates; Jessup et al., 2007) and their canopy tops extend well above the local daytime exobase altitude. They likely arise directly from hot magma (seen in the Pele and Tvashtar calderas) and extend to a height corresponding to inferred lava temperatures of 1200 to 1400 K.

In contrast, Prometheus-class plumes are lower energy and typically extend to less than 100 kilometers as seen in visible wavelengths. They may be more heavily particulate mass loaded, and probably arise from an interaction of surface lava flows with pre-existing volatile frost/ice deposits (Milazzo et al., 2001). The apparent origin of the Prometheus plume itself shifted roughly 80 km between Voyager and Galileo observations (Kieffer et al. 2000), presumably as a lava front advanced, but the plume has been observed to persist to the present as seen in the recent Juno images[1].

Different indications of possible wide-spread outgassing, which could not be directly observed, led to the suggestion of a third class of plumes dubbed *stealth plumes*, which contain gas but very few or no particulates such that they remain undetected in scattered light images (Johnson et al., 1995). Gas emissions from a possible *stealth plume* may have been observed in an eclipse image by New Horizons above the East Girru hot spot, which had no associated dust plume in sunlight (Spencer et al., 2007). De Pater et al. (2020a) propose that the SO emissions observed with the Keck telescope are likely caused by a large number of such stealth plumes.

### 3.1.3 Linking hot spot activity to outgassing at plumes

Io's bulk $SO_2$ atmosphere is relatively stable over timescales of months to years (e.g. Tsang et al. 2013, Giles et al. 2024), even though one of its major sources is volcanic outgassing, which varies stochastically and over much shorter timescales. In addition, there is some evidence that the bulk atmosphere is roughly uniformly distributed between northern and southern mid-latitudes in daylight (Section 3.2.). However, certain species and excited states show emission from localized regions and/or high-temperature gases; these localized regions are assumed to represent volcanic plumes but are not understood.

---

[1] https://www.nasa.gov/missions/juno/nasas-juno-to-get-close-look-at-jupiters-volcanic-moon-io-on-dec-30

An example of such an excited gas is SO, which could be of a volcanic origin. The forbidden $a^1\Delta \rightarrow X^3\Sigma^-$ band complex of SO at 1.7 μm was first detected in 1999 (de Pater et al., 2002); the gas was suggested to being initially at high-temperature (~1500 K) in thermodynamic equilibrium and then cooling adiabatically. Based on the gas temperature and state, the emission was attributed to SO ejected directly from volcanic vents. Later spatially resolved observations found a general lack of correspondence between the locations of SO emission and hot spot thermal emission, suggestive of stealth plumes of SO that are unaccompanied by large-volume lava extrusion (de Pater et al., 2020a). Observations by the James Webb Space Telescope have now finally detected this emission line complex above an active volcanic center (de Pater et al., 2023; Figure 4), supporting a volcanic origin for the excited SO, even though most of the SO plumes are not associated with thermal emission detectable from Earth.

—-

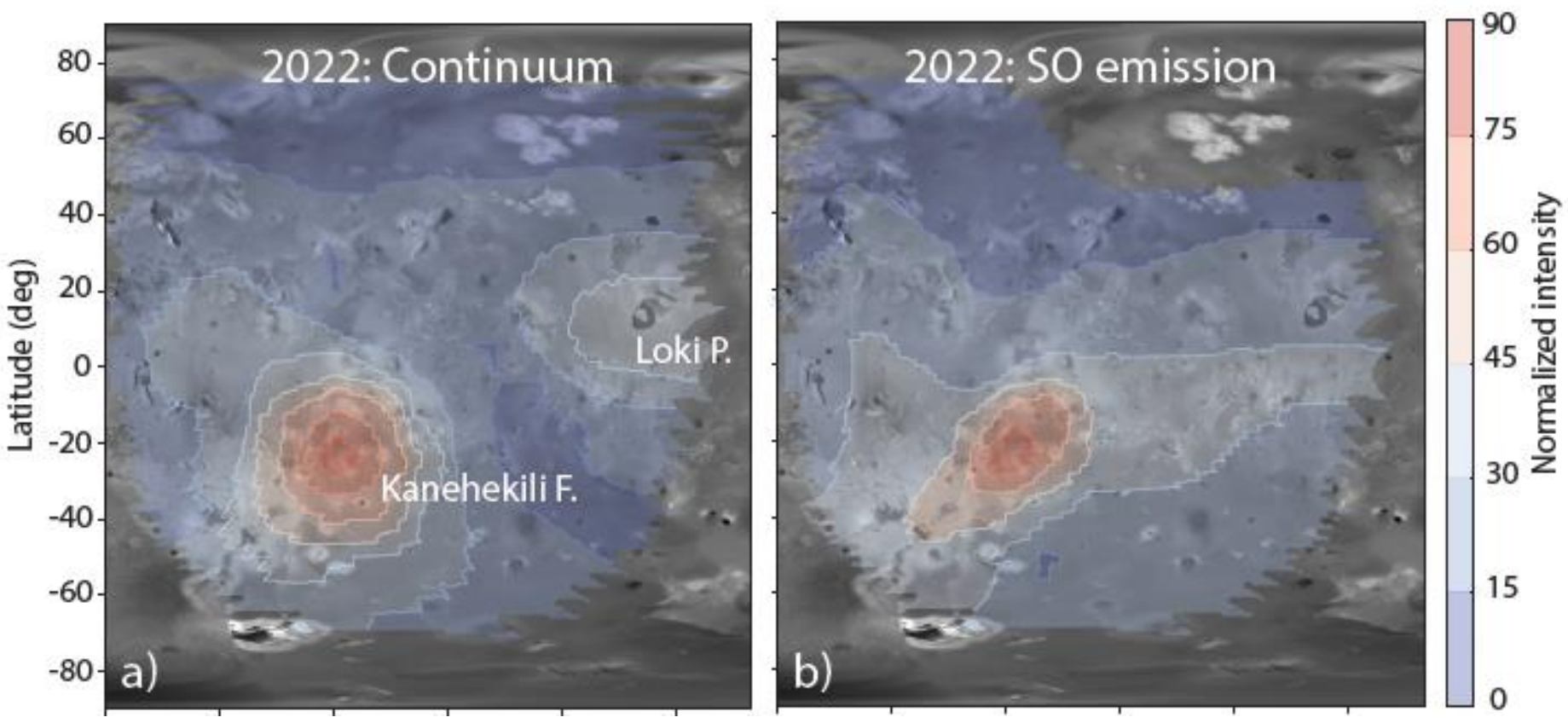

***Figure 4****.Thermal continuum (left) and SO gas emission (right) observed by JWST at 1.7 μm (de Pater et al., 2023). The SO gas emission is localized to Kanehekili Fluctus, which was producing significant thermal emission at the time of observation. Prior detections of the same SO band did not find a clear correlation between SO emissions and active thermal hot spots.*

The case appears to be similar for NaCl and KCl gas. These gases should have at most a lifetime of a few hours in the atmosphere of Io (Moses et al., 2002). They are detected only in localized regions (Redwing et al., 2022; de Kleer et al. 2024) via rotational transitions at millimeter wavelengths. The gas temperature based on both line widths and molecular state populations is much higher than that of $SO_2$ in the same observations. For all practical purposes, NaCl and KCl are solids below 1000 K (Chickos and Acree, 2003). However, when the observations reveal a localized distribution, the NaCl and KCl gases are not spatially-correlated with highly (currently) active hot spot emissions, and simultaneous ALMA sub-mm gas and Keck near-infrared thermal imagery show a lack of spatial correspondence (Figure 5). Thus, if the alkali gases are tracers of plumes, then such plumes are not typically associated with thermal emission at a magnitude that can be seen from Earth. Equally intriguing is the absence of a spatial correlation between NaCl/KCl and $SO_2$ gases, discussed in Section 3.2.2.

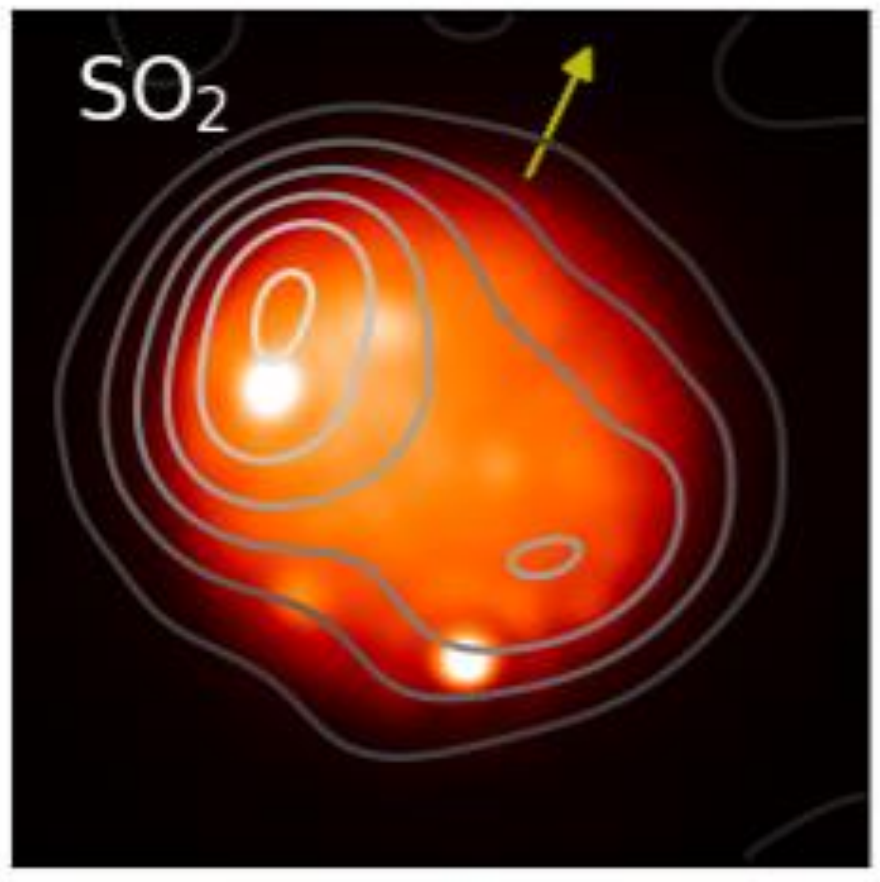

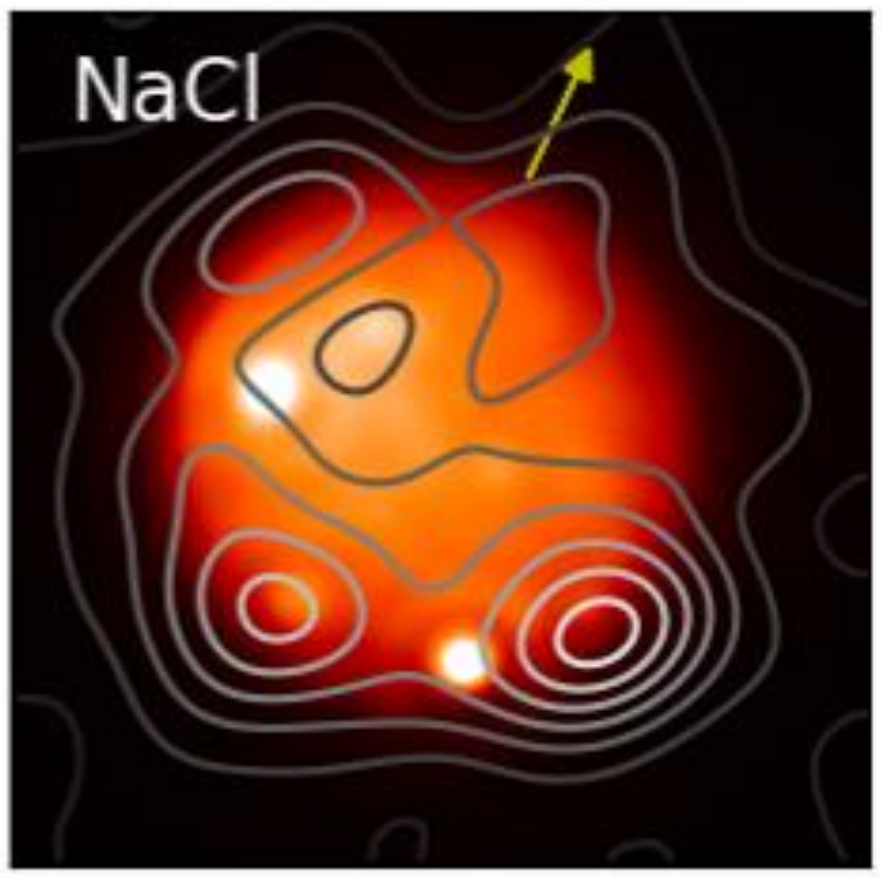

***Figure 5. Simultaneous volcanic thermal emission and gas emission observations.*** *3.8-μm image of Io on UT 2022 May 24 ~15 UT from Keck/NIRC2, with contours overlain for $SO_2$ and NaCl gas distributions from simultaneous ALMA observations (at 430.194 and 428.519 GHz for the two molecules respectively), showing the lack of spatial correspondence between the NaCl gas (presumed to be a tracer of plumes) and active hot spots. The southern hot spot that shows the closest spatial alignment with enhanced NaCl emission is around 49S 106W, where an unnamed patera P197 is located. White represents the peak in the thermal emission, and white contours are the maximum gas densities. The arrow indicates the direction of Io's north pole. ALMA data from de Kleer et al. (2024); Keck data:* [*https://www2.keck.hawaii.edu/inst/tda/TwilightZone.html*](https://www2.keck.hawaii.edu/inst/tda/TwilightZone.html).

However, there are several cases where volcanic sites could be associated with gas output during their thermal eruptions (hot spot activity). The *New Horizons* spacecraft observed emissions from excited plume gases in eclipse and colocated dust plumes over simultaneously imaged hot spots at several locations including Tvashtar Catena with a large plume (Spencer et al., 2007; Roth et al., 2011). Prominent dust plumes were detected in optical images at locations of thermal emission, for example at the volcanoes Loki and Pele as observed by *Voyager* (Strom et al., 1981). The plume gas abundance at the locations, however, was not constrained simultaneously but instead only studied at other times (e.g., Jessup et al., 2007). The lack of observed correspondence between near-infrared thermal emission and gas emission from high-temperature tracers like SO, NaCl, and KCl or plume-only bulk gases thus remains an area where our understanding is incomplete.

#### 3.1.4 Linking hot spot activity to transient torus events

Transient brightening events observed in the systems of the torus plasma and the neutral clouds and nebulae around Jupiter (Sections 3.5 and 3.6) have long been attributed to gas output from volcanic eruptions on Io, although the observational evidence for this link remains tenuous.

The brightening in $S^+$ torus ion and Na neutral emission observed in 1991 was attributed to a putative volcanic outburst on Io (Brown and Bouchez, 1997), which was not directly identified. Mendillo et al. (2004) investigated the correspondence between 3.5 μm hotspot emission on Io's sub-Jovian hemisphere and annual measurements of Na emission from the extended nebula (out to hundreds of Jovian radii) over the period from 1990 to 1998. They found a correlation between Na nebula brightness and activity at Loki Patera, as well as with a thermal outburst event fortuitously detected at Tiermes Patera during this time. However, both datasets are temporally sparse and would generally not be sensitive to variations on timescales of days to weeks. In addition, the hot spot dataset is only sensitive to the sub-Jovian hemisphere and lacks direct spatial resolution, permitting detection of only the brightest eruptions and only when located on the sub-Jovian hemisphere. More recent datasets with much higher cadence and more comprehensive Io surface coverage do not find such a clear correlation, especially with Loki Patera (de Kleer & de Pater, 2016a; Yoneda et al., 2015). A lack of correlation makes sense with our understanding of Loki Patera's activity, which is frequently attributed to the sinking of crust into a magma sea – not by gaseous plume eruptions (Matson et al., 2006; Rathbun and Spencer, 2006). However, it is unclear whether the apparent correlation observed between 1990 and 1998 (Mendillo et al. 2004) was the result of sparse data, or whether Loki Patera (and potentially other IR-bright volcanoes) changed its behavior between the 1990s and the 2000s.

During the spring of 2015, when neutrals, ions and hot spots were all being observed at a higher cadence than ever before, a major brightening was observed in the Na nebula accompanied by an O and S ion and neutral response (Yoneda et al., 2015; Tsuchiya et al., 2018; Koga et al., 2018b). Hot spot activity at different sites including some brighter events were also observed during spring 2015 (de Kleer and de Pater, 2016a). The 2015 torus brightening has most commonly been associated with a large enhancement in thermal emission from Kurdalagon Patera, though the association is complicated and was mostly based on the temporal correspondence. The Na nebula brightening began before near-infrared emissions at Kurdalagon Patera reached a detectable level and right after a moderate brightening was detected at Mithra Patera (see Figure 6). Moreover, Kurdalagon Patera dimmed significantly in the middle of the Na brightening (when Pillan Patera was the most active hot spot), and brightened again two months later without detectable brightening of the Na. Altogether, it is not clear if and how any of the observed hot spot activity during spring 2015 had a (causal) relationship to the 2015 torus event.

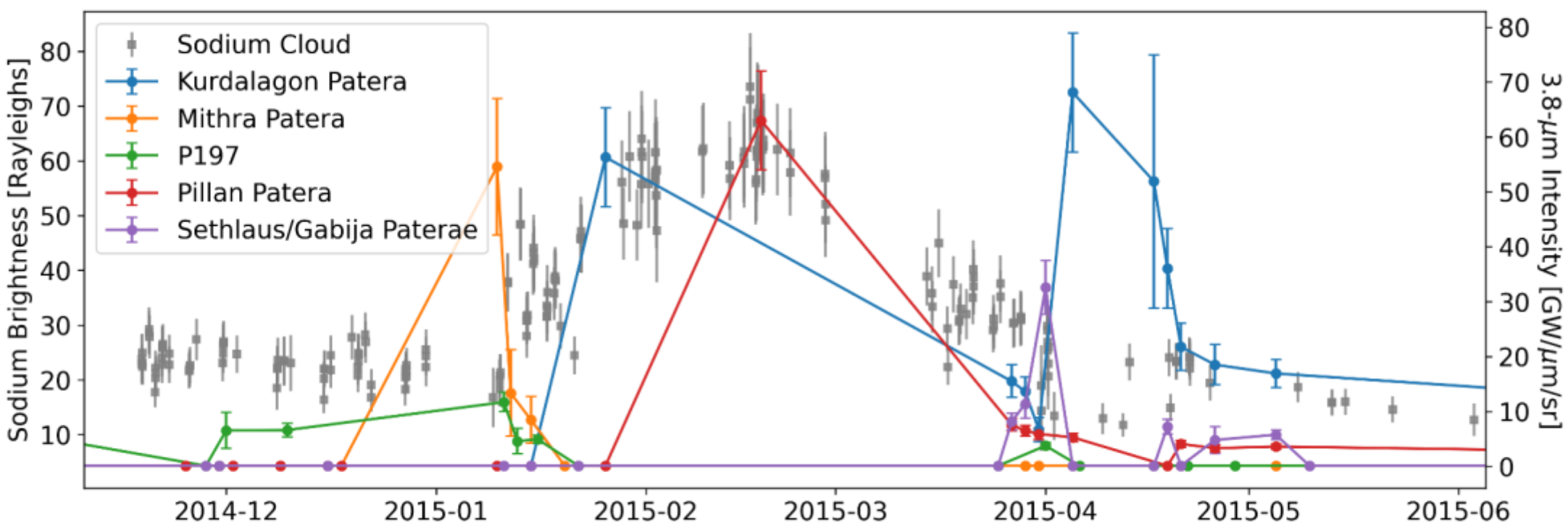

***Figure 6.*** *Timeline of Na nebula emission during the spring of 2015 alongside the thermal emission for several individual volcanoes that could have plausibly contributed. Data from Yoneda et al., 2015; de Kleer and de Pater, 2016a; and de Pater et al., 2016.*

## 3.2 Io's bound atmosphere

Io's atmosphere is unique in several respects. Its composition is dominated by $SO_2$ and contains traces of other volcanic gases, ultimately reflecting its volcanic origin. The dayside atmosphere is in the nanobar pressure range (~$5\times10^{16}$ cm$^{-2}$ gas column density). Its main atmospheric constituent, $SO_2$, is also the dominant surface ice and the range of surface pressures is consistent with expectations based on sublimation equilibrium with the surface. However, Io's atmosphere is "thin" in the sense that sublimation/condensation exchanges are energetically negligible to the surface heating budget. The surface ice temperature is not "buffered" by the atmosphere and can undergo large, insolation-related variations over the globe on timescales from minutes to years, implying large lateral (longitudinal and latitudinal) and temporal variations of pressure and possibly supersonic sublimation winds (Ingersoll et al. 1985). A further unique feature of Io's atmosphere is that it is also directly fed by gases from volcanic plumes, which can interact with the sublimation component and enrich the atmosphere in non-volatile gases (like NaCl or KCl, which have extremely low vapor pressure at Io's temperatures), leading to a presumably non-hydrostatic structure that remains to be characterized.

Although many properties like the vertical structure and global dynamics are still not well characterized, it appears that the average dayside atmosphere is quite stable. The atmospheric state before and during transient events in the torus and neutral clouds is also unknown and thus the role of the atmosphere for these events is not understood. We review the basic characteristics of the atmosphere here. Detailed reviews on Io's atmosphere can be found in de Pater et al. (2023) and Lellouch et al. (2007).

### 3.2.1 Composition

In addition to the major gas $SO_2$, other molecular (SO, $S_2$, NaCl, KCl) and atomic (S, O, Na, Cl) species have been detected. SO and O are expected to be present in significant amounts from photolytic production from $SO_2$ (e.g., Summers and Strobel, 1996). The mixing ratios relative to $SO_2$ are at the 3 to 10% level for SO (Lellouch et al., 1996) and ~10 % for O (Roth et al., 2014). However, SO is also a volcanic gas (Zolotov and Fegley, 1998), and the relative distributions of $SO_2$ and SO mm-wave emissions in sunlight and eclipse (de Pater et al., 2020b) may imply the coexistence of volcanic and sublimation sources. S is a product of SO photolysis and a minor branch of $SO_2$ photolysis, and is also present at ~2% of the abundance of $SO_2$ (Roth et al., 2014). $S_2$ was directly detected on one occasion in transit spectroscopy of Pele's plume on Io's limb against Jupiter, at the level of 8 to 30% of $SO_2$ (Spencer et al., 2000). Atomic Cl is present with a $\sim 5\times10^{-4}$ $Cl/SO_2$ ratio (Retherford, 2003; Feaga et al., 2004). The discovery of chlorine that followed the detection of $Cl^+$ in the Io plasma torus (Küppers and Schneider, 2000) prompted the search for and detection of NaCl (Lellouch et al., 2003) and KCl (Moullet et al., 2013). Their typical abundances relative to $SO_2$ are $\sim 3\times10^{-3}$ and $5\times10^{-4}$, respectively, but these values are derived assuming global $SO_2$ coverage (e.g. Lellouch et al. 2003). NaCl and KCl are the likely sources of the atomic Na and K observed in neutral clouds over many decades. $S_2$, NaCl and KCl have either no solid phase or negligible vapor pressure at Io's surface temperatures (Ewing and Stern, 1974; Chickos and Acree, 2003), so they are most likely of volcanic origin, although surface sputtering may also be a significant source of NaCl and KCl.

### 3.2.2 Horizontal and temporal variability and the volcanic vs sublimation origin

In principle it is possible that all of Io's atmosphere is ultimately of volcanic origin, since the surface frosts that can sustain the atmosphere through sublimation are themselves produced from the accumulation of plume material condensed at the surface. Given that a gas plume can also interact with a "pre-existing" background atmosphere, the distinction between "volcanic" and "sublimation" atmospheres is ultimately somewhat specious. This question can probably be formulated in a slightly more accurate way: what fraction of Io's atmosphere varies in a predictable way with environmental parameters (local time, distance to the Sun, location on the surface); what fraction shows "erratic" variability, associated with volcanic activity; and what are the orders of magnitude of these variations? This issue has been considerably clarified over the last ~20 years, leading to the perhaps unexpected result that Io's atmosphere is generally "predictable," though open questions persist.

*Diurnal vs. longitudinal variability*. Since Io's atmosphere is mostly observed on the dayside, observations typically mix diurnal and geographic variations. However, spatially- and temporally-resolved ultraviolet (UV) spectra indicate that *geographical* variations dominate over *diurnal* variations on the dayside (Jessup & Spencer, 2015; Tsang et al. 2013), even if the latter are still detectable in mid-infrared observations (Lellouch et al., 2015), with the densest atmospheric column above the anti-Jovian hemisphere near longitudes 180 to 220 °W (up to $\sim 2\times10^{17}$ $cm^{-2}$ , e.g., Giles et al. 2024) and a sharp depletion at mid- and high-latitudes. Larger amounts of gas on the anti-Jovian region compared to the sub-Jovian are also evidenced from thermal-IR spectroscopy ($\nu_2$ 19 µm band of $SO_2$, Spencer et al., 2005) and are best interpreted as the effect of the 2h-long eclipses by Jupiter lowering the surface temperature on the sub-Jovian hemisphere (Tsang et al., 2012; Walker et al. 2012).

*Geographical distribution*. Images of Io at Lyman-α, a wavelength at which $SO_2$ gas absorbs, indicate that the atmosphere is mainly confined to latitudes within 30 to 40 °N/S from the equator, with a larger latitudinal extent on the anti-Jovian side, and maximum column densities of $\sim 10^{17}$ $cm^{-2}$ (Figure 7; Feaga et al., 2009; Giono and Roth, 2021). The drop in column density towards the higher latitudes is interpreted as condensation of $SO_2$ towards the poles where surface temperatures are lower. A similar conclusion is reached based on ALMA images of the $SO_2$ mm emission, which appears depleted beyond mid-latitudes (de Pater et al., 2020b). Correlations between the measured $SO_2$ columns in sunlight and volcanic hot spots/plumes are possible but marginal (McGrath et al., 2000; Lellouch et al., 2015; de Pater et al., 2020b).

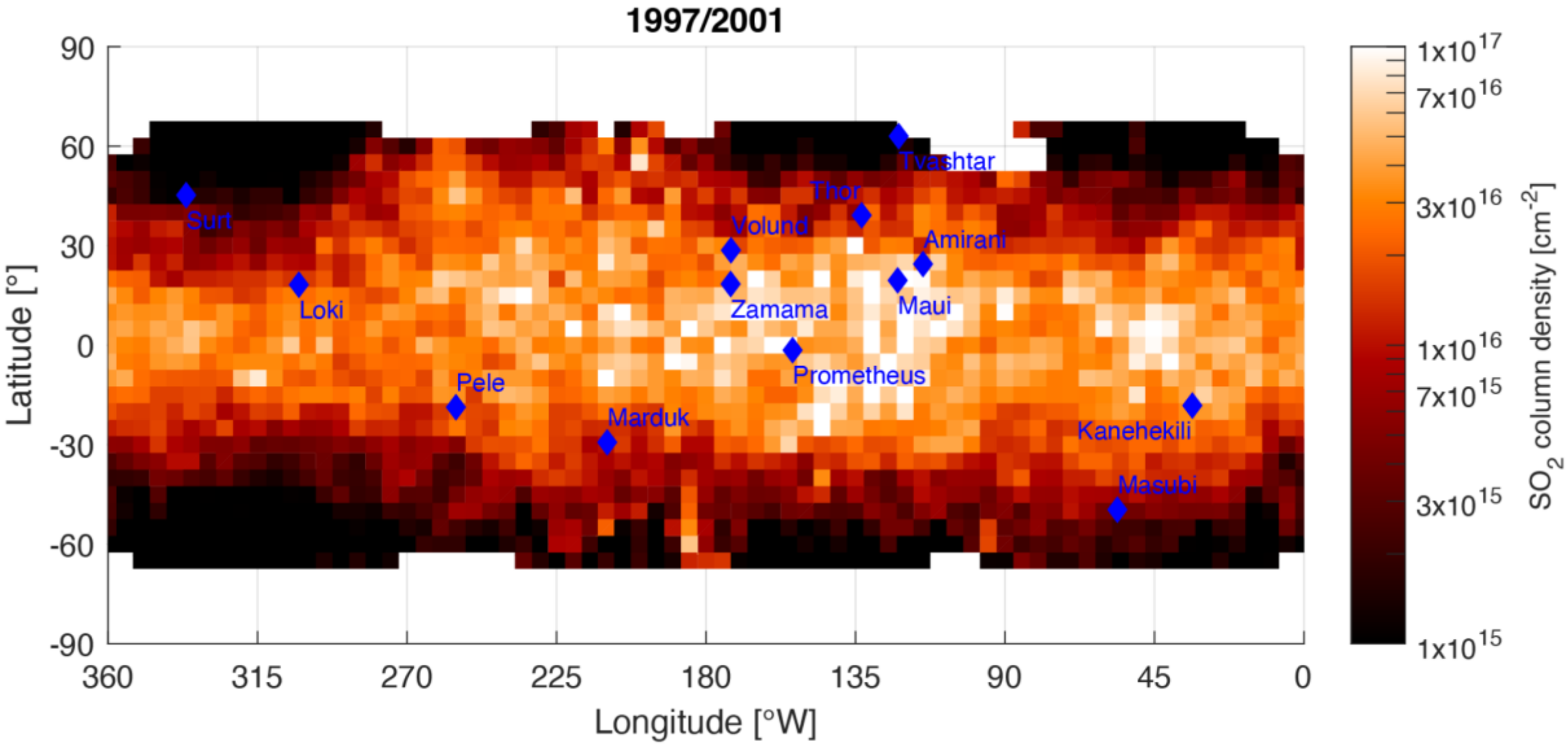

***Figure 7.*** *$SO_2$ column density map inferred from several Lyman-α observations of absorption in the dayside atmosphere. Above 60 °N/S the observations are not sensitive to the low abundances. At the equator even higher column densities are consistent with the data (Giono and Roth, 2021).*

*Variation with heliocentric distance ("pressure cycle")*. Extensive monitoring in the thermal-IR, spanning almost one Jupiter year (11.8 Earth years), indicates a clear anti-correlation between the amount of $SO_2$ gas and heliocentric distance, at least on the anti-Jovian side (Tsang et al., 2012; 2013), but likely also on the sub-Jovian side (Giles et al. 2024). This indicates that the atmosphere responds to surface temperature variations, but the magnitude of the variation (a factor ~2 in pressure from aphelion to perihelion) is somewhat smaller than expected for pure sublimation control, and the data can be empirically modeled as the superposition of a sublimation component, governed by a frost with seasonal thermal inertia of 350 $Wm^{-2}$ $s^{-1/2}$ $K^{-1}$ (MKS) with a time-independent volcanic component of order $6\times10^{16}$ $cm^{-2}$, contributing ~1/3 of the atmosphere when at its maximum (Tsang et al., 2013).

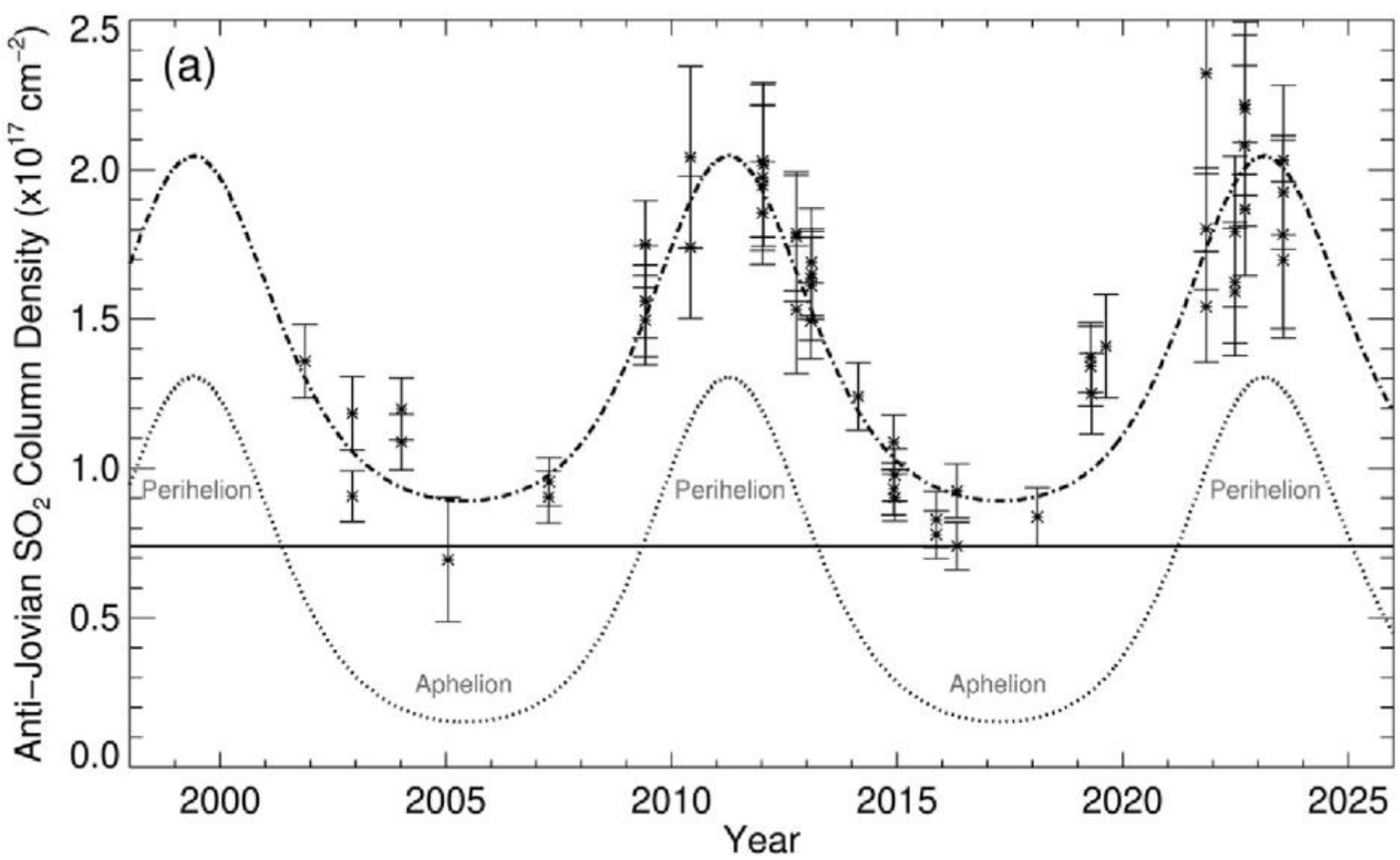

***Figure 8.*** *Seasonal variation of the dayside $SO_2$ column density on the Anti-Jovian side (monitored over almost two Jupiter years or 22 Earth years, from Giles et al. 2024). The dash–dotted line shows the best-fit seasonal model, combining the vapor-pressure equilibrium (dotted line) and a constant component (solid line). There is no evidence for unsystematic, transient changes.*

*Eclipse behavior:* Io's $SO_2$ atmosphere has not yet been detected on the nightside, but the behavior in eclipse, while still uncertain, may be a proxy. Direct observations of $SO_2$ during eclipse yield seemingly contradictory results in the mid-UV (from which Tsang et al. (2015) find no post-eclipse changes) and in the mid-IR (from which Tsang et al. (2016), report an atmospheric collapse during eclipse). In the mm range, ALMA data (de Pater et al., 2020b) indicate that disk-integrated in-sunlight flux densities are ∼2–3 times higher than in eclipse, indicative of a roughly 30–50% contribution from volcanic sources. Maps of Io's $SO_2$ mm emission during eclipse ingress and egress show an overall collapse of the atmosphere, except near known volcanic sites, and a fast reformation time (~10 minutes) after eclipse egress (Figure 9). SO also varies in eclipse in a similar way as $SO_2$ (demonstrating that SO is not a purely volcanic species) but with a longer time constant at egress, which may be consistent with photochemical production from $SO_2$. Although SO is more volatile than $SO_2$, it will still be rapidly removed from the atmosphere because SO is highly reactive with itself on the surface (Lellouch et al., 1996). Finally, NaCl and KCl mm-emission show no significant difference between sunlight and eclipse (in both cases uncorrelated with the $SO_2$ emission), suggesting a purely volcanic origin for these gases, insensitive to the collapse and reformation of the main atmosphere (Redwing et al., 2022).

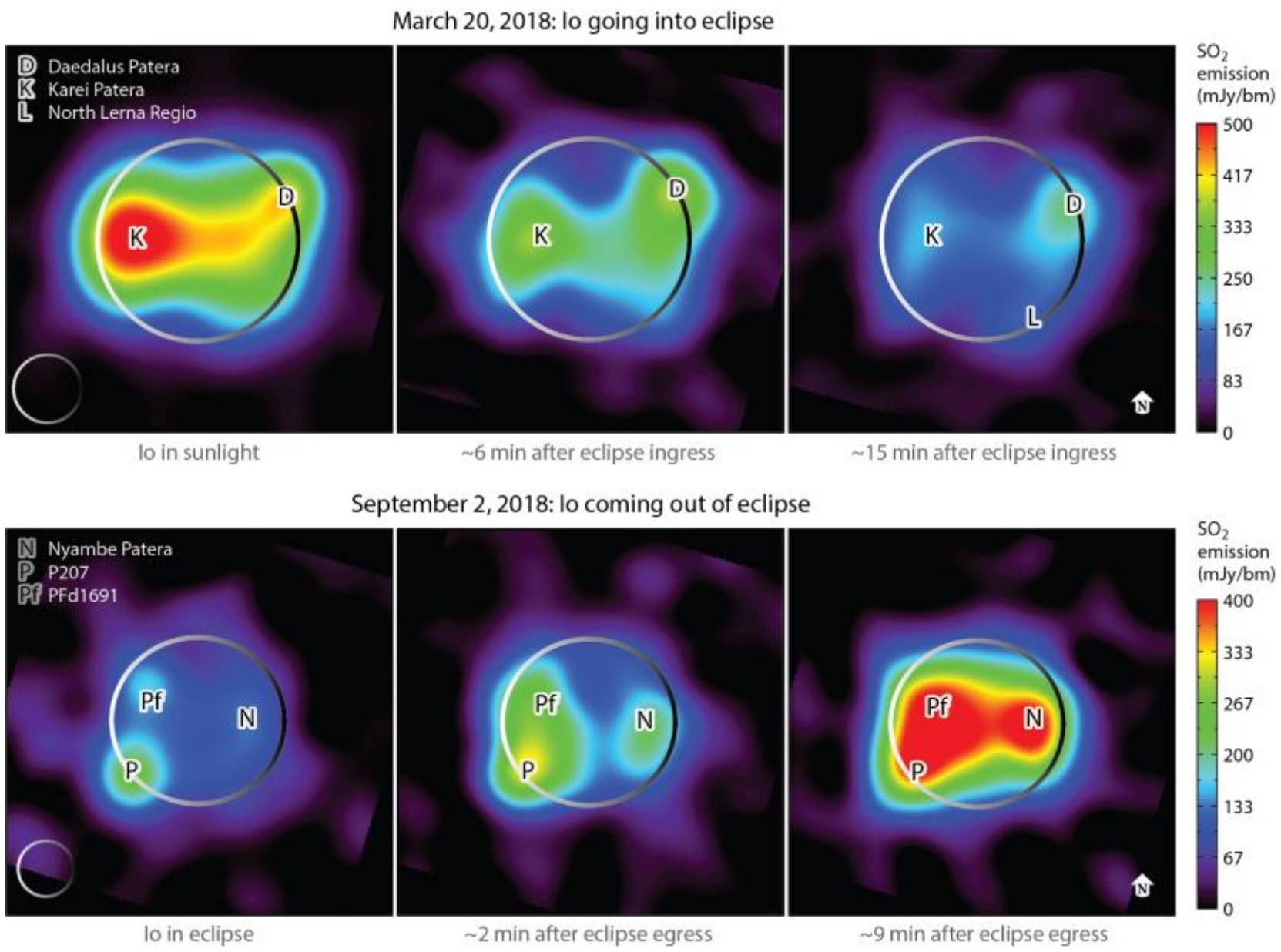

***Figure 9***. *Maps of dayside $SO_2$ emissions while Io moves into and out of eclipse. The emissions clearly decrease in shadow but the remaining $SO_2$ signal suggests a volcanic outgassing contribution between 30% and 50%. The sunlit maps confirm the concentration of the densest atmosphere around the equator. (from de Pater et al. (2021b), based upon observations from de Pater et al. (2020b))*

### 3.2.3 Thermal structure and dynamics

Io's atmospheric vertical structure is observationally uncharacterized, and the degree to which volcanic plumes modify the presumably stable structure associated with sublimation equilibrium is observationally unknown, although model predictions are available (McDoniel et al., 2017 and references therein). Simply measuring the bulk temperature of Io's gas has proven remarkably difficult, as each method has its own limitations and uncertainties. Published results (including mm, thermal-IR, mid-IR and UV observations) range from 110 to 600 K, with a general preference for 200 to 300 K (e.g., Spencer et al., 1995, Lellouch et al., 2003). The most recent (and seemingly most direct) assessments of Io's atmospheric temperature are based on the rotational distribution in mid-infrared bands, and even then, results are not fully consistent: ~110 K from the 19 μm observations (Tsang et al., 2013) and ~170 K from 4-μm spectra (Lellouch et al., 2015). The difference may point to a variation of temperature with altitude as predicted in modeling (e.g., Kumar et al. 1980, Strobel et al. 1994), although temperature retrievals have not been attempted with the observational data. In addition, different gas species yield different temperatures; in sub-mm observations sensitive to multiple rotational lines with different lower energy levels, $SO_2$ is found to be 200-250 K while NaCl gas is measured around 700 K (de Kleer et al., 2024).

Numerous models have been developed to characterize the 3D thermal structure of Io's atmosphere, both for sublimation and volcanic atmospheres. 1D hydrostatic radiative models (Strobel et al., 1994) indicated that Io's atmosphere is likely to be thermally inverted. This is due to solar, plasma, and Joule heating, the latter two effects possibly leading to upper atmosphere temperatures in excess of 500 K or even more. For local pressures exceeding ~10 nbar, a few km deep mesosphere with temperatures a few degrees below the surface temperature may develop in response to $SO_2$ rotational and ro-vibrational cooling. Subsequent work on hydrostatic atmospheres attempted to combine descriptions of vertical transport and horizontal structure, either in a continuum fluid model (e.g., Wong and Smyth, 2000, including also photochemistry) or in a Direct Monte Carlo Simulation (DSMC) approach (e.g., Walker et al., 2010). In these models it is tested how the different effects of surface property variations (e.g. frost abundance), day-night differences and asymmetric plasma conditions affect the resulting atmospheric distribution and temperature.

Io's atmospheric dynamics remain uncharacterized as well. From the theoretical point of view, there is little doubt that given the general dominance of sublimation, the pressure gradient from the warm dayside to the cold night-side must drive a strong day-to-night flow diverging from the region of peak frost temperature / extent, becoming supersonic near the terminator (e.g., Ingersoll et al. 1985; Austin and Goldstein 2000). In addition to this "sublimation wind," Io's atmosphere may be subject to drag due to the plasma torus, which contains ~2000 $cm^{-3}$ ions moving with relative velocities of 57 km/s upstream of Io. Estimates of the drag force suggest that its magnitude is comparable to those of gravity and pressure gradients (Saur et al., 2002). Plasma and sublimation winds tend to add up when Io is at western elongation (both from trailing to leading side) but cancel out each other at eastern elongation (as sublimation winds are inverse). However, the only published observational result (Moullet et al., 2008) which pertains to Io's leading side (eastern elongation), finds that the circulation can be mimicked by a 200±70 m/s prograde zonal flow, in sharp contrast with model predictions.

A relevant time constant for the establishment of a hydrostatic atmosphere is the hydrostatic adjustment time constant. This time constant is equal to the atmospheric scale height divided by the speed of sound and describes the time an imbalance needs to propagate through the atmosphere. Near the surface its value is about 70 s at Io (e.g., Kosuge et al., 2012) suggesting that in the near-surface atmosphere hydrostatic equilibrium prevails.

#### 3.2.4 Plume dynamics

Thermal/dynamical calculations also include DSMC models of volcanic plumes (Zhang et al. 2003, McDoniel et al. 2017, and references therein), either "pure" (i.e., night-side) or in the presence of a background sublimating atmosphere, and account for additional physics such as plume expansion and re-entry shocks, the former effect leading to cold temperatures (20 to 100 K) through most of the plume except in the re-entry region. These simulations include fully-3D simulations (McDoniel et al., 2015; Ackley et al., 2021), unsteady plumes interacting with a changing sublimation atmosphere (McDoniel et al., 2017) or undergoing 3D dynamic pulses (Hoey et al., 2021), and plumes at different locations on Io interacting with impinging streams of Jovian plasma and sunlight (Blöcker et al., 2018; McDoniel et al., 2019). All of these models predict an extraordinarily complex 3D thermal and wind structure for Io's atmosphere, which thus appears critically under-constrained from the observational point of view.

### 3.3 Exosphere and atmospheric escape

Atmospheric escape takes place mostly above a certain level in the upper atmosphere known as the exobase where the transition from a collisional gas to a collisionless gas occurs (Figure 10). Below the exobase the atmosphere can be treated as a fluid, because the average distance a molecule or atom travels before making a collision – the mean free path – is shorter than the smallest macroscopic length scale. The latter is usually defined by the pressure scale height $H$ which characterizes the exponential decay of pressure with altitude.

Above the exobase is a quasi-collisionless region known as the exosphere where the mean free path exceeds the atmospheric scale height. Collisions are sufficiently infrequent that neutral atoms and molecules execute dynamical trajectories that are influenced by mostly Io's and Jupiter's gravitational field.

### 3.3.1 Exobase

Mathematically, the exobase for a hydrostatic atmosphere can be defined where the escape probability for an atom or molecule traveling upward in excess of the escape velocity is $e^{-1}$. This is given by a probability $P$ as

$$P = \int_{exobase}^{\infty} \sigma n(r) \; dr = exp\,(-1/\zeta(r_{exobase})) = exp\,(-1) \;, \qquad (1)$$

where $\zeta(r) = [\sigma\, n(r)\, H(r)]^{-1}$ and $\sigma$ is the neutral-neutral collision cross section with a value for $SO_2$ of $\sigma_{SO2} \sim 1 \times 10^{-14}$ $cm^2$ (Strobel, 2002). Thus the number density at the exobase is $n(r) = 1/(\sigma\, H(r))$. For hard sphere elastic collisions, the mean free path, $\lambda(r)$, is given by

$$\lambda(r) = 1 \,/\, [\sqrt{2}\, \sigma\, n(r)] \,. \qquad (2)$$

This implies that $\lambda(r) = H(r)/\sqrt{2}$ at the exobase defined as in Equation (1). If the probability of escape were 50% (instead of 1/e), the two length scales would be equal which is also often assumed as the definition for the exobase (the altitude where $H(r) = \lambda(r)$). It is important to keep in mind that in reality the exobase is a transition region, rather than a distinct altitude level.

In an isothermal atmosphere, the density n(r) follows $n(r) = n_0\, exp(-\,(r-R_{Io})\,/\,H)$, with the surface density $n_0$ and the moon radius $R_{Io}$ and $r$ measured from the body's center. The altitude $h = r - R$ for a given density n(r) is thus $h = -\,H\, ln(n(r)/n_0)$. The altitude of the exobase according to Equation (1) is then given by

$$h_{exo} = H\, ln(n_0\, H\, \sigma). \qquad (3)$$

In a $SO_2$ atmosphere with a temperature of 120 K near the surface, the nominal scale height is 8.7 km. For a column density of $n_0H = 10^{17}$ $cm^{-2}$ as found in the equatorial atmosphere (Section 3.2), the exobase altitude according to Equation (3) would be at ~60 km.

However, in the upper atmosphere, temperatures can increase significantly due to Joule (Ohmic) heating, essentially changing the altitude profile and increasing the exobase (Section 3.3.3). Values for the exobase altitude inferred by Summers and Strobel (1996) are between 120 km and 500 km (Figure 11).

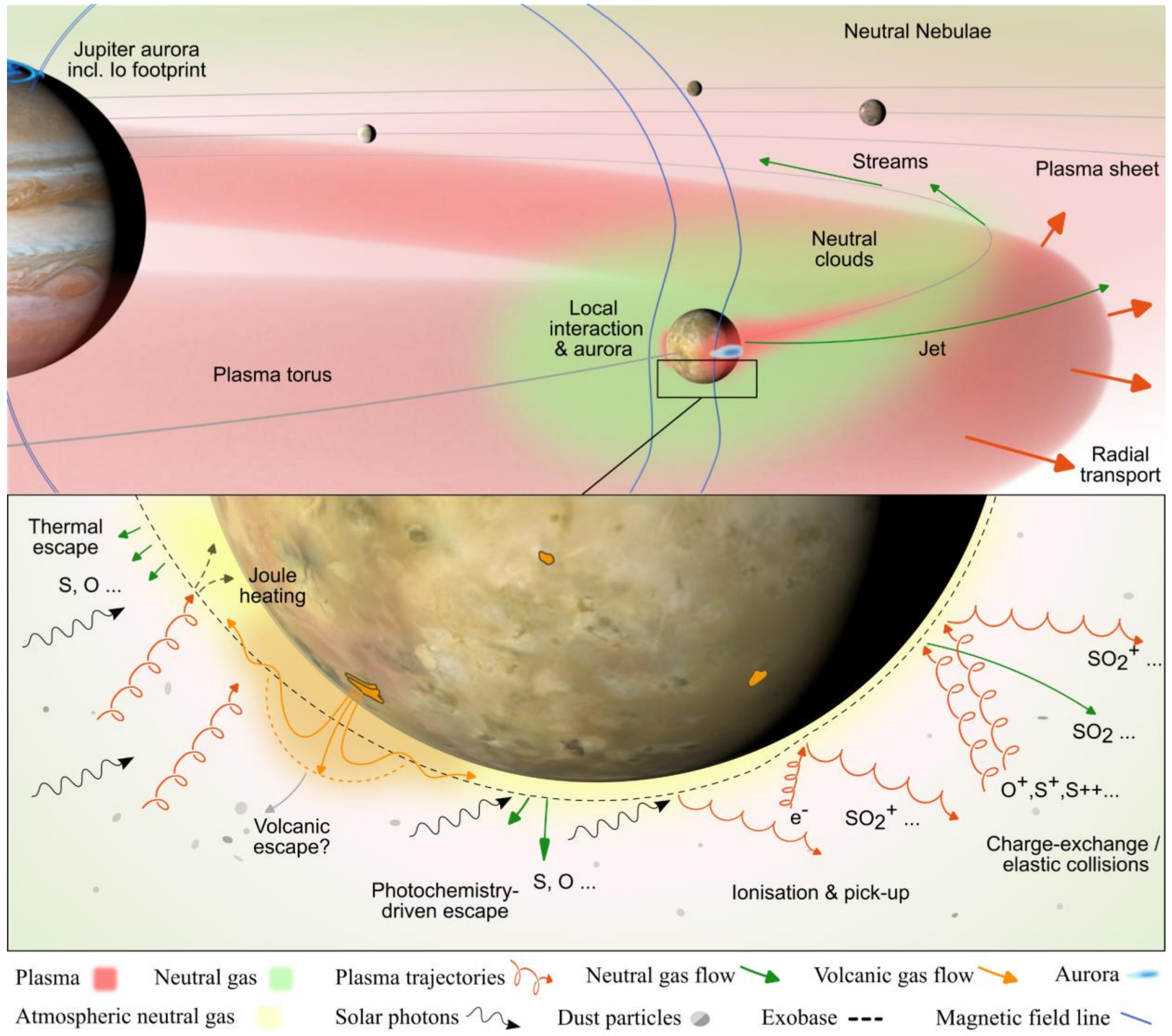

***Figure 10.*** *Overview of escape processes and the magnetospheric environment surrounding Io. Bottom: Various processes near the exobase (dashed gray line) can lead to escape from Io's atmosphere which populates the neutral clouds (slower atomic or molecular neutrals), neutral nebulae (faster neutrals), or the plasma torus (ionized atomic or molecular particles). Top: Ionization of the neutral clouds is the main source for the plasma torus. Loss from the torus is primarily through radial transport to the plasma sheet and other magnetospheric exchange processes. (Credit: Márton Galbács/Lorenz Roth/KTH)*

### 3.3.2 Plume exobase and escape

The exobase is defined for the hydrostatic equilibrium state and, moreover, for a single species represented by a single temperature. A similar definition of a corresponding altitude in a dynamic plume with bulk flow velocity is not possible. McDoniel et al. (2017) showed that plume particles elevated above the exobase of a purely sublimated atmosphere may bounce off the sublimated atmosphere near the exobase when falling back towards the surface (Figure 11, right). In large plumes, the level where upward-moving sufficiently fast particles could escape without collisions is likely above the canopy shock, which is expected to be higher than exobase altitudes found in simulations of the sublimated atmosphere (Summers and Strobel, 1996; McDoniel et al., 2017).

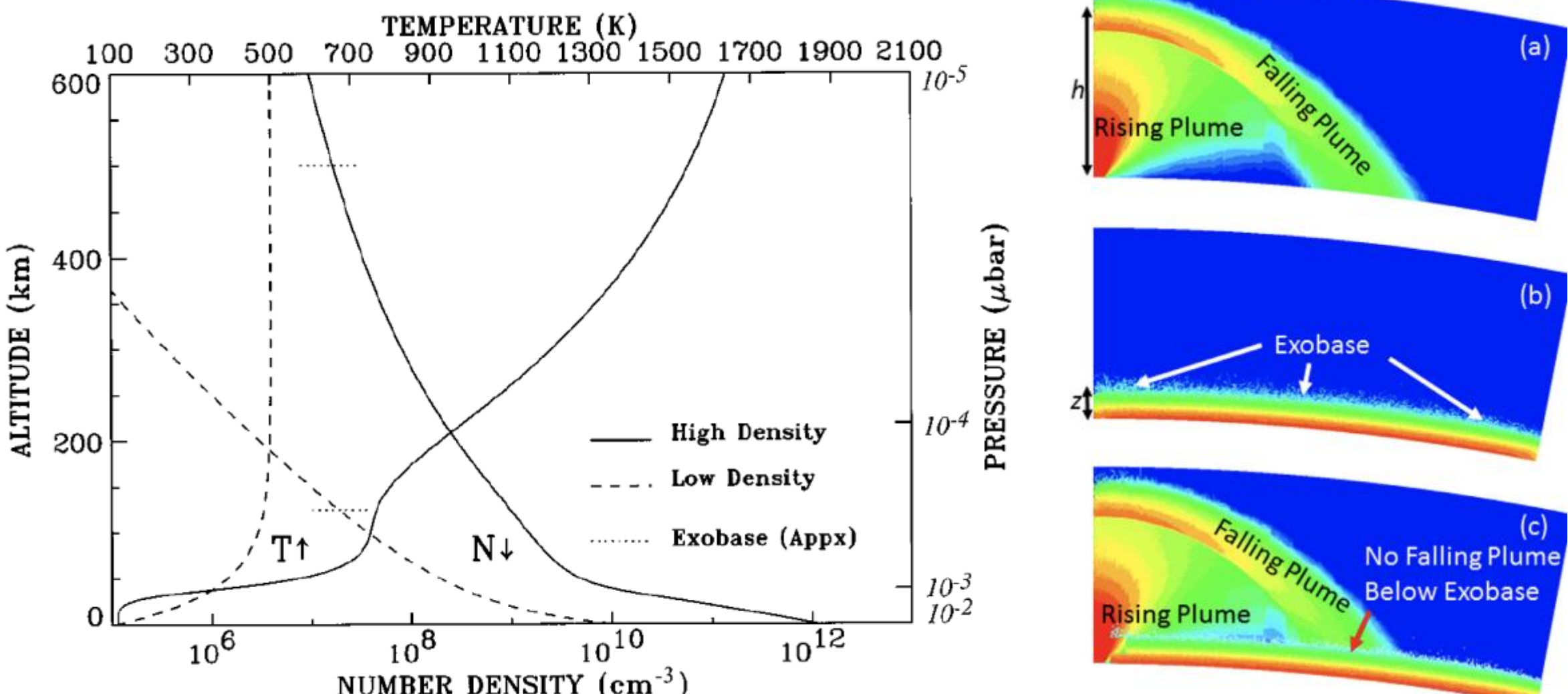

***Figure 11.*** *(Left) Atmospheric density (decreasing with altitude) and temperature profiles with exobase altitudes (horizontal dotted) for two cases, corresponding to two different assumed temperature profiles (hot and cold) from Summers and Strobel (1996). (Right) DSMC modeling results of a large plume rising above the exobase in this simulated (not heated) atmosphere (McDoniel et al., 2017). (Note this may be different for higher exobase cases)*

We also note that the top of Io's atmosphere is not in local thermodynamic equilibrium (LTE) and thus different classes of molecules or atoms may have different temperatures and thus exobase altitudes.

Generally, ejected plume gases do not have sufficient velocities to escape Io's gravity directly. Under ballistic (collisionless) conditions, to reach an altitude of 400 km as inferred for the highest plumes, an ejection velocity of 1.2 km/s is needed. This is still well below Io's escape velocity of 2.56 km/s or the velocity to reach the distance of the Hill radius (the radius where Io's gravity is equal to Jupiter's, near 5.8 $R_{Io}$) of 2.33 km/s. Assuming a Maxwellian velocity distribution with a high core temperature of 800 K around an upward bulk velocity of 1.2 km/s, only fewer than $10^{-5}$ of the intact $SO_2$ molecules reach the escape velocity. Even with an optimistic $SO_2$ plume gas source rate of $10^5$ kg/s, this yields an escape rate of ~1 kg/s, three orders of magnitude lower than the canonical number. Ejection velocity, gas temperatures and $SO_2$ source rates commonly assumed for simulating large plumes like the Pele plume are lower than our assumptions here (Zhang et al., 2003; 2004; McDoniel et al., 2017) and our approximation likely overestimates the escaping fraction. In addition, simulations revealed that the ejected plume gas is effectively slowed by falling gases in the canopy shocks as suggested by Strom and Schneider (1982), likely further reducing the escaping fraction (Zhang et al., 2003).

This situation is vastly different from the Enceladus plume, where the surface gravity is 6% of Io's surface gravity and the fraction of escaping molecules is two orders of magnitude higher than those returning to the surface (e.g., Tian et al., 2007; Villanueva et al., 2023). We note, however, that there may be potential pathways for direct volcanic escape that have not yet been explored, such as the dynamical behavior of volatiles originating from hot surface lavas with temperatures of 1200 K or higher.

### 3.3.3 Thermally-driven escape

In a gravitationally-bound atmosphere with an exosphere, the key non-dimensional parameter governing escape is the Jeans parameter $\lambda_{esc}$, which is defined as

$$\lambda_{esc} = \left(\frac{v_{esc}}{v_{th}}\right)^2 \quad (4)$$

with the escape velocity, $v_{esc}$, and the most probable velocity in the atmosphere, $v_{th}$. Strong escape happens for small $\lambda_{esc}$ parameters near or lower than 1. At Io's surface for T = 120 K, the respective $\lambda_{esc}$ values for O, S, SO, $SO_2$, are 53, 105, 158, 210 implying that the main atmospheric constituents and other volcanic gases are strongly gravitationally bound. Substantial thermal escape is possible only if higher temperatures prevail at the exobase.

Strobel et al. (1994) developed a radiative-thermal heat conduction 1D model for Io's $SO_2$ atmosphere and found that solar heating from the near-IR to UV yielded upper atmospheric temperatures of at most ~ 270 K from non-LTE cooling by $SO_2$ rotational line emissions. Adding plasma heating by impacting thermal torus ions elevated asymptotic temperatures to ~ 700 K. They found the most important heat source to be Joule (Ohmic) heating due to ion-neutral collisions driven by the non-linear Alfvénic electrodynamic interaction of the plasma torus with the sub-nanobar atmosphere raising the temperature an additional 1000 K for an overall exobase temperature as high as 1800 K. For T = 1800 K at 500 km, the respective λ values for O, S, SO and $SO_2$ are 2.8, 5.6, 8.4 and 11, indicating the possibility of significant escape velocities for O and S atoms. Summer and Strobel (1996) show that the effective escape rates strongly depend on vertical diffusion. Escape rates on the order of the canonical rate of 1 tons/s are estimated for O in the case of a high-density atmosphere and high vertical diffusion (Table 1).

It should be noted that Joule heating in Io's ionosphere maximizes when the ionospheric electric field ($E_i$) driving the ions is 0.5 times the external corotation electric field ($E_0$) generated in Jupiter's ionosphere and mapped along magnetic field lines encompassing the Io flux tube (Strobel et al. 1994). The calculation performed by Strobel et al. (1994) for the more realistic 3.5 nbar atmosphere had $E_i = 0.34\ E_0$, which is quite close to the maximum Joule heating rate of $E_i = 0.5\ E_0$. Plasma fluid simulations of the interaction and the ionospheric electric field by Saur et al., (1999) found lower values of $E_i = 0.07\ E_0$ for the conditions and atmosphere considered in their study. This electric field is far from the value of maximum Joule heating ($E_i = 0.5\ E_0$), which means that it is possible to have periods in enhanced Joule heating when $E_i$ increases and obtains values closer to $0.5\ E_0$. The power dissipated through joule heating in the entire atmosphere is suggested to be around ~$4\times10^{11}$ W (Strobel et al. 1994, Saur et al. 1999).

***Table 1.*** *Atmospheric escape rates from different processes derived in different atmosphere studies compared to the canonical number or neutral source rate. We reference here the highest values inferred in each study.*

| **Escape process** | **Highest inferred rates** | **References** |
|---|---|---|
| Direct escape from a single plume | $\leq 10^{27}$ amu/s | Section 3.3.2 |
| Thermal (Jeans) escape | $3 \times 10^{29}$ amu/s [low density atmosphere]<br>($1 \times 10^{28}$ O/s and $5 \times 10^{26}$ S/s)<br><br>$2 \times 10^{30}$ amu/s [high density atmosphere] | Summers and Strobel (1996) |

| | (7 x $10^{28}$ O/s and 2 x $10^{28}$ S/s) | |
|---|---|---|
| Photochemistry triggered non-thermal escape | 5 x $10^{28}$ amu/s<br>(2 x $10^{27}$ O/s and<br>7 x $10^{26}$ S/s) | Huang et al. (2023) |
| Non-thermal escape from plasma-atmosphere collisions | ~3 x $10^{30}$ amu/s<br>(primary neutrals are $SO_2$ from ion-neutral collisions; smaller fractions for O, S, SO) | Dols et al. (2008), Bagenal and Dols (2020) |
| **For comparison** | | |
| *Neutral source rate (canonical number)* | ~1 tons/s, or ~$10^{30}$ amu/s<br>($1.6\times10^{28}$ $SO_2$/s, or<br>$3.1\times10^{28}$ S/s, or<br>$6.3\times10^{28}$ O/s.) | E.g., Broadfoot et al. (1979), Delamere et al. (2004), Hikida et al. (2020) |

### 3.3.4 Non-thermal escape

Non-thermal escape occurs when atoms and molecules are created with excess translational energy in chemical reactions or through dissociation and ionization by photons or charged particles. In a recent study on non-thermal escape induced by photochemistry, Huang et al. (2023) adopted three models from Summers and Strobel (1996). Two of the cases assume a somewhat too dense atmosphere and the third case a too dilute atmosphere when compared to average observed $SO_2$ column densities of $(1–6)\times10^{16}$ $cm^2$ (see Section 3.2). Their calculations suggest escape rates driven by photochemistry of (1.1 to 2.0)$\times10^{27}$ $s^{-1}$ for O and $(1.5−6.7)\times10^{26}$ $s^{−1}$ for S. These rates are still about a factor 10 too small to supply the canonical escape rate and fuel the plasma torus with $10^{30}$ amu $s^{-1}$ (Table 1). In addition, a fraction of the loss from the atmosphere does not feed into the torus, further increasing the deficit. In their Table 2, Huang et al. (2023) state that the non-thermal escape rate of O atoms driven by photodissociation is only 1.6 times larger for the thick atmosphere (case A) than the thin atmosphere (case C). This would suggest that a hundred times thicker atmosphere hardly changes the escape rate and that most of the escape originates from the top column density of $10^{16}$ $cm^{-2}$ of the $SO_2$ atmosphere.

A potentially efficient way to remove $SO_2$ from the upper atmosphere/exosphere is by collisions of Io plasma torus atomic ions ($O^+$ & $S^+$) with $SO_2$ imparting translational energy to $SO_2$ followed more probably by impact dissociation to SO and O with sufficient velocities to escape. In the case of Io, this process (called atmospheric sputtering) is complicated by the asymmetric electrodynamic interaction of Io's conducting ionosphere with the surrounding plasma which results in the reduction of Io's effective cross-sectional area by divergence of the upstream torus ions around Io. The effective area is reduced by the ratio $E_i/E_0$ and estimated to be approximately 0.07 (e.g., Saur et al., 1999). Performing a sputtering calculation by the method of Haff and Watson (1979) yields a sputtering rate for Io's $SO_2$ atmosphere of ~$10^{28}$ $s^{-1}$ with mostly SO and O products under the assumption that the torus ions impact $SO_2$ with the full corotation relative velocity of 57 km/s, ignoring the plasma flow slowing and diversion. The torus atomic ions can also charge exchange with $SO_2$, resulting in fast O and S atoms which escape and $SO_2^+$ which will be accelerated in a reduced ionospheric electric field to acquire initially a reduced $E \times B$ drift velocity and a gyro velocity treated as a perpendicular temperature (Dols et al., 2008; 2012). The interaction with the surrounding plasma torus including its effects on atmospheric loss are discussed more in the following Section 3.4.

## 3.4 Electrodynamic interaction, plasma-neutral collisions, and the related atmospheric loss processes

The plasma in the Io torus is magnetically coupled to Jupiter and thus rotates with the same angular velocity as Jupiter's ionosphere corresponding to a period of 9 hours and 55 minutes, significantly shorter than Io's orbital period of 42.5 h. The plasma is therefore rotating faster than Io and overtakes the moon with a relative velocity of 57 km/s. The fast-moving plasma interacts with Io's atmosphere and surface, which causes a large variety of plasma and atmospheric effects that contribute to mass loss from Io. Reviews on this plasma interaction are presented in, e.g., Kivelson et al. (2004), Bagenal and Dols (2020, 2023), Saur et al. (2004, 2021). A pre-Galileo analysis on losses due to various ion collisions is presented in Sieveka and Johnson (1984).

Various types of collisions of the torus plasma ions and electrons with Io's atmosphere lead to an exchange of matter, momentum and energy between the ionized and neutral gases as depicted in Figure 10. These collisions slow down the plasma in Io's ionosphere and its vicinity. The modified and slowed plasma around Io generates plasma waves traveling away from Io, a slowed wake downstream of Io, and draped magnetic field lines around Io. The most important wave mode excited by the interaction is the Alfvén mode which travels along Jupiter's field lines towards Jupiter in the northern and southern direction (see pink structures in Figure 12, left).

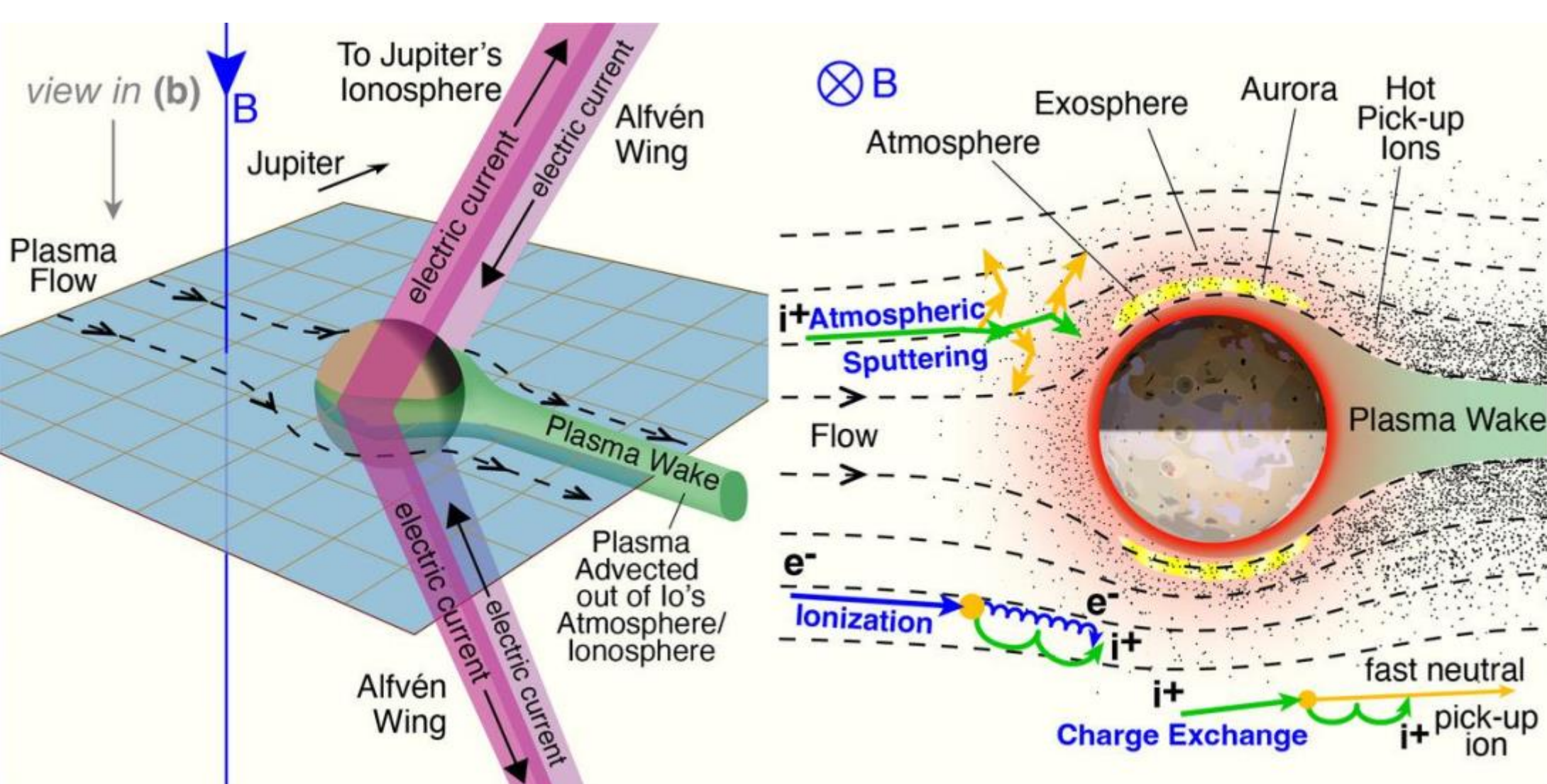

***Figure 12**. (Left) 3D sketch of the plasma environment around Io. (Right) Processes in Io's atmosphere in the plane perpendicular to Jupiter's background magnetic field; the top of the figure is towards Jupiter**.** Credit: S. Bartlett adapted from Bagenal and Dols (2020).*

### 3.4.1 Plasma-neutral collisions as primary loss process

The aforementioned collisions are the engine of Io's plasma interaction, in the sense that they cause momentum exchange between the fast moving magnetospheric plasma and Io's neutral atmosphere. The momentum exchange perturbs the plasma flow and slows it down within Io's atmosphere. The velocity perturbations and associated perturbations in the magnetic field and pressure are propagated by magnetohydrodynamic (MHD) wave modes outside of Io’s atmosphere (for details see reviews in Kivelson et al. 2004, Saur et al. 2004). These collisions are also likely the primary reason for the loss of Io’s atmosphere into the torus and will be reviewed in this subsection. The loss of $SO_2$ from Io’s atmosphere occurs in various collisional pathways that also include dissociation and ionization into sulfur and oxygen neutrals and ions (e.g.,Thomas et al., 2004; Nerney and Bagenal, 2020).

The collisions causing loss from Io’s atmosphere can be subdivided into ion-neutral, electron-neutral, photon-neutral and neutral-neutral collisions. In Figure 13, the total rates for a set of important collisions within Io's atmosphere are displayed as a function of Io's atmospheric surface density with an atmospheric scale height assumed to be 100 km (lower scales could not be resolved numerically). The column density of Io’s atmosphere at low to mid latitudes lies in the range of $(1–10) \times 10^{16}$ $cm^{-2}$ (Section 3.2) corresponding to a surface density of $(1$ to $10) \times 10^{15}$ $m^{-3}$ (Figure 13, x axis) for a scale height of 100 km. Under the assumption that Io's atmosphere consists of $SO_2$ only, a model by Saur et al. (2003) finds that the dominant collision process in Io’s atmosphere is elastic collisions of torus and ionospheric ions assumed to be $SO_2^+$ with $SO_2$ (Figure 13). In this calculation elastic collisions (see Glossary) include charge exchange collisions. Due to velocities of the ions of tens of km/s, elastic collisions generate neutrals with velocities larger than Io’s escape velocity (~ 2.56 km/s at Io’s surface). The possible generation of multiple subsequent collisions of recoiling neutrals in Io's atmosphere is referred to as atmospheric sputtering (Haff and Watson, 1981). If the subsequent path of these neutrals does not go below the exobase (Section 3.3), then the neutrals can escape Io with a large likelihood. These elastic collisions are therefore a main loss process of Io's atmosphere populating the Io torus with neutrals (Saur et al., 2003; Dols et al., 2008; Blöcker et al., 2018).

Other processes are electron-impact dissociation, electron-impact ionization and photodissociation (Figure 13). Photoionization plays a smaller role at Io. Photodissociation does not affect the plasma interaction.

The electron-impact ionization rate does not grow linearly with increasing neutral density because the total amount of electron energy available for ionization is limited by the amount of electron energy available in the torus electrons upstream of Io. The simulations in Saur et al. (2003) include potential negative feedback, arising from the increased diversion of the incoming plasma flow around Io and increased electron cooling for an increase in atmospheric column density. However, their results show that the total elastic collision rate scales approximately linearly with increasing neutral density and thus not all processes can be assumed to be affected by negative feedback. Neutrals which are ionized by electron impact turn into plasma and are subsequently accelerated by the local electromagnetic forces. These accelerated ions and electrons are subsequently advected out of Io’s atmosphere into the plasma torus.

The electron ionization process is energetically limited and is significantly less frequent than the elastic collision and photodissociation. It is estimated, based on the Galileo flyby in Io’s wake, at ~ 300 kg/s (Saur et al., 2003; Dols et al., 2008, Bagenal 1997). Although this local mass loading directly populates the torus, this rate is significantly smaller than the torus neutral supply rate of ~ 1 ton/s. Consequently, most of the mass that leaves Io’s atmosphere is in the form of neutrals. Most of these neutrals have a velocity larger than the velocity to reach the Hill sphere (>2.33 km/s) but smaller than the escape velocity from the Jupiter system at Io’s orbit (<25 km/s), and the escaping neutrals feed the extended neutral clouds. A small fraction of the neutrals, e.g., those generated in the high-velocity flank regions have velocities larger than 25 km/s and can gravitationally escape the Jupiter system. Charge exchange of fast neutrals with ions can lead to fast ions whose kinetic velocity will primarily go into gyration.

Another important aspect of the collisions of the magnetized torus plasma with Io's atmosphere is, in addition to the momentum exchange, the energy exchange, i.e., the heating of the neutral atmosphere. The heating occurs in the form of plasma and Joule heating (Vasyliunas and Song, 2005; Saur et al., 1999), which can significantly increase the temperature and thus the scale height of the atmosphere and ionosphere (Strobel et al., 1994) possibly leading to increased thermal escape into the torus. The heating rates are model-dependent and currently no consensus on the true thermal escape rate exists.

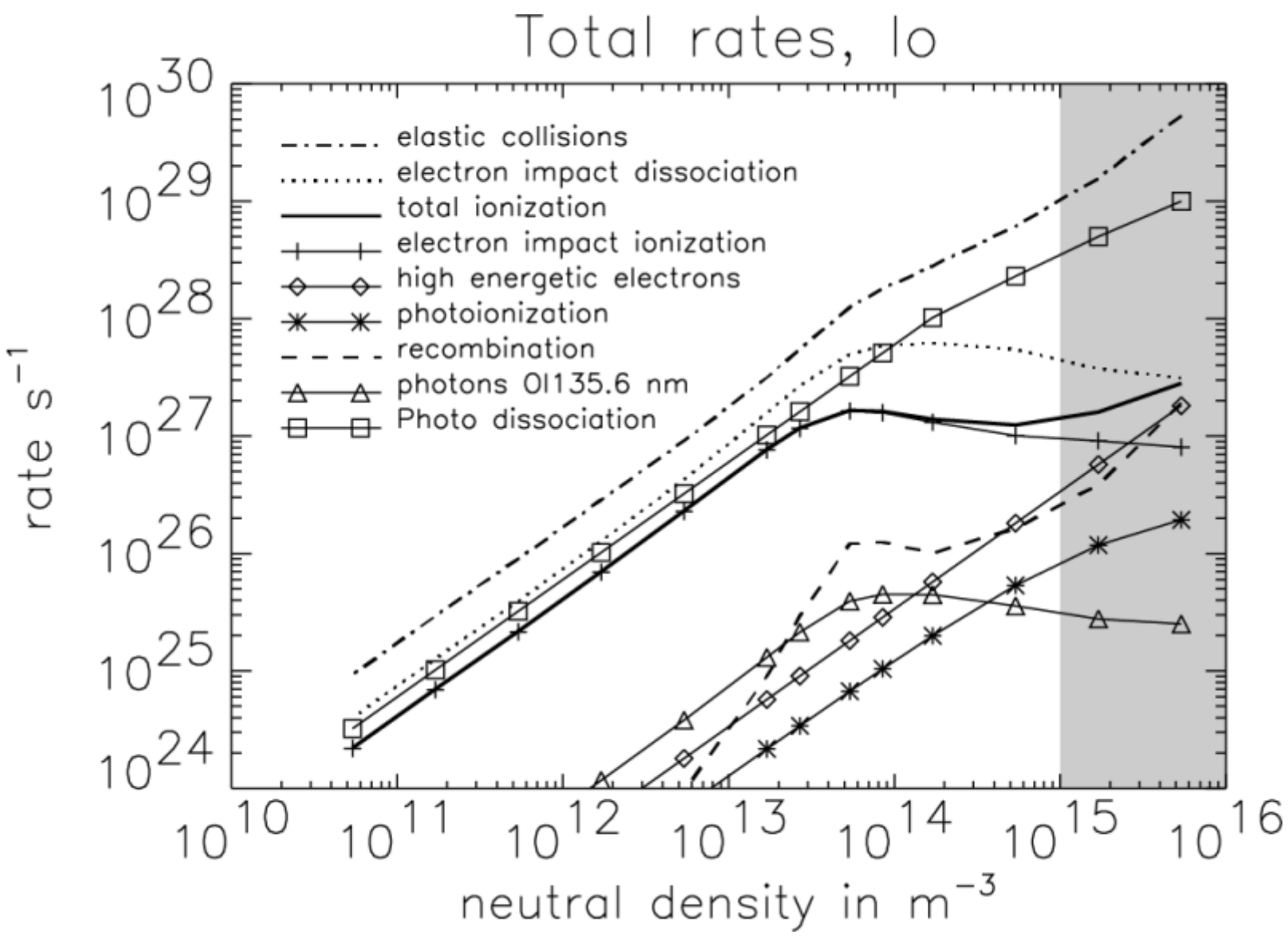

***Figure 13***. *Various total rates within Io's atmosphere as a function of atmospheric content in units of surface density (adapted from Saur et al. (2003)). The range of commonly accepted equatorial atmospheric densities is shown by the shaded gray area.*

Because of the limited capabilities of the Galileo plasma instrument, the detailed composition of ions that leave Io is still undetermined and can only be addressed through numerical simulations. Summers and Strobel (1996) propose a 1D photochemistry model of Io's atmosphere and ionosphere where the ionospheric composition depends on the atmospheric model assumed. Using a multi-species chemistry model, Dols et al. (2008, 2012) propose that the incoming S and O ions from the torus are quickly removed by charge exchanges with the $SO_2$ atmosphere, leaving $SO_2^+$ and $SO^+$ as the dominant new ions. We note here that the charge exchange process for $S^+ + SO_2$ is, however, endothermic (~2eV) and thus inefficient.

The role of individual volcanoes in Io's plasma interaction has been studied by Roth et al. (2011) and Blöcker et al. (2018). Taking Tvashtar and Pele as examples for large plumes, Blöcker et al. (2018) found that plumes modify the total production and collision rates in Io's atmosphere by only <3%. This is primarily due to the low gas content of the plumes when compared to the global atmosphere. This indicates that individual volcanoes may only weakly influence the loss rate from Io's atmosphere to the torus.

Constraints on the ion composition around Io are available through observations of ElectroMagnetic Ion Cyclotron waves (EMIC) measured by the magnetometer of the Galileo spacecraft. Certain anisotropic ion phase space distributions typically associated with larger perpendicular temperatures compared to parallel temperatures generate positive growth rates for EMIC waves. Pickup ions also cause such anisotropies and can generate ion cyclotron waves at frequencies close to the ion gyrofrequency of the pickup ion, which thus constrains the charge-to-mass ratio of the ions. Based on an analysis of several Galileo Io flybys, $SO_2^+$, $SO^+$, $S^+$, and possibly $H_2S^+$ (or more likely $^{34}S^+$) have been detected downstream of Io and on the flanks seven or more $R_{Io}$ away from the moon, and further some localized $S^+$ emissions were detected downstream of Io (Russell and Kivelson, 2000, 2001; Huddleston et al., 1998; Blanco-Cano et al., 2001). The detection of cyclotron waves indicates pickup processes far downstream of Io in a putative extended corona.

### 3.4.2 Remote observations of the local plasma-atmosphere interaction

Electron-impact excited emission from Io's atmosphere provides a diagnostic means to investigate the structure of Io's atmosphere and its ion loss into the torus. This emission is often referred to as "auroral emission" (see glossary) and can be observed in the UV from Earth (e.g. Roesler et al. 1999, Figure 14 left) or at visible wavelengths by spacecraft cameras when Io is eclipsed by Jupiter (e.g. Geissler et al. 2004). Such remote observations provide significant information about the state of the atmosphere and plasma interaction, although they do not directly monitor the rate and variations of Io's neutral losses.

The cross sections for electron-impact ionization of $SO_2$, S and O have very similar energy dependencies as the cross sections for electron-impact excited UV emission from these species (e.g., Saur et al., 2003). Thus, the UV emission from Io's atmosphere is a direct monitor of electron-impact ionization in Io's atmosphere. Io's auroral emission observed in the UV and at visible wavelengths shows two bright spots near the limb of Io at Io's magnetic equator, defined as the plane perpendicular to Jupiter's magnetic field through Io's center (see Figure 14 left, Roesler et al., 1999; Retherford et al., 2000; Geissler et al., 2004; Roth et al., 2014; 2017). The physical reason is that the convection pattern of plasma through Io's atmosphere and electron heat flux along Jupiter's field lines control the transport of electron energy into Io's atmosphere (Saur et al., 2000; Roth et al., 2014; 2017). The heat flux also explains why the northern or southern hemisphere facing the center of the torus is brighter in UV than the opposite one (Retherford et al., 2003; Roth et al., 2014). Analysis of observations taken over four years (Roth et al., 2014) showed that the variations in the UV emissions can be explained solely by changes in the plasma environment and collapse of Io's atmosphere during eclipse. Variations caused by a change of the global atmospheric density putatively caused by sporadic volcanic eruptions were not detectable, supporting the hypothesis of a stable atmosphere (Section 3.2).

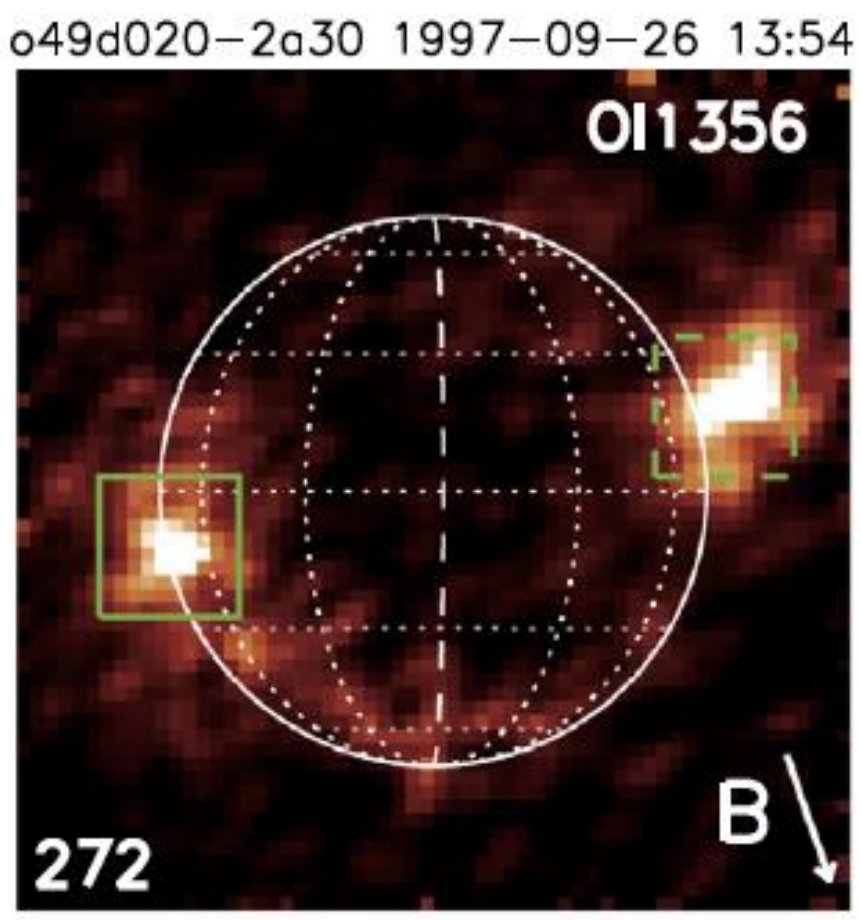

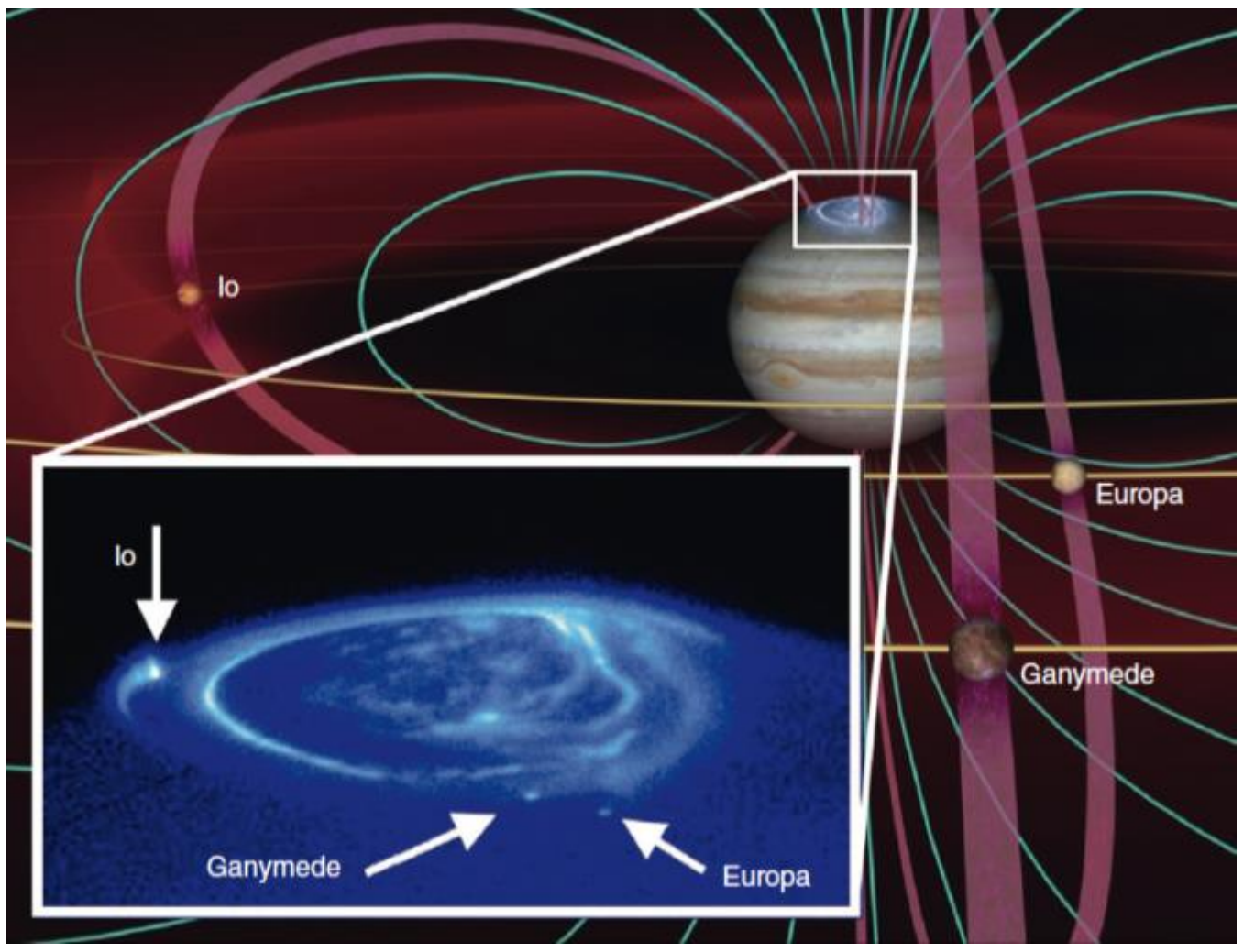

***Figure 14.*** *(Left). Local UV emission of the oxygen O I 135.6 nm line taken with the Hubble Space Telescope (HST). The emission is dominated by two bright spots near Io's magnetic equator, i.e., perpendicular to Jupiter's background magnetic field B. The number at the lower left corner describes the sub-observer longitude (from Roth et al., 2014). (Right) Sketch of the interactions of Jupiter's moons with the magnetosphere. Turquoise lines display Jupiter's magnetic field lines, purple tubes show Alfvén wings connecting the moons with Jupiter. The inset on the lower left shows HST observations of the auroral footprints of the moons in Jupiter's atmosphere resulting from particle acceleration within the Alfvén wings (Image Credit: J. Spencer and J. Clarke).*

### 3.4.3 Io footprint in Jupiter's atmosphere as diagnostic for Io local interaction

We consider the variability of the brightness of Io's footprint in Jupiter's aurora an indirect tool to study Io's atmosphere and its supply to the magnetospheric environment, in particular because it relies on complex acceleration processes along the Alfvén wings (Hess et al., 2013; Szalay et al., 2018; Saur et al., 2013). Based on measurements by the Juno Ultraviolet Spectrograph (UVS), Hue et al. (2019) found that the brightness of Io's footprint does not significantly change when Io passes through eclipse. This may imply that Io's interaction and the power transmission is more strongly saturated than expected, i.e., a change in the atmosphere density does not change the power transmission (Blöcker et al., 2020). An alternative explanation would be that the atmosphere collapses less than derived from other observations. As Io's auroral footprint is also affected by the state of the torus (mostly via plasma density and wave travel times), the footprint as well as other features of the Jovian aurora are expected to reflect torus variability, which is discussed more in Section 3.7.2.

## 3.5. Neutrals from Io in Jupiter's magnetosphere

Neutrals are continuously lost from Io into its local environment due to a variety of mechanisms. Io's orbital speed is ~17 km/s and at its semimajor axis of 5.9 $R_J$ (1 $R_J$ = 71,492 km), the local escape speed from Jupiter is ~25 km/s. As neutrals lost from Io by atmospheric sputtering processes typically leave the moon's exobase with excess speeds relative to Io on the order of a few km/s (with most under 1 km/s; Smyth and Marconi, 2003), the majority of neutrals lost from Io remain gravitationally bound to Jupiter, populating co-orbiting neutral toroidal (or partial toroidal) clouds in the vicinity of Io's orbit (Smyth and Combi, 1988a; 1988b;

Smyth and Marconi, 2003; Smith et al., 2022). The bulk of the mass in these clouds is in the form of sulfur and oxygen, which are subsequently ionized to form most of Jupiter's magnetospheric plasma mass (e.g., Bagenal and Dols, 2020, and Section 3.6).

As most of the material from Io that is ultimately supplied to the plasma torus and plasma sheet in the magnetosphere comes from the neutral clouds (and not from direct ionization at Io), they carry important information about the interaction of Io with the Jovian magnetosphere and the exchange of mass. The clouds themselves also have substructure in their local densities, a consequence of the characteristics of their sources and sinks. We begin by highlighting observations of sodium (Na) and potassium (K), the two minor species most readily detectable remotely providing important tracers of Io's neutral species' evolution. We then highlight the current understanding of Io's sulfur (S) and oxygen (O) clouds and how they provide the seed for Jupiter's magnetospheric plasma. We also explain that while transient changes of Na have been observed, it is not clear yet how these are related to volcanism and the bulk neutral environment.

Previous reviews on the neutral clouds can be found as part of chapters by Schneider and Bagenal (2007) and Thomas et al. (2004).

### 3.5.1. Sodium and Potassium Clouds

Although a minor species in Io's atmosphere ( ~ a few percent), sodium is the species in the neutral clouds escaping Io that is most readily observed due to its significantly larger cross section to scatter light compared to other species. Resonance scattering of solar photons brightly illuminates Na clouds at optical wavelengths and these clouds exhibit distinct substructures in their density distributions. The sodium cloud orbiting Jupiter is densest in a region both leading and trailing Io in its orbital path, which is termed the "banana" and described below. Additionally, "streams" and "jets" of Na have been observed and show a clear relation to the Jovian magnetic field, which is indicative of ion chemistry (Figure 15). Na escapes from Io into the (neutral) "banana cloud" at rates of (1 to 9) × $10^{26}$ atoms $s^{-1}$, and the total Na escape rate from Io including ion loss is (3 to 25) × $10^{26}$ atoms $s^{-1}$ (Wilson et al., 2002).

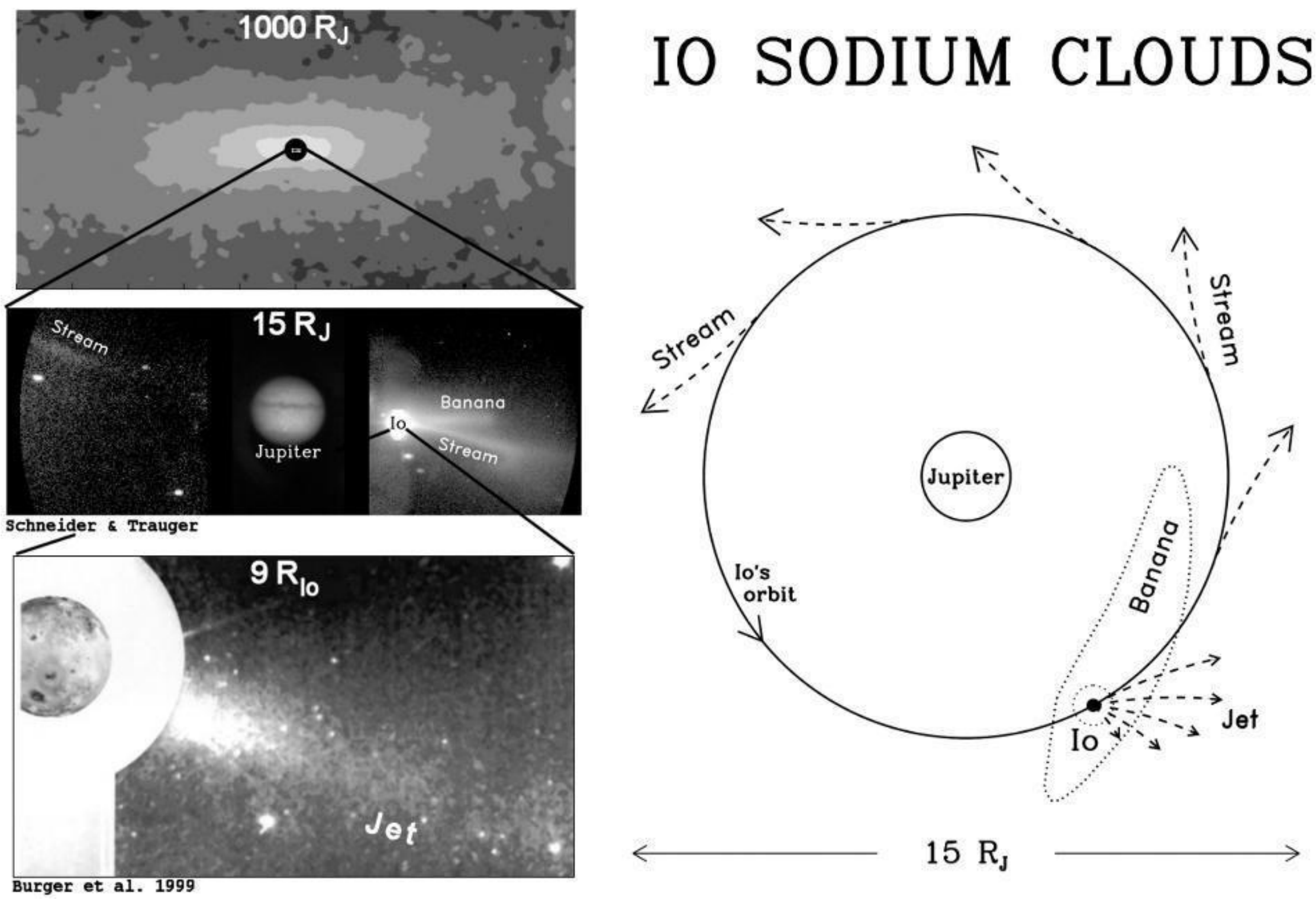

***Figure 15.*** *Images of Io's sodium cloud features, with labels identifying their different spatial scales. Adapted from (Burger et al., 1999; Mendillo et al., 1990; Schneider et al., 1991)*

*Banana Cloud.* Sodium gas that is either sputtered or chemically created near Io's exobase at a velocity exceeding 2.3 km/s can escape the moon's gravity. This yellow (due to the emission line at 589.0 nm) sodium gas cloud is curved in shape leading and trailing Io in its orbit; it is aptly termed the "banana." This cloud is shaped by the angular momentum of the escaping gas relative to Io, and to a lesser degree by solar radiation pressure. Gas escaping from Io's leading hemisphere exceeds Io's orbital velocity and thus drifts radially outward from Jupiter. Gas escaping Io's trailing hemisphere has a lower orbital velocity than Io and falls inward toward Jupiter. Near Io's orbit, the core electron temperature of the Io plasma torus is comparable to the 5.14eV sodium ionization potential, and electron temperature increases radially from Jupiter (Bagenal, 1994). Hence, electron-impact ionization truncates the banana cloud outward of about 6 Jovian radii, while sodium gas radially inside of Io's orbit is preserved as neutral. The precise contours of where Na ionization occurs are complicated by the 0.13 $R_J$ dawnward offset of the Io plasma torus (e.g., Schmidt et al., 2018) and by the collisional cooling of the electron population within the denser gases near Io and its wake (e.g., Dols et al., 2012).

*Jet and Stream.* Structures of neutral sodium in the center left panel of Fig. 15 are tied to Jupiter's magnetic field, appearing distinct from the banana cloud in Io's orbital plane. These clouds must be produced through ion chemistry, whereby species are first ionized and then neutralized. Either charge exchange or dissociative recombination are viable as neutralization mechanisms, but the neutralization of atomic ions via recombination with a free electron is clearly too inefficient. Dissociative recombination of a molecular ion is considered more plausible than charge exchange as the source for the neutral jet and stream, and $NaCl^+$ has come into focus as a likely chemical pathway considering that NaCl outgassing rates are sufficient to supply the escape rates into the cloud (Lellouch et al., 2003) and NaCl is a major constituent of Io's volcanic dust (Postberg et al., 2006).

$NaCl^+$ was initially thought to be the result of direct photoionization of NaCl gas, however, solar photons of sufficient energy to ionize NaCl would also dissociate the molecule (Heays et al., 2017). As a different reaction is necessary to form $NaCl^+$ from NaCl, Schmidt et al. (2023) proposed that the primary pathway may be $SO_2^+$ or $SO^+$ ions charge exchanging with NaCl. The relevant $SO_2^+$ or $SO^+$ ions must be produced by photoionization of $SO_2$ and SO; they cannot be attributed to electron-impact ionization, since the source rates of Io's Na respond strongly to the moon's ~2-hour passage through Jupiter's shadow (Grava et al., 2014; Schmidt et al., 2023), while the torus' ionization of Io's atmosphere would persist in the absence of sunlight. This leaves open the question of whether Io's ionosphere is predominantly sourced by electron impact or by photoionization: the former is expected from calculations of the plasma torus electron-impact ionization of $SO_2$ (Saur et al., 1999; 2002, Section 3.4), but the strong sodium response to the eclipse phase is evidence for the latter. The link between the jets and brightenings of the plasma torus or sodium nebula remains unclear (De Becker et al. 2023).

The stream and jet are distinct in that the stream emanates from the diffuse plasma torus, while the jet emanates from Io itself. Ions in the jet are formed by chemistry tied to Io's ionosphere, where the lifetime for dissociative recombination is less than 2 minutes. Sodium can be seen as a stream oriented along the torus equator more than 1 million km from Io (Schneider et al., 1991, Figure 15), which corresponds to a plasma transport time of several hours. It remains uncertain if the stream feature is produced by the same chemical reaction as the jet; charge exchange with Na ions in the torus to produce fast neutral Na cannot be ruled out.

*Extended sodium nebula.* The extended sodium nebula formed by ion chemistry in the Io-Jupiter interaction is one of the largest structures in our solar system. At times, its observable diameter can exceed 1000 $R_J$, an angular size of ~5.5° (which is about twelve times the diameter of a Full Moon) as viewed from Earth (Wilson et al., 2002). The distant Na nebula is thus a good target for monitoring variability in the Io-Jupiter system using small coronagraphs designed for wide-field, low surface brightness measurements (Mendillo et al., 2004). Transient brightness increases of this extended nebula have been observed relatively frequently in various studies. The long-term study by Mendillo et al. (2004) identified weaker brightenings in 1990, 1991 and 1997 and stronger brightenings in 1995 and in 1998. Later on, Yoneda et al. (2009, 2015) reported brightenings in 2007 and 2015 and for the latter a simultaneous change in the oxygen neutral cloud and plasma torus were monitored by Hiaski (Section 3.6). Most recently, Morgenthaler et al. (2019) reported a transient event in the sodium nebula in 2018. Hence, the sodium nebula appears to intermittently undergo transient brightenings.

*Potassium.* The structure of the potassium clouds around Io is similar to that of the sodium cloud, where again the cloud leading Io drifts inward towards Jupiter and extends farther in longitude than gas on the trailing hemisphere (Trafton, 1981). Neutral potassium has been measured with high Doppler shifts indicative of ion chemistry, but the relative strength of its fast component is weak as compared to sodium (Schmidt, 2022; Thomas, 1996). Disk-resolved ALMA measurements show that the NaCl/KCl ratio is in the 3.5 to 10 range (Redwing et al., 2022), marginally lower than the Na/K ratio of 7 to 13 in Io's exosphere (Brown, 2001), suggesting that Na may escape more efficiently than K. Io's extended fast chlorine cloud presents a challenge for remote sensing measurements and remains unconstrained, but its presence can be inferred since dissociative recombination of $NaCl^+$ would also impart kinetic energy to Cl, albeit at a lower energy due to its higher mass.

### 3.5.2. Bulk neutral clouds

Neutrals leave Io's atmosphere both in molecular form as $SO_2$ and SO (and possibly others) and in their atomic constituents S and O. The bulk atomic species have been constrained observationally to some extent, although not as much in detail as the sodium clouds. Molecules in or near Io's orbit have not yet been measured directly and our understanding relies mostly on modeling work.

*Neutral cloud modeling.* Smith et al. (2022) carried out 3D numerical simulations of the $SO_2$, SO, S and O neutral clouds (constrained by the line-of-sight UV oxygen emission that Hisaki observed) along the orbit of Io on the dusk and dawn ansae at 5-6 $R_J$ and 6-7 $R_J$ distances as a function of Io's phase (Koga et al., 2018a). Figure 16 shows the modeled neutral densities of these species, each of which exhibits a core, dense region near Io and various degrees of azimuthal symmetry.

The Smith et al. work builds upon previous models that have studied the formation of Na, O, S, $SO_2$, and SO neutral clouds and corona (e.g., Wilson and Schneider, 1999; Wilson et al., 2002; Burger and Johnson, 2004; Smyth et al., 2011; Smyth and Marconi, 2003; 2005). Such models prescribe a flux of particles (here $SO_2$ and O, as there are no Hisaki constraints on S and SO) from the exobase and follow the particle trajectories under the gravity of the moon and Jupiter. They prescribe a source velocity distribution at the exosphere peaking at a low velocity of 0.5 km/s (< 2.33 km/s, the Hill sphere escape velocity at Io's surface) with an incomplete collisional cascade tail (Smyth and Combi, 1988a). The resulting neutral clouds for each species are shaped by the interactions with the surrounding plasma (ionization and charge exchange), photo-ionization and photo-dissociation.

Smith et al. (2019) concluded that the Hisaki observations are consistent with the prescription of two separate exospheric neutral sources: O at ~ 200 kg/s and $SO_2$ at ~ 400 kg/s, which escape preferentially from the upstream sub-Jovian hemisphere. These results provide a unique constraint on Io's atmospheric escape processes that have yet to be addressed. The O neutral cloud extends over the whole Io orbit, dominating the $SO_2$, S and SO clouds, which are also more limited in their extension along Io's orbit.

*Constraints on molecular clouds.* There is some indirect evidence for the abundance of the molecular clouds. Freshly created pickup ions from Io's neutrals perturb their local magnetic environment and create ion cyclotron waves with frequencies corresponding to the period of these ions' gyromotion about the magnetic field. *Galileo* measured these ion cyclotron waves at the $SO_2^+$, $SO^+$ and $S^+$ gyrofrequencies mainly within 20 $R_{Io}$ radially outward from Io and in a smaller region inward, with a north-south extent of ~1 $R_{Io}$. Ion cyclotron wave growth is caused by a ring distribution in phase space (Huddleston et al., 1998; Warnecke et al., 1997), which is only possible if fresh pickup ions complete their gyro period without colliding.

The presence of ion cyclotron waves at large distances from Io thus implies that the source of fresh pickup ions must be a very extended neutral exosphere. Crary and Bagenal (2000) found that the wave amplitude could be explained by atmospheric escape rates of 1-3.5 x $10^{27}$ $SO_2$ / s, but that only 10% of these molecules were ionized by the time they reached Io's Hill radius at 5.8 $R_{Io}$. Such a broad distribution is no surprise for atomic fragments, but it is less obvious how the heavy and cold molecular species, $SO_2$ and SO, could extend for thousands or even tens of thousands of kilometers from Io.

Russell and Kivelson (2000) suggested that the molecular ion cyclotron waves *Galileo* measured are best explained as ions that have been neutralized and re-ionized. Molecules in Io's upper atmosphere are ionized by UV photons or electron impact and gyrate perpendicular to the local magnetic field. Gyrating ions then charge-exchange with neutrals and spread out in a fan radially outward since the electric field associated with the corotation of the torus plasma initially accelerates the ions away from Jupiter. These neutral molecules far from Io are then re-ionized, producing the waves observed by Galileo. Dols et al. (2012) modeled plasma and field data from the five Io encounters of the Galileo spacecraft and inferred an extended molecular corona consistent with this multistep chemistry (see their Figure 19). Charge exchange efficiency is dependent on both the energy level and relative velocity, but symmetric charge exchange reactions, e.g., $SO_2^+ + SO_2 \Rightarrow SO_2$ (fast) $+ SO_2^+$ provide a plausible chemical pathway to produce an extended molecular corona (Dols and Johnson, 2023).

*Atomic clouds.* Electron-impact dissociation and photo-dissociation breaks down Io's molecular corona into extended atomic O and S coronae and unbound neutral clouds that orbit Jupiter. Wolven et al. (2001) carried out a survey of O and S emissions within ~10 $R_{Io}$, which showed each to be more extended than Io's sodium corona. They estimated an O column of (1 to 2) $\times 10^{14}$ $cm^{-2}$ at 1 $R_{Io}$ from the surface. The lifetime of sulfur against electron-impact ionization is much shorter than O, but not as short as sodium and potassium (at just a few hours in the densest region of the torus). For this reason, the average S abundance of ~6 $cm^{-3}$ (Durrance et al., 1995) is well below the 30 to 35 O atoms $cm^{-3}$ in the clouds along Io's orbit (Skinner and Durrance, 1986; Lagg et al., 1998).

The sodium banana cloud has a measurable extent roughly 90º in longitude leading Io and the timescale to reach such distance is about 3 days at Io's escape velocity. The O lifetime against electron-impact ionization is at least 20 hours, and so it persists long enough to form a neutral torus wrapping entirely around Jupiter (Burger and Johnson, 2004), as shown in Figure 17 (Smith et al., 2019). Brown (1981) showed that faint electron-excited [O I] 630 nm could be seen remote from Io, leading it by nearly 90º along its orbital path. Thomas (1996) confirmed a similar [O I] brightness of 8.8R at the ansa just trailing Io. In both cases, the Doppler shift of this emission feature was at Io's orbital velocity. This establishes that it could not be produced by O originating from an ionic chemical pathway, which would instead reveal Doppler shifts related to the ion gyro motion.

Dissociative recombination of molecular ions could potentially produce fast atomic O and S neutrals analogous to the Na jet, but this has been difficult to establish given sparse observational evidence.

Charge exchange with the torus, rather than electron-impact ionization, was shown to be an important loss pathway for oxygen atoms from the clouds (Nerney and Bagenal, 2020). Singly charged oxygen is predominant in the torus at Io's distance and the $O^+ + O \Rightarrow O + O^+$ reaction is efficient (McGrath and Johnson, 1989), but charge exchange losses do not supply new ions to the torus. From ultraviolet monitoring with Hisaki, Koga et al. (2018a) estimated 80 O atoms/$cm^3$ in the densest regions of the torus at 5.7 $R_J$, twice the oxygen density that they determined at Io's orbit. While this is seemingly at odds with the non-detection of fast oxygen by resolved spectroscopy, the broad range of velocities in the torus would thinly spread the faint emission line, making it more challenging to distinguish than a narrowly peaked emission line of the same brightness.

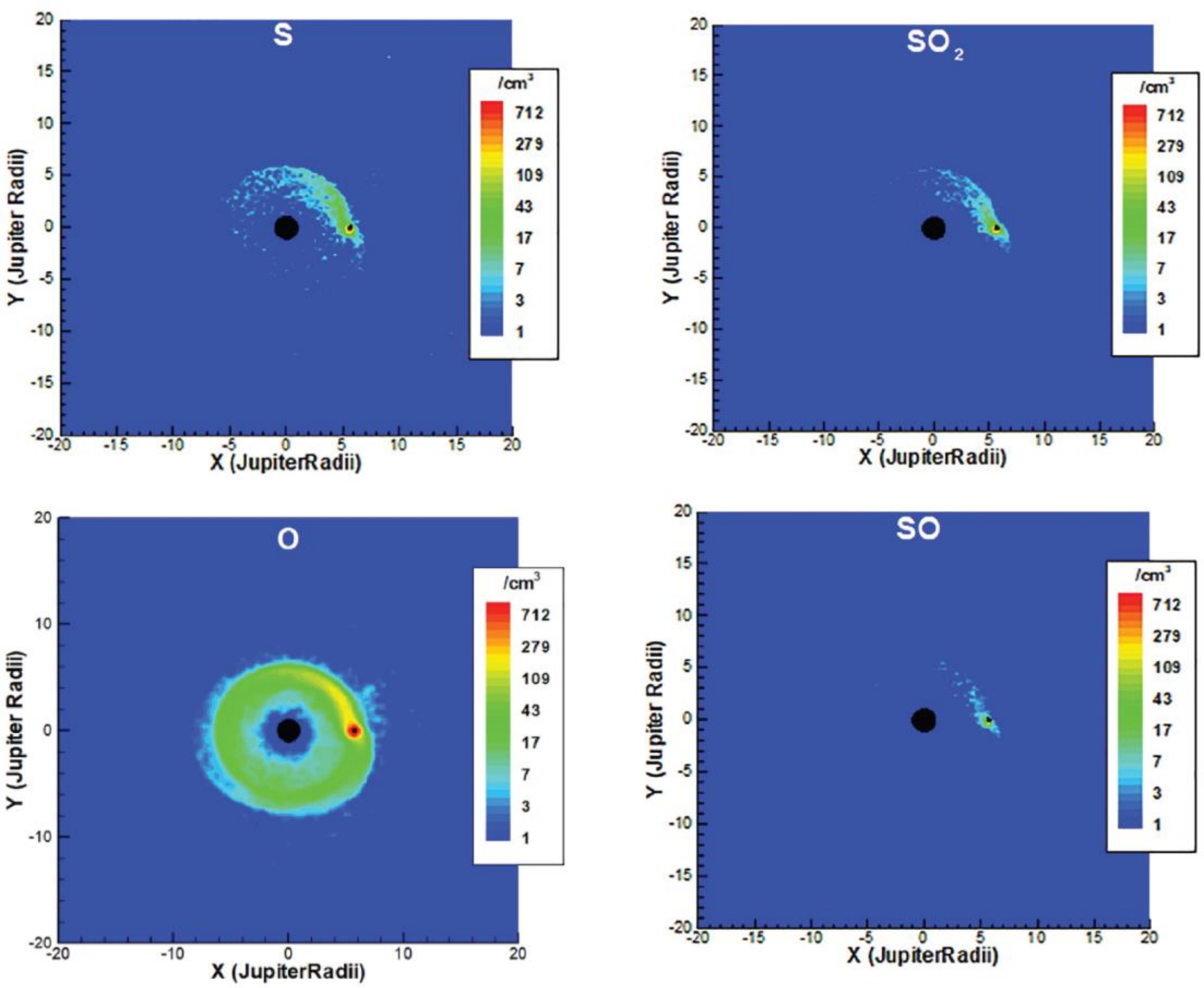

***Figure 16.*** *Modeled $SO_2$, SO, S, & O neutral toroidal clouds in the vicinity of Io's orbit from a perspective looking down onto the orbital plane (Smith et al., 2022). Due to the longer lifetime, oxygen atoms populate the complete orbit around Jupiter, with O densities exceeding the density of S, $SO_2$, and SO (except for the region very close to Io hardly resolved here).*

### 3.5.3. Transient changes in the neutral environment and connections

The transient brightenings regularly observed in the sodium nebula (Section 3.5.1) are commonly interpreted as being triggered by a change in volcanic activity at Io. So far, there is evidence for a transient brightening of the bulk neutral species only from the 2015 Hisaki observations of the O emissions from the neutral cloud (Koga et al., 2018a) which show near-simultaneous changes in O and Na (Yoneda et al. 2015). In other cases, an enhancement in torus ion emission was observed close in time to the onset of the sodium enhancement (Brown and Bouchez, 1997; Yoneda et al., 2009; Tsuchiya et al. 2018 and Koga et al. 2019).

If the sodium brightenings are indeed caused by volcanic activity, it is not known what volcanic event exactly may trigger them. It is also unclear if the sodium transient events are always connected to changes in the bulk neutral and plasma torus. Emission in the sodium nebula scales with the rate at which Io-genic sodium is neutralized, and this does not necessarily constitute a proxy for Io's bulk outgassing rates. Disk resolved measurements show that gas-phase NaCl and KCl salts and $SO_2$ gas are not (always) co-located above Io's surface (Redwing et al., 2022). A possible interpretation they propose is that plumes at low latitudes precipitate hot material onto frosts, vaporizing copious $SO_2$, while plumes at high latitudes may produce less $SO_2$ vapor and have a higher relative abundance of gaseous salts.

## 3.6. Plasma torus and sheet, energetic particles

Since the discovery of sulfur emissions (Kupo et al., 1976) observations of the plasma torus have been made by ground- and space-based telescopes, and *in situ* and remote spacecraft measurements. These observations show that most of the time the Io plasma torus appears overall stable over weeks or months, although the amount of material supplied directly or indirectly from Io to the magnetosphere is expected to vary on different time scales. Significant transient changes in the torus on time scales of days up to 2 months were inferred from different observations. It is often suggested that these changes in the torus are triggered by changes in mass supply from Io. To understand the feedback mechanisms that generally stabilize the structure and density of the Io plasma torus but sometimes allow transient changes, we need to understand the mass and energy balance in the Io plasma torus and how the system responds to changes in the source of material.

Reviews of the plasma torus can be found in Bagenal and Dols (2020), Thomas et al. (2004) or for a past-Voyager perspective in Strobel (1989).

### 3.6.1. General description of the Io plasma torus and energetic particles

Spatial distributions of the plasma torus are complex, and have characteristic radial, longitudinal and latitudinal structures in the density and temperature (e.g., Thomas et al., 2004). The radial distribution of the plasma torus consists of the *cold torus*, the *ribbon*, and the *warm torus*. The cold torus (< 5.4 $R_J$) is a region narrowly confined to the centrifugal equator in latitude, with high plasma densities. It is the only region of the torus where the ion and electron temperatures equilibrate; both populations have energies of <2 eV (Bagenal, 1994). This dense cold feature is consistent with the lack of fresh pickup ions (a few hundred eV) and a very slow inward transport time (Herbert et al., 2008) . The ribbon is a radially narrow structure located just inside Io's orbit (~5.6 to 6 $R_J$). Since the time scale of plasma transport in this region is still slow, the ions in the ribbon here are cooled from their few hundred eV pickup energy to ~20 eV through radiation. Plasmas in the warm torus (~6 to 7 $R_J$ and beyond) are thermalized (ion ~100 eV, electron ~5 eV) and contain newly ionized plasma that moves outward on timescales of tens of days.

The outward transport of plasma in the warm torus is due to the centrifugal-force-driven instability (Siscoe and Summers, 1981). As plasma transports slowly outward, the azimuthal plasma flow is accelerated by transporting the planetary angular momentum through a magnetosphere-ionosphere coupling current system (Cowley and Bunce, 2001; Hill, 1979). The Io-genic plasma is heated in the magnetosphere, and the energetic sulfur and oxygen ions become primary contributors to the plasma pressure in the plasma disk (Mauk et al., 2004).

Plasma in the torus is generated at Io's orbital distance through ionization of the neutral clouds (Section 3.5) and by ionization and pick-up from Io's atmosphere (to a smaller extent, Section 3.4). As the plasma from the torus is transported outward, ultimately becoming part of the plasma sheet in the outer magnetosphere. The plasma in the disk is finally released from the magnetosphere toward the tail region through reconnection (e.g., Kivelson and Southwood 2005, Hill 2006). The bulk convection of this material through the magnetosphere produces a dawn-dusk electric field (Ip & Goertz 1983, Barbosa and Kivelson 1983). This offsets the entire plasma torus dawnward resulting in adiabatic heating of plasma on the dusk side. The measured

UV brightness asymmetry (Murakami et al., 2016) and ribbon positions (Schmidt et al., 2018), agree on a mean field strength of 3.8 mV/m, with a spread of 1-9 mV/m that is dependent on the solar wind and plasma convection rates. Plasmoid ejection via the Vasyliunas type reconnection (Vasyliunas, 1983) is thought to be the predominant process to release mass from Jupiter's magnetosphere (McComas et al., 2007). However, the communication of plasmoid losses back to the torus at the Alfvén speed occurs on a comparable to the Jovian rotation period, and so it is challenging to connect events in the torus and events in the magnetotail unambiguously.

### 3.6.2 Stability of the Io plasma torus

The stability of the Io plasma torus in response to variable input has been discussed based on either the regulation of the escape of material from Io ("supply-limited") or the regulation of the loss from the plasma torus ("loss-limited") (Brown and Bouchez, 1997).

The system is supply-limited if an increase in plasma density in the torus causes a decrease in the escape of material from Io's atmosphere. For example, an increase in plasma precipitation into Io's atmosphere increases the ionospheric conductivity, which causes more of the plasma flow to deflect around Io, reducing plasma precipitation and subsequent atmospheric escape. It is necessary to investigate the evolution of Io's atmosphere and ionosphere and the associated changes in satellite-plasma interactions and mass exchange (Section 3.4) to better understand the supply-limited scenario.

The system becomes loss-limited if the increase in plasma supply to the torus leads to an increase in plasma loss from the torus. The centrifugal-force-driven interchange instability can become unstable if the outward gradient of plasma mass density increases due to an increase in the plasma source from Io, which is feasible with the loss-limited process. A challenge for this interchange instability may be due to heavily-loaded flux tube fingers with very small longitudinal width and thus difficult to detect even with orbiting spacecraft (Yang et al., 1994). According to the Cassini and Hisaki observations (Section 3.6.3), the loss rate from the plasma torus increases as the neutral source rate in the torus increases, which agrees with the loss-limited scenario. Also, some of the key features (e.g., changes in the radial gradient of plasma density) have been observed by the spatially resolved observations of the plasma torus (Hikida et al., 2020; Tsuchiya et al., 2018; Yoshioka et al., 2018).

### 3.6.3 Transient changes in the torus

In the last two decades, clear and significant changes in the Io plasma torus on timescales of weeks to months have been detected twice from ultraviolet (UV) observations made by Cassini/UVIS (UV Imaging Spectrograph) and the Hisaki spectroscope in 2001 and 2015, respectively. Studying these events allowed significant updates on how the system responds to changes.

***2000/2001 event.*** Observations of torus emissions made during the Cassini flyby of Jupiter (October 2000 to March 2001) showed short-term variations of the torus over a ~4-month period. The measurements of the emissions of all major ionized species allowed estimation of the density, composition, and temperatures in the plasma torus (Steffl et al., 2004a; 2004b). Delamere et al. (2004) modeled the changes seen in the Cassini data (Figure 17), inferring that the neutral source rate for torus supply changed from >1.8 tons/s to 0.7 tons/s, i.e., it changed by a factor of >2.5. The putatively increased dust rate as diagnostic for an enhanced volcanic activity before the Cassini UVIS measurements as invoked by Delamere et al. (2004) is, however, based on measurements with very large uncertainty and may be an effect of the unfavorable observing conditions only, see Section 3.8 for details.

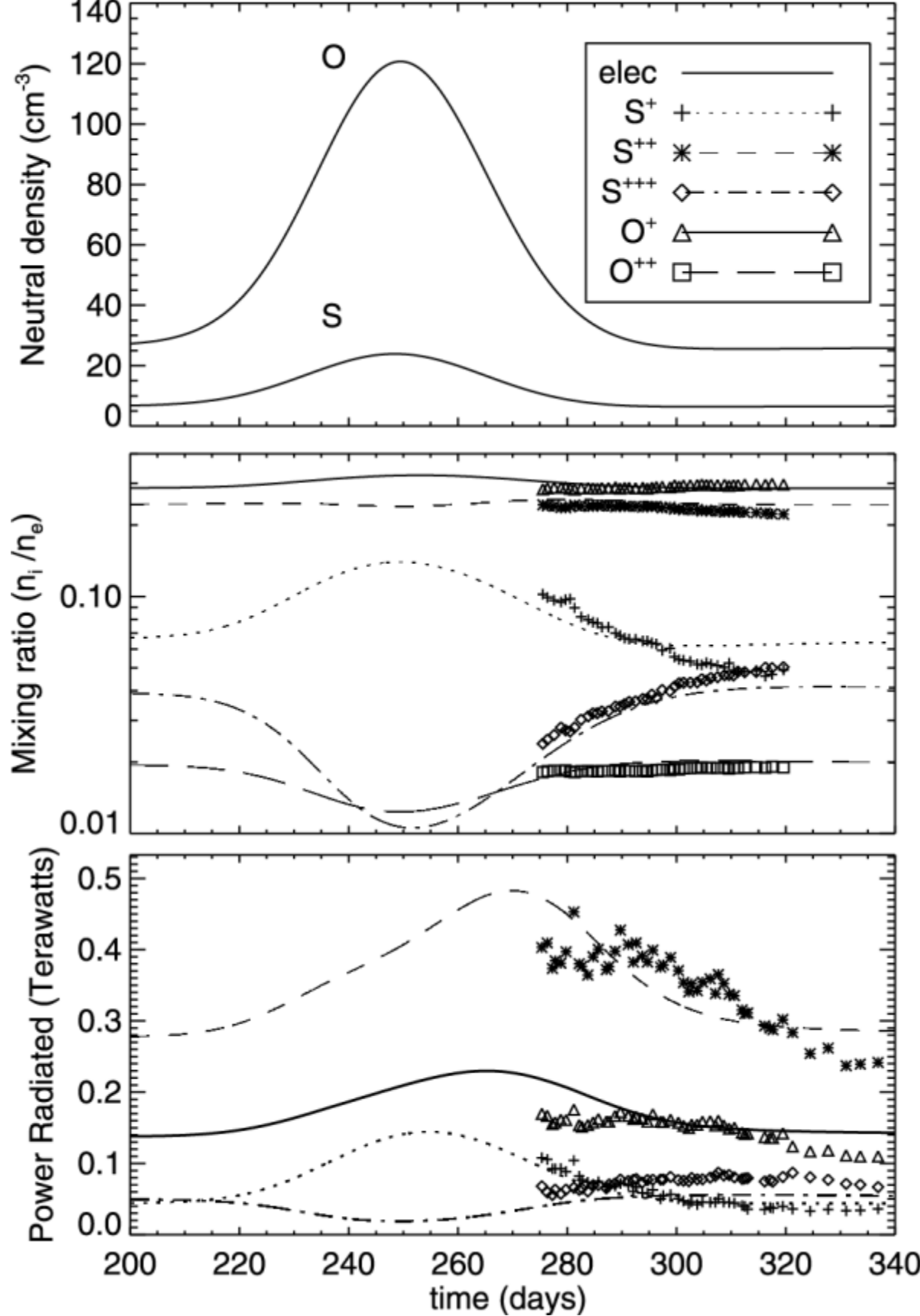

***Figure 17.*** *Neutral densities, relative ion densities and emitted power for different torus ion species modeled and inferred from Cassini UVIS observations from Delamere et al. (2004), figure 9. The declining emission intensities are consistent with a transient enhancement (before the measurements started) in neutral source rate.*

*2015 event.* The Hisaki satellite has been conducting long-term monitoring of Io plasma torus since December 2013 and has captured the response of the magnetosphere to the increase in neutrals from Io in early 2015 (Kimura et al., 2018; Koga et al., 2018a; Tao et al., 2018; Tsuchiya et al., 2018; Yoshikawa et al., 2017; Yoshioka et al., 2018). Ground-based sodium observations increased in brightness during a period from mid-January to March 2015 (Yoneda et al., 2015). Hisaki identified not only an increase in ion emissions but an increase in neutral oxygen atom emissions around Io by a factor of 2.5 which is correlated well with the increase in sodium emissions for this event (Koga et al., 2018a). The changes in neutral gas emissions and subsequent changes in singly- and multiply-ionized species suggest that the supply of neutral species and the subsequent plasma supply to the magnetosphere increased over a period of a few weeks (Figure 18). There were suggestions that specific detected hot spots (e.g., an outburst at Kurdalagon) triggered this event, but the relationship between hotspots and changes in the neutral cloud and torus is completely unclear (Section 3.1) and connections made were purely based on temporal coincidence of hot spot detections with onset of the gas emission increase. During the other observing seasons of the Hisaki satellite, a relatively stable torus with only smaller variations or long-term trends were measured (Tsuchiya et al., 2019; Roth et al., 2020).

*Voyager event 1979.* In a re-analysis of UV observations, Delamere and Bagenal (2003) found that the torus underwent a change between the Voyager 1 flyby in March 1979 and later Voyager 2 flyby in July 1979. The change in torus emissions was on a similar scale to that found in Cassini data and the authors derived a neutral source rate of 0.8 tons/s for Voyager 1 and 2.4 tons/s for Voyager 2 (Table 2). This event was not connected to specific observations of volcanic activity or the sodium nebula.

### 3.6.4 Mass and energy flow in the Io plasma torus

A physical chemistry model has enabled us to investigate mass and energy flows through the Io plasma torus (Copper et al., 2016; Delamere and Bagenal, 2003; Delamere et al., 2004; 2005; Hikida et al., 2020; Nerney et al., 2017; Nerney and Bagenal, 2020; Yoshioka et al., 2018). Figure 20 shows the derived mass and energy flow based on different plasma torus measurements. Hot electrons and pickup of fresh ions resulting from electron-impact ionization and charge exchanges are the main sources of energy for the torus. Mass loss from the plasma torus is caused by fast neutrals and outward plasma transport. In general, the fast neutral contribution is larger than plasma transport. Detailed mass and energy flows for each process are described in Nerney and Bagenal (2020).

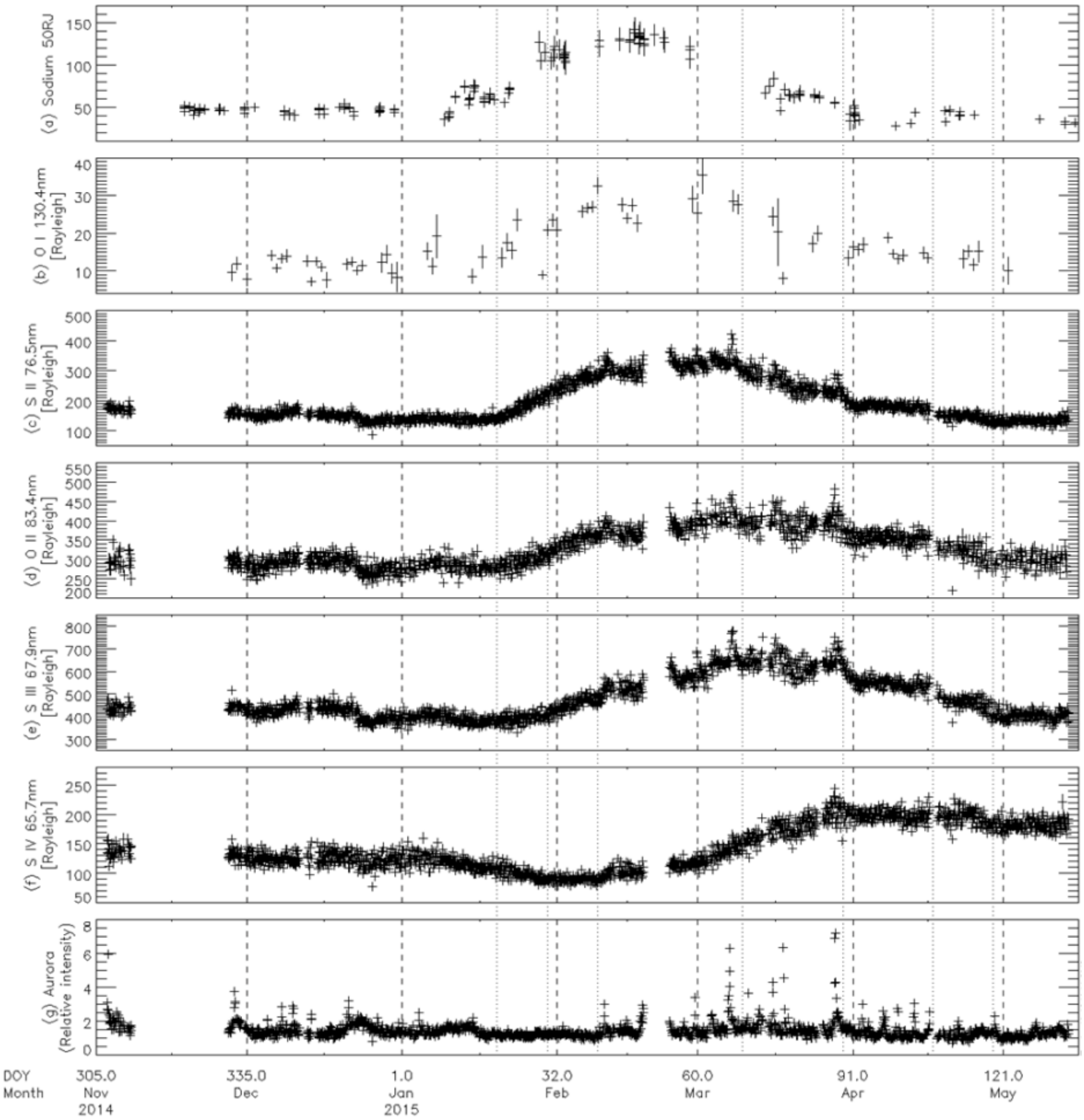

***Figure 18.*** *Io plasma torus and Jovian UV aurora variability from the end of November 2014 to the middle of May 2015. (a) Optical emission in the sodium nebula in Rayleigh units, (b–f) neutral oxygen and ion (OI 130.4nm, S II 76.5 nm, O II 83.4 nm, S III 67.9 nm, and S IV 65.7*

*nm) brightness. (g) Brightness of Jupiter's aurora from 124 to 145 nm, relative to System-III longitude-dependent brightness averaged over 2014-2015 (Tsuchiya et al. 2018).*

*Source location of the ions.* One of the remaining issues is whether the ion source location is close to the immediate region around Io (atmosphere and corona) or the neutral clouds far from Io. Simulations and Galileo measurements suggest that only a small fraction of ions (~20%) are fed directly into the torus from Io (Bagenal et al. 1997, Saur et al. 2003, and Section 3.4). The supply from the neutral clouds is not well characterized, because the bulk (S and O) neutral clouds themselves are not characterized in detail by observations due to their dim intensity (Section 3.5). Koga et al. (2018b) estimated the source rate of $O^+$ from their O neutral clouds at 400 kg/s, confirming the importance of the remote source.

*Neutral source rate and transport timescales.* Delamere et al. (2004) found that the total radiated power from the torus increases by only 25% in response to the factor ~3 change in the neutral supply rate they had derived (Section 3.6.3), and argued that the energy input is diverted by increased losses from the torus through fast neutral and outward plasma transport besides the radiation. Yoshioka et al. (2018) deduced radial distributions of the plasma torus in the spring of 2015 and revealed a higher neutral source rate (~3 tons/s) and a 2-4 times faster outward transport timescale (~10 days) than those during a quiescent period (~0.7 tons/s and ~34 days, respectively). Hikida et al. (2020) derived a time series of the neutral source rate and the transport timescale and showed that the transport timescale decreased soon after the source rate increased (Figure 19, right). The neutral source rate and radial transport time scale derived from various measurements are summarized in Table 2. The source rate varies between 0.7 tons/s and 3.1 tons/s. The loss timescale of the outward transport increases (decreases) as the neutral source rate decreases (increases). Koga et al. (2019) compared the time variations of the oxygen atom emission with that of the oxygen ion and argued that the lifetime of the oxygen ion decreased to ~20 days during the active period, which was about half of that in the quiescent period. These results suggest that the Io plasma torus is consistent with a loss-limited system (see Section 3.6.2).

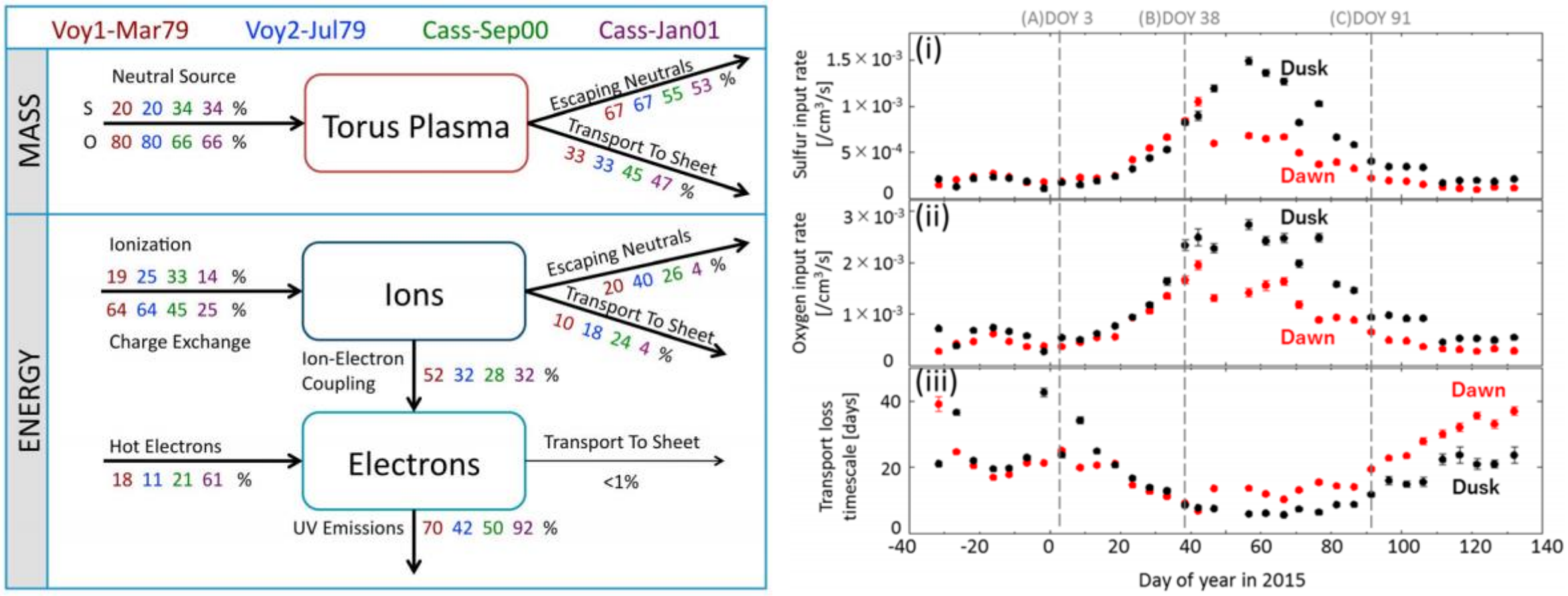

***Figure 19.*** *(a) (left) Schematic diagram of the mass and energy flow through the Io plasma torus (Bagenal and Delamere, 2011). (b) (right) The neutral source rate and outward transport loss timescale derived from the Hisaki observations in 2015 (Hikida et al., 2020).*

*Hot electrons.* The physical chemistry model also shows that a small fraction of the hot electrons is an important source of energy for the torus (Delamere and Bagenal, 2003; Delamere et al., 2005; Nerney and Bagenal, 2020) representing 0.3 to 1% of the torus thermal electron population. The source mechanism of these hot electrons remains undetermined. There are two hypotheses for the origin of this hot population. One is local heating within the magnetic flux tubes connected to the plasma torus (Coffin et al., 2022; Copper et al., 2016; Hess et al., 2011). The Galileo spacecraft observed a suprathermal electron population in the inward moving flux tube (Frank and Paterson, 2000). The beams of supra-thermal electrons within the flux tube suggest the low-altitude acceleration region.

Another idea is that the hot electrons are injected from outside the torus. Yoshioka et al. (2014; 2018) and Hikida et al. (2020) showed that the hot electron density decreases gradually with decreasing radial distance despite the short collisional cooling time scale, suggesting that global inward transport of flux tubes containing hot plasma continuously supplies hot electrons to the plasma torus. Assuming that the cooling time is determined by the Coulomb coupling between hot and core electrons, the timescales for the inward transport across the torus were estimated to be 16 ± 3 h (~2.5 km/s) during the stable torus period and 9.4 ± 1.0 h (~4.3 km/s) during the enhanced torus period. These values are in agreement with those estimated from Galileo *in situ* measurements (Hikida et al. 2020). Yoshikawa et al. (2016; 2017) and Suzuki et al. (2018) found a short-lived brightening of the plasma torus following the transient auroral brightening. The torus brightness increased by no more than 10%, and the brightening does not last long (< 24 h), indicating that the contribution of the transient event is too small to sustain the plasma torus radiation. This suggests that the hot electron injection into the plasma torus is maintained in a steady manner.

Both ideas of the source of hot electrons assume that the electrons are contained in an inward moving flux tube. Flux tube interchange motion is one of the accepted processes that transport Io-genic plasma outward and hot magnetospheric plasma inward although the spatial structure and temporal evolution of the exchange process have not yet been determined. Further observations are needed to characterize the interchange process in the inner magnetosphere and to clarify the origin of the hot electron population.

#### 3.6.5. Energetic ions

*Role for heating and stabilization.* Energetic ions have been discussed in terms of heating sources and stabilization mechanisms for the Io plasma torus. Schreier et al. (1998) considered hot ion populations diffusing inward as an external energy input to the torus. Based on measurements of the hot ion population made by the Galileo spacecraft (Mauk et al., 2004), the density of the hot ions is insufficient to explain the thermal electron temperature in the torus (Delamere et al. 2005). The outward gradient of plasma mass density in the plasma torus is sufficient to develop the interchange instability. It has been argued that there must be some regulating processes for the outward transport of Io-genic plasma to keep the torus structure stable (e.g., Thomas et al., 2004). One possibility proposed is the "ring current impoundment" (Siscoe et al., 1981; Southwood and Kivelson, 1987), where the outward gradient of torus density is balanced by an opposite pressure gradient of energetic plasmas that surrounds the torus. Mauk et al. (1998; 2004) showed that the hot plasma pressures that can impound Io-genic plasma were substantially depleted during the Galileo mission, as compared with those during the Voyager era. Ongoing Juno observation of energetic particles inside the Europa orbit will provide an opportunity to measure the hot plasma pressure and investigate whether it has a role to impede the outward transport of the Io-genic plasma.  An alternative mechanism for impeding the outward transport is "velocity shear impoundment" (Pontius et al., 1998). The velocity shear of the corotation lag in the Io plasma torus (Brown, 1994) creates vortex flow patterns in the torus though nonlinear mode coupling of the fluid, which controls the radial transport of the plasma (Hiraki et al., 2012).

*Diagnostic for the neutral environment.* The depletion of energetic ions in the plasma torus is thought to be related to the neutral cloud in the region between Europa and Io's orbits. The neutral clouds around Io and Europa are important for loss of energetic ions through charge exchange interaction (Mauk et al., 2003; 2004). Lagg et al. (1998) proposed charge exchange with neutrals as an explanation for energy dependent losses of energetic protons measured in Io's orbit, suggesting that the ion dropouts could be a diagnostic for the neutral density in the neutral cloud. However, Mauk et al. (2022) argued, based on a neutral cloud model by Smith et al. (2019), that near Io's orbit charge exchange with low energy ions would dominate over charge exchange with neutrals.

Observations of energetic ions are a useful tool to study the interaction between moons and magnetospheric plasma as well as their neutral environment. Huybrighs et al. (2024) show that dropouts of energetic protons (~100 keV) are present during close Io flybys of Galileo. A particle-tracing model demonstrates that the dropouts outside of ~0.5 Io radii are likely dominated by charge exchange with Io's atmosphere. The dropout structure is sensitive to the density and three dimensional structure of the atmosphere. Thus, measurements of energetic protons provide an additional diagnostic to investigate Io's atmosphere's structure, near Io.

***Table 2.*** *The neutral source rate and radial transport time scale derived from various measurements (Hikida et al., 2020)*

| Observing facility – period | Neutral S source rate (atoms/$cm^2$/s) | Neutral O source rate (atoms/$cm^2$/s) | Mass source rate (tons/s) | Outward transport time (days) | Refs. * |
|---|---|---|---|---|---|
| Voyager 1 – Mar. 1979† | $\sim 2 \times 10^{-4}$ | $\sim 8 \times 10^{-4}$ | ~0.80 | ~50 | (1) |
| Voyager 2 – Jul. 1979† | $\sim 6 \times 10^{-4}$ | $\sim 24 \times 10^{-4}$ | ~2.40 | ~23 | (1) |
| Cassini – Oct. 2000 | $\sim 6 \times 10^{-4}$ | $\sim 11 \times 10^{-4}$ | ~1.8 | ~27 | (2) |
| Cassini – Jan. 2001 | $\sim 2 \times 10^{-4}$ | $\sim 4 \times 10^{-4}$ | ~0.7 | ~64 | (2) |
| Hisaki – Nov. 2013 | $(1.4 \pm 0.3) \times 10^{-4}$ | $(3.4 \pm 1.0) \times 10^{-4}$ | $0.70 \pm 0.33$ | $34 \pm 7$ | (3) |
| Hisaki – Jan 2015 (DOY3) | $(1.8 \pm 0.1) \times 10^{-4}$ | $(4.6 \pm 0.2) \times 10^{-4}$ | $0.70 \pm 0.02$ | $24.4 \pm 0.6$ | (4) |
| Hisaki – Feb 2015 (DOY38) | $(8.3 \pm 0.2) \times 10^{-4}$ | $(20 \pm 0.6) \times 10^{-4}$ | $3.13 \pm 0.09$ | $8.9 \pm 0.1$ | (4) |
| Hisaki – Feb 2015 (DOY52) | $(7.6 \pm 0.9) \times 10^{-4}$ | $(13 \pm 2.1) \times 10^{-4}$ | $3.0 \pm 0.3$ | $9.9 \pm 0.9$ | (3) |
| Hisaki – Apr 2015 (DOY91) | $(3.1 \pm 0.1) \times 10^{-4}$ | $(8.00.2) \times 10^{-4}$ | $1.22 \pm 0.03$ | $15.7 \pm 0.4$ | (4) |

* (1) Delamere & Bagenal (2003), (2) Delamere et al. (2004), (3) Yoshioka et al. (2018), (4) Hikida et al. (2020)

† The Voyager numbers may need to be adjusted for updated calibrations of the UVS instrument as reported by Queremais et al. (2013)

## 3.7 Jupiter's aurora and connections to the Io torus

Various auroral features have been identified and characterized by observations as diagnostics for the state of the magnetosphere. The main auroral structures—as seen from mid-latitudes towards the poles of Jupiter—are the footprint aurora (Io, Europa and Ganymede) and the low-latitude emission, the main auroral emission, and finally the polar emission. The main aurora is additionally subdivided into three different zones (Mauk et al., 2020). In this section, we mainly focus on auroral observation and modeling studies related to the magnetospheric mass balance and changes in the torus environment. For general details of the auroral process and dynamics, see review papers by Badman et al. (2016), Grodent (2015), or Mauk et al. (2020). Io's auroral footprint is discussed in Section 3.4, as it relates to the local interaction of Io's atmosphere with the surrounding magnetosphere.

### 3.7.1. Main emission and mass balance

The rotational motion of out-flowing Io plasma in Jupiter's magnetosphere is considered to be maintained by the transfer of angular momentum from the gas giant to the plasma itself (e.g., Hill, 1979). In the standard picture before the Juno mission, the main aurora was suggested to be related to the quasi steady-state field-aligned current system produced in such angular momentum transfer (e.g., Cowley and Bunce, 2001; Hill, 2001). Theoretical and numerical models indicate that the location of the main aurora would be shifted toward lower latitude in the case of increased plasma mass loading to the torus, because the momentum transfer is supposed to occur efficiently over a more limited radial distance of the equatorial magnetosphere (e.g., Nichols, 2011; Nichols and Cowley, 2005; Tao et al., 2010; Ray et al., 2012). Nichols (2011) found that the correlation or anti-correlation of the field-aligned current with the mass-loading rate would depend on the assumption in the model, i.e., whether the cold plasma density depends on the mass-loading rate or not. However, recent Juno observations complicate this paradigm, and suggest that the main aurora is predominantly caused by broad-band bi-directional electron beams, which can deposit up to 3000 $mW/m^2$ (Mauk et al., 2017; 2020; Salveter et al., 2022). These electron distributions may be generated by highly time-variable, turbulent electric currents and fields caused by the radial transport constantly perturbing the magnetosphere (Saur et al., 2018). The ionospheric Alfvén resonator is proposed to produce additional high frequency waves (Lysak et al., 2021). A simulation study considering dispersive scale Alfvén waves shows that a large ratio between the torus and high-latitude densities can act to enhance the broadband aurora (Damiano et al., 2019).

### 3.7.2. Auroral signatures connected to transient torus events

Some aurora observations are believed to be connected to events in the torus and at Io. Bonfond et al. (2012) found that the main aurora expanded to lower latitudes - up to equatorward of the Ganymede footprint location - and that the occurrence rate of large equatorward isolated auroral features increased during a period in May and June 2007, close in time to a brightening in the sodium nebula (Yoneda et al. 2009). The increased occurrence of equatorward isolated features was attributed to injection, replacing a large amount of outward-moving heavy flux tubes with flux tubes sparsely filled with hot plasma. The Io footprint aurora disappeared (power <1 GW) compared to other footprint observations at similar Jovian System-III longitude at Io, where the power is usually around 3-6.5 GW (Bonfond et al., 2012). For the same event, the activity of hectometric radio emission (HOM), which is an indicator of Jupiter's auroral particle acceleration, is decreased (Yoneda et al., 2013).

Bonfond et al. (2012) interpreted the auroral characteristics observed in 2007 as the result of an increase in mass loading triggered by Io. We note, however, that a larger survey of Jupiter's aurora by Grodent et al. (2018) found that the aurora revealed similar features in 18,5% of all observed cases in a period between November 2016 and July 2017, during which both the sodium nebula and torus ion emissions were constantly at a low and stable level (Roth et al., 2020).

During the transient changes in the Io torus in January-March 2015 (Section 3.6), simultaneous monitoring of plasma torus emission and polar-integrated auroral spectra showed interesting responses indicating magnetospheric dynamics (see Figure 18, bottom panel with auroral intensity). Auroral sporadic enhancements lasting less than ~10 h were sometimes observed, followed ~7-20 h (average 11 h) later by sporadic enhancements of the ion brightness in the plasma torus (e.g., Yoshikawa et al., 2016). The sporadic auroral enhancements would represent transient energy input to the ionosphere and were linked with auroral signatures of injections between the main oval and the Io footprint (Kimura et al., 2015), which may have been driven by reconfigurations in the outer magnetosphere, as shown in Figure 20.

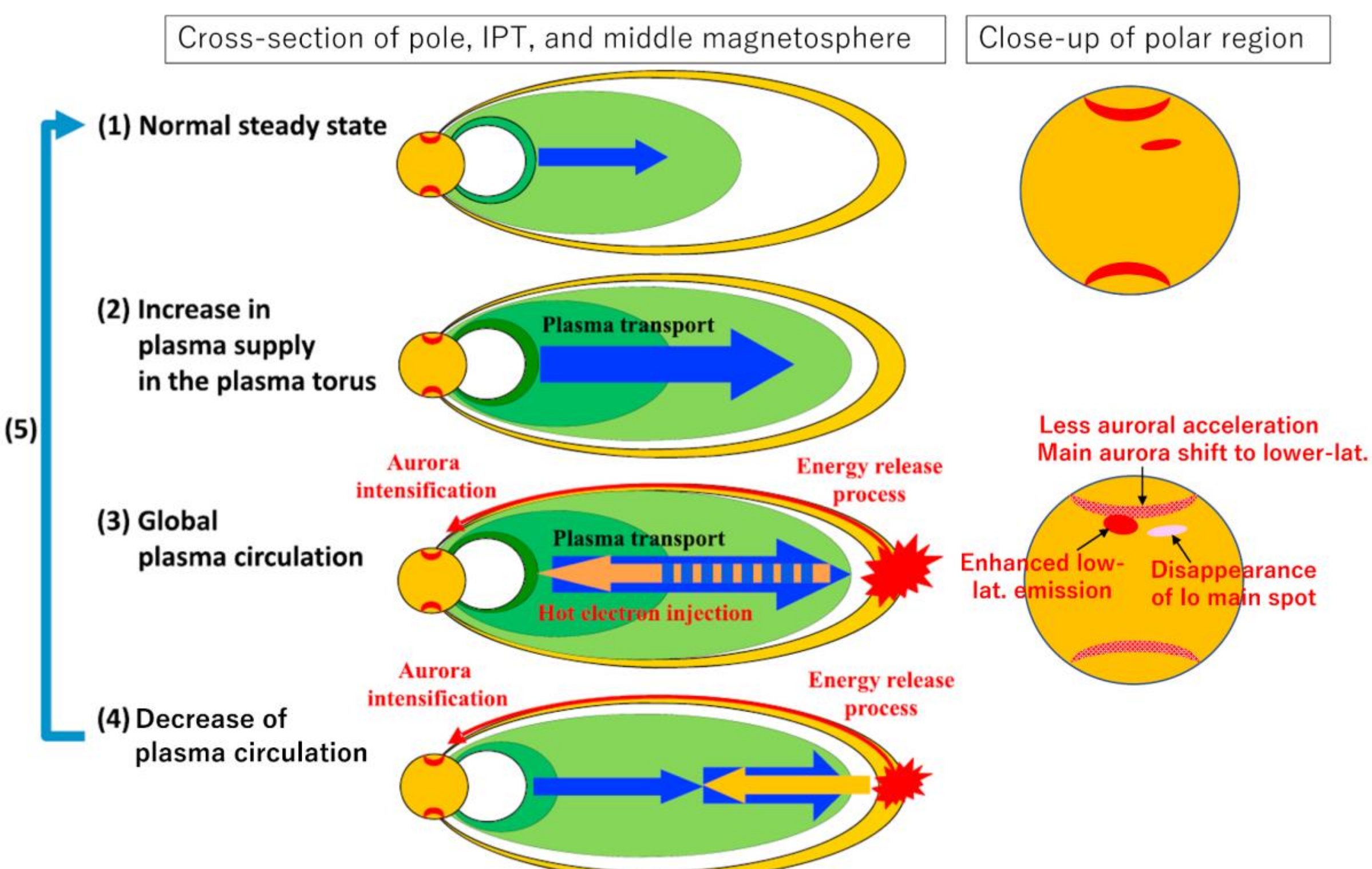

***Figure 20.*** *Schematic illustration of the effect of Io's volcanic activity enhancement on the Jovian magnetosphere divided into five time-phases (left part, from Tsuchiya et al., 2018). From normal steady state (1), increase in plasma supply to the plasma torus as the phase (2), thermal plasma originating from Io (dark and light green areas) extends, followed by enhanced outward transport (blue arrow) of Io-genic thermal plasma (3) and inward injection of hot plasma (orange arrow) (4). Then return to the normal steady state (5). Several auroral variations during the event time are shown (right part). Change of the magnetic structure is not shown for simplicity. See the text and Tsuchiya et al. (2018) for further details.*

The changes in the torus emissions indicate an enhancement of the hot electron population in the inner magnetosphere. Pairs of these intensifications were frequently identified from ~20 days after the start of torus S+ emission increase and until the decrease in emissions to the common lower level. After that only the auroral intensification continued (Tsuchiya et al., 2018). The ~11 h time delay of a torus brightening from a corresponding aurora intensification did not change compared with the lower standard torus state (Yoshikawa et al., 2017). Auroral sporadic intensifications are much larger and more frequent during the enhanced torus emission interval (Kimura et al., 2018; Tao et al., 2021). A change in the auroral spectrum during the enhanced torus emission interval indicates a decrease in auroral electron energy and higher density magnetospheric source plasma in the middle magnetosphere (Tao et al., 2018).

## 3.8 Dust from Io

Io is a persistent source of dust in the Jovian magnetosphere. Grains released from Io, either ejected via impact bombardment from interplanetary dust grains or volcanic activity, become charged and experience the force of Jupiter's gravity and electromagnetic fields. Since the discovery of Io's strong volcanism in 1979 (Smith et al., 1979; Morabito et al., 1979), it has been proposed that dust grains from volcanic plumes are injected continuously into Jupiter's magnetosphere through electromagnetic forces (Johnson et al., 1980; Morfill et al., 1980). The first observational evidence for the dust particles was provided when the Ulysses spacecraft flew by Jupiter in 1992 and the onboard dust detector measured periodic bursts of sub-micrometer dust particles within 1 AU from Jupiter. These dust particles were measured in dust streams radiating from the direction of Jupiter, indicating that the periodic bursts of dust come from the Jovian system (Grün et al., 1993a; 1993b). Somewhat similar to the readily observable trace species like sodium, dust is used to probe for variability in the Io-Jupiter system and was sometimes even connected to volcanic activity. We focus here on the smaller dust grains (~0.01 µm), which were found to trace back to Io, while the larger grains are found to originate from a variety of sources in the Jovian system (e.g., Graps et al., 2000, Liu and Schmidt, 2019). A general review of the Jovian dust environment can be found in Krüger et al. (2004).

### 3.8.1. Galileo dust measurements and possible connections to volcanic activity

More detailed insight into the Jovian sub-micrometer dust environment was given by long-term in situ dust measurements of the Galileo spacecraft mission. The frequency analysis of the Galileo dust detector (DDS) data by Graps et al. (2000) led to the direct evidence of Io's volcanoes being the main dust source in the Jovian magnetosphere. Impact ejecta from Io was ruled out as a dominant source of dust for the dust streams (Krüger at al., 1999). Furthermore, it was shown that the stream particles are strongly coupled to Jupiter's magnetic field (Grün et al., 1998).

Using Galileo's measurements of the Jovian dust streams as a monitor for its volcanic activity, Krüger et al. (2003a) conducted a study to examine the orbit-to-orbit variability of the dust emission and link it to the volcanic activity on Io. The eruptions of large Pele-type plumes are expected to contribute most to the dust escape on Io (Krüger et al. 2003b) as only they may be able to accelerate the dust grains to high altitudes so they can escape Io's gravity (Johnson et al. 1980, Ip 1996). The temporal coverage of direct sightings of plume activity during the Galileo mission is very limited (McEwen et al., 1998; Keszthelyi et al., 2001; Geissler and McMillan, 2008) and therefore makes it complicated to correlate the sightings to dust observations. A better time coverage of plume activity is provided by observations of surface changes due to eruptions (Geissler, 2003), but these surface changes do not provide a precise date of the eruption.

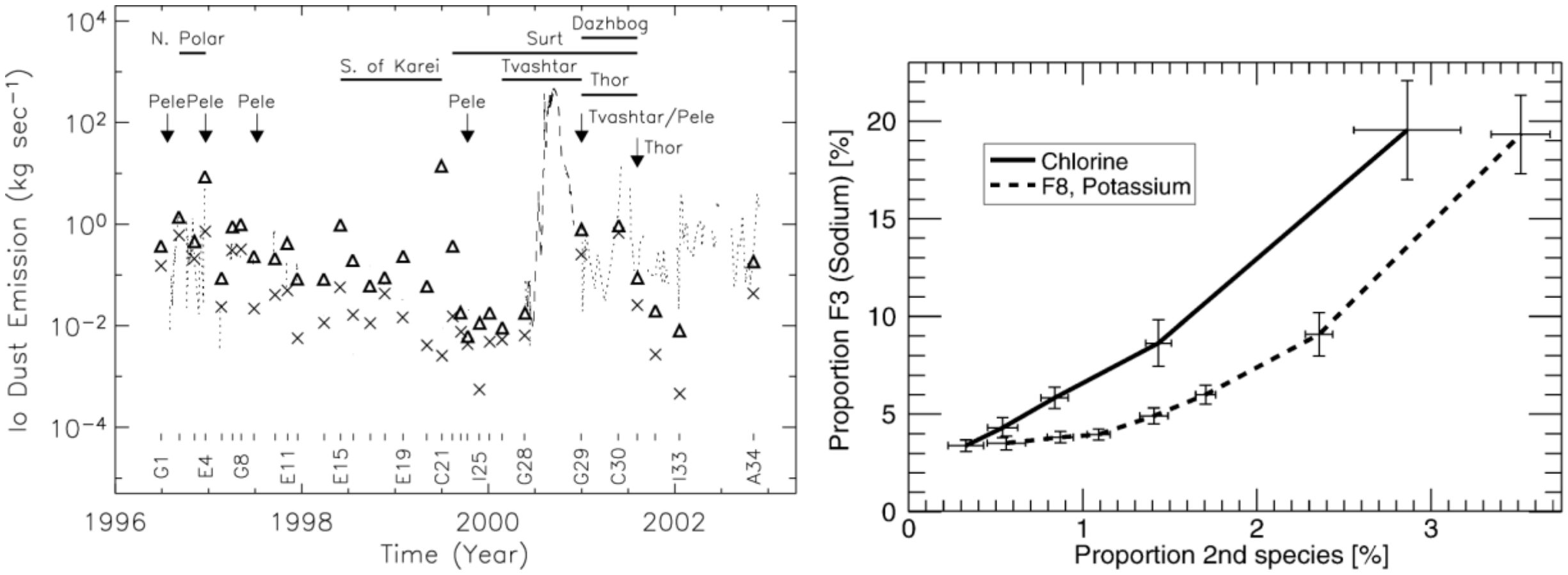

***Figure 21.*** *(Left) Calculated dust emission rate of Io using Galileo observations. Triangles and crosses denote the maxima and minima derived from measurements when the Galileo spacecraft was at distances of 13–30 $R_J$ to Jupiter, respectively. The dashed line is for the G28 orbit when Galileo was at distances of 30–280 $R_J$, dotted lines show the remaining orbits with distances of 30–400 $R_J$. Horizontal bars indicate periods when large-area surface changes occurred on Io, arrows indicate individual plume sightings. Note that the duration of the eruptions is not known. Galileo flybys are indicated at the bottom. From Krüger at al. (2003b). (Right) The clear correlation of the Na+ (measured in data feature F3 on y axis) with Cl+ (black solid) ions in the Cassini measurements suggest NaCl as a major dust component (from Postberg et al., 2006).*

Figure 21 (left) shows the derived minimum and maximum emission rates by crosses and triangles, respectively, for measurements taken when the spacecraft was at distances between 13 and 30 $R_J$. Horizontal bars represent periods when large-area surface changes were observed (Geissler, 2003). Arrows show the time of individual volcanic plume sightings, note that the duration of the eruptions is not known. After ejection from Io, the escape of the dust particles from the torus is influenced by the dawn-to-dusk asymmetry of the plasma torus as grains are charged and experience electromagnetic forces. Due to the different charging conditions at dawn and dusk, grains on the dusk side preferentially escape with timescales $\lesssim$1 hour, while grains on the dawn side reside longer in the torus, escaping with timescales of ~1 day (Horányi et al., 1997). After grains leave the torus, they take several hours to travel to a distance of 30 $R_J$ (Krüger et al., 2003b). Therefore, the particles arrive within 1-2 days at the Galileo spacecraft for the derived dust emissions shown in Figure 21 (left). Krüger at al. (2003b) derived a typical average dust emission rate of 0.1 to 1 kg/s in their most precise measurements. These rates imply that the dust constitutes only about 0.01 to 0.1% of the total mass (assuming the canonical number of 1 ton/s for the loss of mass from Io) (Krüger et al., 2004).

In many cases, the time of the giant plume eruptions match the time periods when increased dust emissions were detected suggesting that dust measurements may provide an effective monitor of Io's volcanic activity (Krüger at al. 2003b). However, the total duration of the eruptions is not known, and the lack of a plume detection does not mean there is no ongoing eruption. These two uncertainties complicate this interpretation.

Converting the local dust fluxes measured by Galileo to estimates of total dust output from Io requires assumptions on their outward radial transport. Therefore, measurements farther from Io have more uncertainty with respect to estimating total Io dust emission. Hence, the large dust emission rate of about 100 kg/s (dashed line between G28 and G29 labels in Figure 21) should be accepted with caution because during orbit G28 Galileo was located far away (about 280 $R_J$) from Jupiter (Krüger et al., 2003b).

### 3.8.2. Cassini dust measurements: Composition of the dust particles

While the Galileo measurements provided a long duration of dust measurements, it lacked the ability of further characterizing the dust particles' composition. The measurements by the Cosmic Dust Analyser (CDA) onboard the Cassini spacecraft taken during the Jupiter flyby in 2000 provided first constraints on the dust particle makeup (Postberg et al., 2006). Sodium and chlorine ions were the most detected species from the dust and their correlation (Figure 21, right) suggested sodium chloride (NaCl) to be the primary dust particle constituent. In addition, sulfur-bearing as well as potassium-bearing components were identified. Postberg et al. (2006) interpret the primarily alkali composition of the dust as an indication that >95% of the measured particles originate from Io and its volcanoes.

The Cassini measurements started on September 4 in 2000, potentially capturing the end of the putative enhancement in the Galileo data around September (Figure 22). However, Cassini was at a large distance from Jupiter (>1 AU, on approach) at this time and an anomaly in the Cassini dust counts for this time is not mentioned in Postberg et al. (2006).

Finally, based on observations of a particular feature in sodium gas emissions Grava et al. (2021) showed sodium atoms may be sputtered from charged dust grains escaping from Io. This result supports the hypothesis that dust particles may be an important carrier of alkalis that ultimately populate the neutral clouds and extended neutral nebulae (Section 3.5).

# 4. Connections in the system and transient events: What we know and what we do not know

## 4.1 Current understanding for normal (stable) conditions

Altogether, there is a qualitatively consistent understanding of how material from Io feeds into and is distributed over the Jovian system for stable conditions. "Normal conditions" refers here to a stable torus as observed by near constant emissions and in-situ measurements over several weeks to months (a Jupiter observing season is ~6-8 months per year for Earth-bound observations) and that otherwise no unusual conditions are observed in the magnetosphere, like a brightness increase in the neutral clouds or nebulae.

Material is ejected from volcanic sites from the subsurface, delivering volatiles to the atmosphere and surface. Sublimation of surface frost deposits (50% - 80% atmosphere source) and the direct outgassing at volcanic sites (20% - 50% atmosphere source) sustain Io's atmosphere. The atmosphere reveals strong lateral (longitudinal and latitudinal) and diurnal density variations but appears to have a stable averaged $SO_2$ abundance on the dayside (Section 3.2). Despite potentially varying volcanic outgassing, the atmospheric stability is likely maintained by the effects of the sublimated fraction (which maintains vapor pressure equilibrium) and possible mutual effects between outgassed and sublimated gases. The bulk $SO_2$ atmosphere is then eroded primarily from the interaction with the surrounding plasma. This creates new torus ions locally at Io (roughly 200-300 kg/s) and ejects atomic and molecular neutrals into the neutral clouds in and near Io's orbit (ionized later in the torus leading to supply of fresh torus ions) and into the extended neutral nebulae (never added to plasma torus), see

Section 3.4. All other processes that allow volatiles to escape Io and be added to the neutral clouds or plasma torus are at least an order of magnitude lower and are thus expected to be only secondary contributions to the supply of new ions into the torus (Section 3.3, Table 1).

Electron-impact ionization of the bulk neutral cloud gases constitutes the main production of plasma sourced into the plasma torus (Section 3.5). Finally, there is a net radial outward transport of plasma (on a time scale of 10-60 days, Section 3.6 and Table 2), which feeds the Io-genic material into the outer torus and then the plasma sheet, which extends far out into the magnetosphere. At the radial distance where essential momentum input is required to maintain corotation of the plasma, field-aligned currents lead to energy transport processes along the magnetic field lines causing the main emission in Jupiter's aurora (Section 3.7).

The potential positive feedback on the mass supply from Io that would be expected because the loss depends on the torus plasma density (via collisions of plasma with the atmosphere and neutral clouds) is constrained by one or several limiting mechanisms: The outward transport was shown to be faster during times of enhanced torus density which suggests that a loss-limited mechanism is effective (Section 3.6). The diversion of the incoming plasma due to the plasma-atmosphere interaction can work as a balancing factor by limiting the supply, although simulations suggest only minor effects (Section 3.4).

Although it is still not fully understood which processes drive the mass transport through the magnetosphere, there is a relatively consistent picture of the mass fluxes, pathways and time scales of mass transfer in the Io-Jupiter system for the stable conditions. The limiting mechanisms that maintain the stability of the torus density should also constrain the effects on the torus of changes at Io.

## 4.2 Canonical number for mass supply

The mass rate of ~1 tons/s was first derived by Broadfoot et al. 1979 based on the assumption that the power radiated away in the extreme-UV (EUV) and far-UV (FUV) is balanced by energy input from the pickup of freshly produced ion, which are entrained in the local bulk plasma flow at the local velocity. Hence, the canonical number is the rate of neutrals (kg/s or particles/s) removed by the interaction with the Io plasma torus (primarily electron-impact ionization and charge exchange) from the neutral clouds and Io's corona. In modeling papers it is called the torus *neutral source rate* or the neutral source strength (e.g., Delamere et al. 2004), because it describes how many neutrals must be re-supplied to the neutral clouds to balance the loss to the torus.

The electron-impact ionizations of the neutrals supply additional plasma to the torus without plasma losses in the same processes. This is thus net *mass-loading* of the torus. Charge exchange results in a new slow ion and converts an "old" fast-moving torus ion into a fast neutral that leaves the system. Both ionizations and charge exchanges contribute to the supply of energy to power the torus UV emissions and sustain the torus ion and electron temperatures. (Hot electrons also make a significant contribution to the energy input to the torus, see Section 3.6.)

In equilibrium, the neutrals removed from the neutral clouds are resupplied from Io. To contribute to the neutral clouds, neutrals from Io must reach a sufficient velocity to overcome Io's gravitation and at least reach the Hill sphere, where particles could continue on orbits bound to Jupiter. At the surface this velocity is 2.33 km/s. The escape to infinity, the escape velocity is 2.56 km/s. Neutrals ejected at lower speeds supply a corona that remains bound to Io. Neutrals ejected at speeds faster than the Jovian escape velocity at Io's orbit (25 km/s in Jupiter's reference frame, while Io's orbital velocity is ~ 17 km/s) escape the Io system on hyperbolic trajectories and do not provide neutrals to the neutral clouds or plasma torus. Instead, they contribute to the formation of the nebulae. In addition, some of the material likely migrates radially inwards (also forming the cold torus).

Thus, the canonical rate (or *neutral source rate*) of the plasma torus does not equal the full mass loss rate from Io but instead represents a lower limit for Io's total neutral loss and the energy needed to support the UV power radiated by the plasma torus. The full mass loss rate from Io's atmosphere consists of the neutral source rate, the rate of ionization of the atmosphere in the local slowed plasma flow at Io, and the rate of fast neutrals being ejected from Io into regions beyond the neutral clouds (and possibly beyond the magnetosphere).

## 4.3 Transient events in the plasma torus, neutral clouds and nebula, and aurora

As reviewed in Section 3, several phenomena observed in the Jovian system indicate significant transient changes in the magnetosphere and are often explained by some change in volcanic activity. It is argued that this volcanic event enhances the mass output from Io over some short period of time. Primarily, these are observations of

(1) significant changes in *plasma torus* UV emissions, or

(2) an increase in the brightness of the *sodium cloud* or *sodium nebula*, or

(3) a particular morphology or periodic intensifications of the *Jovian aurora*.

A change in the bulk (oxygen or sulfur) neutral cloud, which has been observed for O in the 2015 plasma torus event, would be a diagnostic as well but has never been detected independently of a detected change in torus plasma density.

Table 3 lists all events published in the literature of significant changes in the bulk torus (1) and one event in 2007 where an increase in the sodium nebula (2) was observed as well as a change in the auroral morphology (3). Events where only an increase in sodium nebula brightness was reported (e.g., Wilson et al., 2002; Mendillo et al., 2004; Morgenthaler et al., 2019) or only auroral signatures potentially indicative of enhanced mass output from Io were detected are not listed because of the following reasons. Brightenings of the sodium nebula were observed relatively frequently (about 7 observed instances reported since 1990) and the abundances and pathways of the trace species sodium may not be representative for the bulk mass abundance and transfer in the system (Section 3.5). Thus, Na changes may not always coincide with changes in the bulk torus and reconfigurations of the magnetosphere. Jupiter's aurora is shaped and affected by various magnetospheric and external processes, and we therefore consider it not a reliable diagnostic for changes triggered by Io. The caveats with sodium and aurora observations as diagnostics are discussed in more in Section 4.4.

### 4.3.1 Time scales of transient events

Out of the five events listed in Table 3, two relate to measurements of spacecraft visiting Jupiter: Voyager 1 and 2 (1979) as well as Cassini on the inbound and outbound leg (2000/2001). In these cases, the timeline of the variations in the torus is difficult to determine. The flybys of Voyager 1 and 2 happened ~4 months apart and the change (increase from Voyager 1 to 2) in the torus emissions as inferred by Delamere and Bagenal (2003) thus must have happened in between the flybys. The Cassini torus UV observations revealed a decrease in emissions from the start of the observations over a period of about 50 days. However, the timelines of the increase and the high emission phase were not observed and could only be projected in simulations (Figure 18). In both cases modeling suggests a change in the net supply by approximately factor 3 (Delamere and Bagenal, 2003; Delamere et al., 2004).

The event observed by Brown and Bouchez (1997) suggests a period of about 25 days of increasing torus sulfur ion emissions followed by a declining phase of roughly 50 days (Figure 2, left). The simultaneously monitored sodium cloud (banana) emissions seem to increase much more rapidly (within <10 days). Due to the relatively large statistical spread of the observed brightnesses and gaps in temporal coverage, these inferred times have some uncertainty. For the 2015 event, the plasma torus, neutral oxygen cloud, and sodium nebula emissions were monitored at high cadence. In this case, both the neutral oxygen cloud emissions and the sodium nebula (up to 50 $R_J$) followed a similar timeline with an increase phase (including possibly a high stable phase) of around 50 days, as well as a declining phase of ~40 days. Hence, the total transient event in the neutrals lasted for about 3 months. For the singly charged torus ions ($S^+$), the onset is close to the onset for the neutrals due to the short lifetime in the neutral clouds and the declining phase is somewhat longer. The cadence of production of multiply charged ions accounts for the lag of their emissions.

***Table 3.*** *Major transient events in the magnetosphere reported in the literature. Events where only a brightening of the sodium nebula or a possibly diagnostic change in auroral signatures were observed are not listed.*

| Year | Detected changes | Facility / Observable | Period / Length | | Supply change | Comments, References |
|---|---|---|---|---|---|---|
| **2015** | Transient brightening in emissions from neutrals and plasma | Hisaki / UV torus ion and neutral emission | January - May 2015 | ~4 months | Transient increase by factor ~4 | Best monitored torus event so far. Often related to hot spot at Khurdalagon but evidence for relation is lacking. E.g., Yoshikawa et al. (2017); Kogal et al. (2018a), Yoneda et al. (2015); and Section 3.5 |
| | Transient brightening of emissions from Na nebula | Telescope at Mt. Haleakala / Na optical | January - April 2015 | ~3 months | | |
| | Auroral signatures indicative of magnetospheric dynamics | Hisaki / Jupiter UV aurora intensity | | | | |
| **2007** | Transient brightening of emissions from Na nebula | Telescope at Mt. Haleakala / Na optical | in May 2007 | ~1 month | n/a | Putatively connected to the Tvashtar plume observed by New Horizons. Second event that was connected to Tvashtar, but possibly only a coincidence since in both cases a spacecraft with imaging capability happened to image a plume. Yoneda et al. (2009); Bonfond et al. (2012) |
| | Expanded main emission and equatorward features in aurora | HST, Jupiter UV aurora imaging | May/June 2007 | | | |
| **2001** | Change in torus density and charge states between on inbound Cassini measurements (stable low emissions on outbound trajectory) | Cassini UVIS, UV torus ion emissions | Oct - Dec 2000 | 2 months | Decrease by factor ~3 | Putatively connected to Tvashtar plume observed by Cassini. The dust measurements are uncertain and were not connected to Tvasthar or any particular volcanic event in the original paper, see Section 3.8 (Krüger et al., 2003b) Delamere et al. (2004) |
| | Potential increase in dust flux | Galileo dust detector, sub-µm dust particles | Sep 2000 | | | |

| Year | Detected changes | Facility / Observable | Period / Length | | Supply change | Comments, References |
|---|---|---|---|---|---|---|
| 2015 | Transient brightening in emissions from neutrals and plasma | Hisaki / UV torus ion and neutral emission | January - May 2015 | ~4 months | Transient increase by factor ~4 | Best monitored torus event so far. Often related to hot spot at Khurdalagon but evidence for relation is lacking. E.g., Yoshikawa et al. (2017); Kogal et al. (2018a), Yoneda et al. (2015); and Section 3.5 |
| | Transient brightening of emissions from Na nebula | Telescope at Mt. Haleakala / Na optical | January - April 2015 | ~3 months | | |
| | Auroral signatures indicative of magnetospheric dynamics | Hisaki / Jupiter UV aurora intensity | | | | |
| 2007 | Transient brightening of emissions from Na nebula | Telescope at Mt. Haleakala / Na optical | in May 2007 | ~1 month | n/a | Putatively connected to the Tvashtar plume observed by New Horizons. Second event that was connected to Tvashtar, but possibly only a coincidence since in both cases a spacecraft with imaging capability happened to image a plume. Yoneda et al. (2009); Bonfond et al. (2012) |
| | Expanded main emission and equatorward features in aurora | HST, Jupiter UV aurora imaging | May/June 2007 | | | |
| 1992 | Transient changes in sulfur ions torus and neutral sodium nebula emissions | Telescope at Lick Observatory | Mar - May 1992 | 2-3 months | Transient increase by factor ~2 | First published observational evidence for short-term changes. No independent observations. The Galileo mission arrived 3 years later. Brown and Bouchez (1997) |
| 1979 | Change in the torus density and charge states between Voyager 1 and Voyager 2 flybys | Voyager 1 in-situ plasma; Voyager 1 and 2 plasma UV emissions | Voy1: Mar 1979 Voy2: Jul 1979 | 4 months or less | Increase by factor ~3 | Re-Analysis of Voyager UV data in 2003. Delamere and Bagenal (2003) |

The length of the declining phase of the transient events in torus and neutral gas is consistent with a period of around 1-2 months, somewhat longer than but similar to the timescale for the outward radial transport (Section 3.6). The length of the increase period is usually associated with the length of a putative change in supply from Io but may also relate to the timescales of the atmosphere (lifetime of ~10 days, Section 3.2) or of a transient mechanism that increases the loss from Io until a new limit and equilibrium are reached.

The 2007 observations show a relatively short transient enhancement of the sodium nebula for only 10 days. The observed aurora changes are first seen during this 10 day period and continued thereafter for at least a few days (Bonfond et al., 2012). Given the uncertainties in the relation of the sodium and auroral features to the bulk neutral gases and plasma torus, it is not worth estimating or interpreting time scales for this event.

### 4.3.2 Inferred changes in the neutral source rate for the torus

Through modeling of the mass and energy flows in the torus, effective supply rates as well as transient changes in these rates were inferred for the three events where UV torus emission enhancements were monitored (Table 2, Section 3.6). According to the modeling, the total mass supply rate under normal conditions is mostly around 0.7 tons/s, so somewhat lower than the canonical number of 1 tons/s. During transient events an increase of a factor of 3-4 is derived for the three cases with highest neutral source rates around ~3 tons/s.

The enhancement in neutral oxygen emissions around Io's orbit observed by Hisaki for the 2015 event is key evidence that changes in the torus are preceded by a change in the bulk neutral clouds, at least in the one case for which monitoring of neutral oxygen emissions exists (Section 3.5). This supports the hypothesis that a change of the supply of neutrals from Io to the neutral clouds precedes and possibly causes the transient changes in the plasma torus and magnetosphere.

We note again that these quantified inferred changes relate to the supply rate of material to the bulk (sulfur and oxygen ion) plasma torus (and for 2015 to the oxygen neutral cloud). The total atmospheric loss rate does not necessarily change by exactly the same factor, as other loss processes, like through fast neutrals (to outer magnetosphere or beyond) or through local ionization at Io, may behave differently (see purple arrows in Figure 22). However, for triggering a change in neutral cloud and torus supply rate of factor 3-4, a substantial change at Io would in any case be required.

It is currently not understood how the mass loss from Io to supply the torus can change significantly and explain observed changes of the plasma torus and neutral clouds. The hypothesis of a significant transient increase of mass loss from Io is in fact difficult to reconcile with the current understanding of the atmosphere and escape from it. In the next section, we discuss what such a change may imply for the loss processes from Io and the lack of evidence for aperiodic changes in the atmosphere, and we summarize caveats with the assumptions about transient changes triggered by Io.

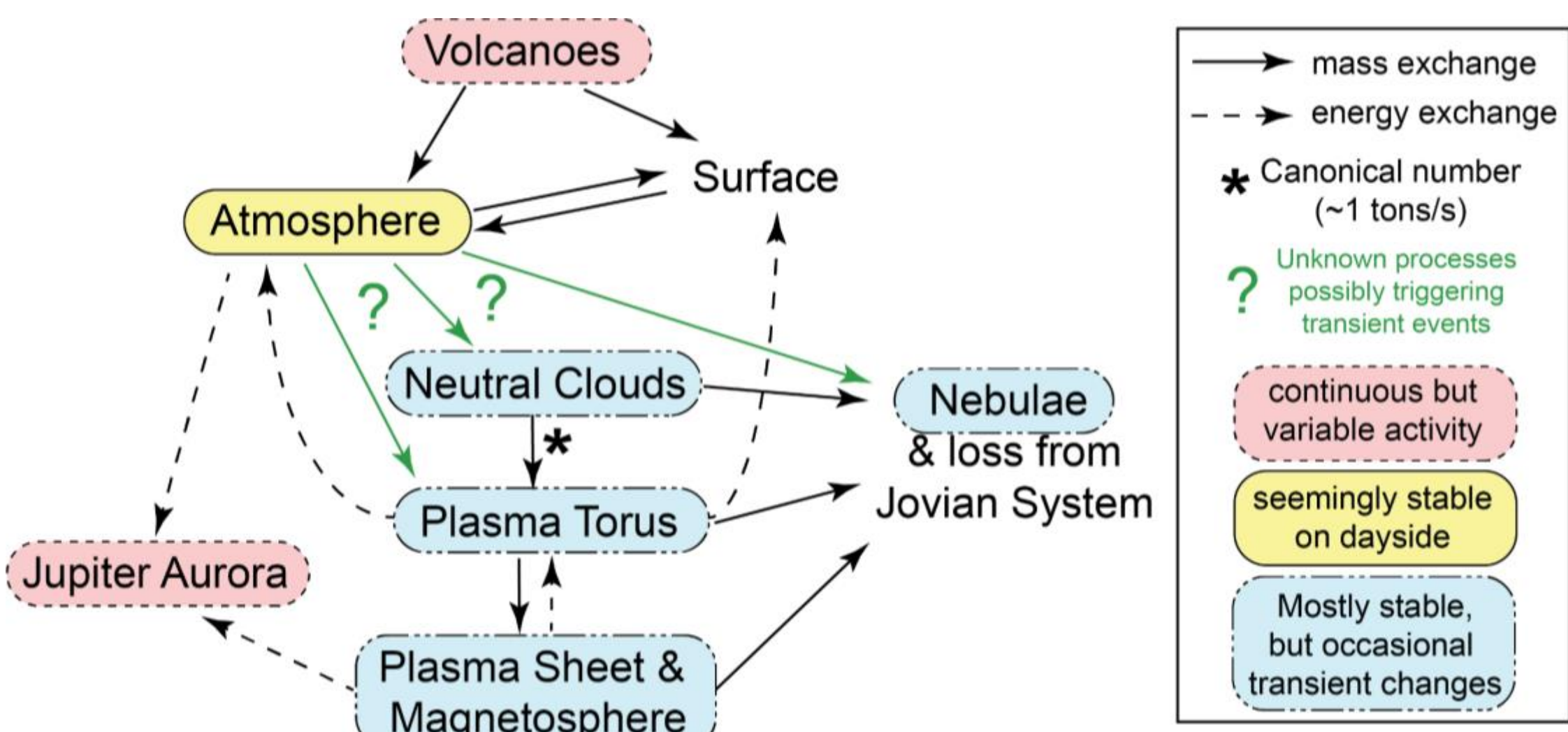

***Figure 22.*** *Schematic depiction of causal connections in the Io-Jupiter system. Solid arrows show connections that include flow of substantial mass. Dashed arrows indicate connections primarily through energy exchange (e.g., sputtering by energetic particles, injections of hot plasma into the inner magnetosphere, or energization for powering aurora). The asterisk indicates the mass transfer, for which the early studies (Section 2.1) constrained the rate (~1 ton/s) that has become the canonical number (Section 4.2). The stability of Io's atmosphere and the processes possibly enabling large changes in the atmosphere loss (green arrows) are key factors in the connecting chain that are not understood.*

## 4.4 Gaps in understanding, contradictions, and inconsistencies

### 4.4.1 Significant increases in atmospheric loss inconsistent with current understanding

It can be assumed with some confidence that the volatiles from Io that supply the torus must first populate the moon's atmosphere. As summarized in Sections 3.3 and 3.4, the loss from the atmosphere to neutral clouds and local ionization and pick-up into the torus is primarily due to collisions of the magnetospheric plasma with the atmosphere or neutral clouds. All other processes are likely insufficient to maintain a supply rate to neutral clouds and ultimately (or directly) to the plasma torus on the order of the canonical value of 1 tons/s. Importantly, direct escape from outgassing plume neutral gases seems far too low (Table 1) to cause an enhancement of several tons per second.

Plasma collisions at or near the exobase (which possibly may be at the surface at some locations like on the night side) most effectively provide momentum to the molecules or atoms to escape from Io's gravity. The effectiveness of these losses largely depends on the mass and energy flow of the corotating plasma that interacts with the atmosphere near the exobase. The characteristics of this exobase, like its altitude and variability, may thus play a key role for the atmospheric loss processes to supply the neutral and plasma environment.

Using a simple thought experiment by Schneider et al. (1989) which relates Io's atmospheric mass loss to a change in exobase altitude, we will show that the observed changes in torus supply rate require changes in the exobase, temperature and column density of Io's atmosphere that are not consistent with an apparently stable dayside atmosphere. For the purpose of the experiment we assume that changes of factor 3 (similar to the inferred changes) in the torus supply rate (neutral source rate) are triggered by similar changes in Io's atmospheric mass loss. Schneider et al. (1989) suggested that a change in mass loss from the atmosphere or mass supply to the torus $\dot{M}$ may be proportional to the cube of the radius of the exobase $r_{exo}$, so

$$\dot{M} \propto r_{exo}^{3} \qquad (5)$$

The exobase radius can be expressed as the moon radius plus the exobase altitude $h_{exo}$,

$$r_{exo} = R_{Io} + h_{exo} \qquad (6)$$

Assuming the collisions between plasma and atmosphere happen exactly at the exobase, the surface area of this "exobase sphere" is proportional to the radius squared. In other words, the higher the exobase, the larger the body of plasma that is intersected by the neutral atmosphere cross section. In addition, the higher the altitude of the exobase, the lower is Io's gravity, and thus the easier a particle escapes. The $r$-dependence of the decreasing escape velocity gives the third power in the proportionality.

A change of supply of factor 3 within weeks as suggested by e.g., the Hisaki results (e.g., Hikida et al., 2020) would increase the altitude of the exobase from an assumed $h_{exo}$ = 120 km (see Section 3.3, low density case of Summers and Strobel, 1996) to over 1000 km (8-fold exobase altitude increase), according to Equation (5).

Now considering an isothermal atmosphere with fixed temperature and thus scale height, the increase in exobase altitude can be related to an increase of surface density $n_0$ through Equation (3). The 8-fold increase in $h_{exo}$ requires an exp(8)=3000 times higher surface density $n_0$. For comparison, the exobase for the high density atmosphere case with a column density of $10^{18}$ $cm^2$ of Summer and Strobel (1996) is at 500 km as compared to their low density atmosphere ($8\times10^{15}$ $cm^2$) exobase at 120 km.

We can also assume a fixed surface density and increase the scale height H and thereby the atmospheric column density, again for an isothermal, exponential atmosphere. The 8 times higher exobase would imply a ~5-fold temperature increase and thus a 5-fold increase in column density, as we had assumed a fixed surface density.

A transient, strong increase of the upper atmosphere temperature may potentially be caused by a period of significantly enhanced Joule heating. The available power for Joule heating in the corotional electric field at Io is likely not fully used under standard interaction conditions (see details in Sections 3.3.3) and thus a change of atmospheric conditions has the potential to lead to an increase of Joule heating.

Overall the strong changes (>3 orders of magnitude in density, or by a factor of ~5 in temperature) are in stark contrast to the observational findings of a stable dayside atmosphere (Figure 22). The lack of evidence for major changes in the dayside atmosphere is discussed in the following section.

#### 4.4.2 Lack of observational evidence for transient changes in the atmosphere

While Io's atmosphere reveals clear lateral (in longitude and latitude) and temporal (day-night, eclipse passages, seasons) variability (Section 3.2), even a small change in the global atmospheric abundance not related to these systematic variabilities has never been measured with certainty. This means there is no observational evidence that the dayside atmosphere density undergoes unsystematic, transient changes. The longest observational coverage of the dayside $SO_2$ abundance came from mid-infrared observations (22 years) and revealed only seasonal variability on the order of factor ~2 due to the changing sublimation with changing heliocentric distance of Io (Tsang et al., 2013; Giles et al. 2024; Section 3.2, Figure 8).

Various atmospheric temperatures between 110 and 600 K were inferred from different methods. However, when the same method is used for temporally separated measurements similar temperatures are found (Section 3.2). Hence, there is no observational evidence for significant temperature changes in the atmosphere so far. We note, however, that the temperature of the upper atmosphere is not probed by most atmospheric observations, as they are sensitive to the bulk atmosphere. Therefore, a transient change that takes place only in the uppermost atmospheric layers, due to e.g., Joule heating, would remain undetected in common atmosphere observations. Jeans escape varies exponentially with the Jeans escape parameter $\lambda_{esc}$. Therefore, a significant increase in the upper atmosphere temperature may lead to significant escape of the lighter and thus mostly atomic species (Section 3.3.3).

Recent observational results suggest that volcanic outgassing is a relevant source for the atmosphere in addition to sublimation of surface frost (e.g., de Pater et al., 2020b), but massive gas plumes that produce densities much higher than the average equatorial dayside atmosphere density have never been seen. There is some evidence for $SO_2$ and other gases in volcanic plumes (Section 3.1) but the abundances above plume locations are similar to generally inferred abundances in the equatorial dayside atmosphere. Thus there is no evidence (yet) for events of extreme outgassing or any other transient change in the atmosphere that would suggest an Io-triggered change of the atmospheric loss by a factor of 3-4 that is derived from the change in the neutral source rate.

### 4.4.3 Commonly assumed but unconfirmed correlations and connections

The lack of understanding of the role of the atmosphere as well as of observations of time-varying atmospheric events is a key missing part to understanding the system as a whole. This missing link raises some doubts about the connections of other parts in the system and we want to point out some weak or not yet substantiated points often made in arguments on the connection of Io's volcanic activity to the torus and magnetosphere:

A. *Global state of Io's volcanic activity undefined.* Despite the much higher cadence in monitoring of thermal emissions from Earth in the last decade and new results from the Juno mission, the existing observations of thermal emissions from Io do not provide evidence for globally different states of volcanic activity at Io at different times (Section 3.1). Hence, the often cited "volcanically active" and "volcanically quiet" periods cannot be defined or derived from actual observations of volcanic activity. This nomenclature is a concept that was invented for explaining the different supply rates to the neutral clouds and plasma torus or other changes in the magnetosphere. There are strong increases of thermal emissions observed at volcanic sites, dubbed *outbursts*, that have not, so far, been correlated with changes in the magnetosphere.

B. *Large plumes are seen in most close-up spacecraft images but are hardly observable from Earth*. Often, imaging observations of large plumes like Pele or Tvashtar taken during spacecraft flybys like those of Cassini or New Horizons are considered as evidence for a particular volcanic event. However, (large) plumes are seen in almost all spacecraft images (mostly taken at high phase angles) of Io, but remote observations from Earth at low phase angle are difficult and only allow faint detections of large (known) plumes (Jessup and Spencer, 2012). Hence, cadence or activity cycles of such large plumes are not established but instead the cadence of plume detections is determined by the availability of spacecraft imaging data suitable for plume detections.

C. *Complex and unclear connection between hot spots and outgassing.* Hot spot activity is not necessarily connected to outgassing and thus does not provide a diagnostic for volcanic gas input to the atmosphere (or to the neutral clouds and torus). This applies even for the presence of volcanic gases like NaCl. In addition, Galileo data showed (and recent Juno data confirmed, e.g., Zambon et al., 2023; Davies et al. 2024) that the hot spots detectable from Earth are only the brightest and there are many more small sites with enhanced thermal emissions undetectable from Earth. Furthermore, the correlation of thermal hot spots and sodium trace gas suggested in the study of Mendillo et al. (2004) has been questioned based on new observational insights (Section 3.1).

D. *Unclear connection between sodium and bulk gases.* The pathways of alkali compounds through the system are likely quite different from the pathways of bulk gases. The alkali compounds (e.g. NaCl, KCl) are sourced to the atmosphere primarily from volcanic outgassing while $SO_2$ gas in the atmosphere is sublimated to at least 50% from surface frosts. The escape processes for alkali compounds and their daughter species may be different from the escape processes for bulk $SO_2$ and daughter species. In addition, the high velocity (larger or near Jupiter escape velocity of 25 km/s) particles that source the nebula (sodium or any other) likely originate from different processes than those sourcing the neutral clouds and ultimately plasma torus. Therefore, the variation observed in the sodium nebula may not be coupled to changes in the neutral clouds and torus.

E. *Unclear connection between dust in the Jovian system and volcanic eruptions.* Dust streams measured in and beyond the Jovian magnetosphere have been associated with dust in Io's plumes and thus volcanic activity. The dust particle trajectories, the flux variability and composition of the dust stream identify Io as the source and suggest volcanic origin of the particles. However, as for the gaseous trace species, the connection of abundance and variation of dust and of the bulk gases ($SO_2$ in the atmosphere, S and O neutrals and ions in the magnetosphere) in the system is unclear. There seems to be a wide range of dust to gas ratio in plumes including dust-free "stealth" plumes (Section 3.1) and the escape processes of the dust from Io are not yet well understood. The mass rate of dust lost from Io is 3-4 orders of magnitude lower than the neutral source rate for the torus. The Galileo dust measurements did not provide evidence for a temporal connection of the intensity of the magnetospheric dust streams to volcanic hot spot detections. The putative dust increase in 2001 suggested to be connected to a torus change has a large observational uncertainty because of the large distance of the Galileo spacecraft to the inner magnetosphere.

F. *Auroral features connected to injections possibly triggered by Io are frequent.* Jupiter's aurora is shaped and affected by various magnetospheric and external processes and the connection to the mass output from Io is rather indirect (Section 3.7). The morphological features in the main emission that were suggested to be connected to Io mass output enhancements appear more frequently than other transient events (several times per year as compared to once in several years). Unfortunately, there are no aurora imaging observations from the year 2015 during the strong and well monitored transient torus and neutral cloud enhancement.

#### 4.4.4 Conclusion and open questions

Although there is evidence that the neutral gas in the magnetosphere and the plasma in the Io torus occasionally undergo transient changes, it is not known how they are triggered or caused. The idea that volcanic activity at Io causes large scale changes in the magnetosphere is therefore a hypothesis with many unknown elements that yet needs to be substantiated. While there are many open questions about the details of each of the parts in the system reviewed in Section 3, we provide here a list of overarching questions either on the workings of the system or on the diagnostics commonly used:

1) How do thermal eruptions relate to volcanic outgassing? In particular, what types of volcanoes or styles of activity directly produce gas (and dust) and how much? How much gas is released before, during, and after a thermal eruption phase?
2) Is it possible that local outgassing at a volcanic site significantly changes the overall loss of neutral gases (or dust) from Io? If so, what effect, if any, does latitude, longitude, or time of day have on whether outgassing products are lost from Io?
3) Does the global atmosphere undergo significant transient changes, possibly preceding and triggering the transient events in the torus and magnetosphere? If so, what causes these significant global changes?
4) Can Io's mass loss to the neutral clouds and plasma torus be enhanced significantly without significant changes in the bound atmosphere?
5) What is the composition of neutral and ionized gases lost to the environment? In particular, how much is lost in molecular vs atomic form? What is the fraction and composition of the ions directly supplied from Io (Io's ionosphere) to the torus?

6) Is every brightening of the sodium nebula accompanied by changes in the neutral clouds and plasma torus?
7) Is the dust input from Io to the magnetosphere correlated with the gas supply?
8) What physical processes trigger and affect auroral phenomena during transient torus enhancements? Specifically, is the location shift of aurora solely achieved by variation of mass loss outflow, or do other quantities (e.g., large scale magnetospheric flow variabilities, electron temperatures or the Pedersen conductivity in Jupiter's ionosphere also contribute? How does the morphology and brightness of the main emission evolve over a transient torus enhancement event like the one observed in 2015?

# 5. Future observations and methods

The previous sections have shown that there are still many unknowns in the Io-Jupiter system and specifically several open questions about specific aspects on the supply of mass from Io to the magnetosphere. For advances in understanding the complete system, it will require many advances on these individual aspects and questions which likely can be achieved through a variety of remote observations, in-situ measurements and theoretical or modeling efforts.

## 5.1 Spacecraft measurements

There are three planetary missions targeting the Jupiter system that may provide measurements relevant to the topic. The NASA Juno spacecraft carried out close flybys at Io in its extended mission which will end in 2026. Later in the 2030s, both NASA's Europa Clipper mission and the Jupiter Icy Moon Explorer (Juice) of the European Space Agency (ESA) will orbit Jupiter for several years targeting primarily the planet's large icy moons. Although the latter two missions will not come close to Io, they will provide valuable remote data and in-situ data about Jupiter's magnetosphere. Finally, a mission dedicated to Io would potentially allow a major leap forward.

### 5.1.1 Juno

The NASA Juno spacecraft went into orbit around Jupiter on 4 July 2016 and the ~5-year primary mission was designed for 35 perijove passes. The spacecraft's polar elliptical orbit precesses such that the orbital distance at which Juno crosses the equatorial plane evolves inwards.

In the extended mission's additional 43 orbits, these crossing points reached the orbital distances of the Galilean satellites and opportunities became available to observe the moons (including Io) up close. The recent observations by Juno have provided visible and thermal images (from JunoCam and JIRAM respectively) that show a large number of higher temperature areas on the surface (e.g., Davies et al., 2024). Furthermore, Juno is conducting 15 flybys within 150.000 km of Io between April 2022 and May 2025, as part of this extended mission. Of those flybys, the two closest occurred at an altitude of slightly below 1500 km:

- PJ57 Io: 2023-12-30 08:36
- PJ58 Io: 2024-02-03 17:48.

During the Io flyby of PJ (perijove) 57 the spacecraft passed above Io's north pole near close approach, and at PJ58's Juno transited south of Io's near-wake environment.

Juno's plasma and particle instrumentation were designed to observe in Jupiter's auroral regions and not in the high density and high radiation environment of Io's orbit. However, Juno can still contribute to improving our understanding of the spatial and energy distribution of the ion species near Io. The Jovian Auroral Distributions Experiment (JADE), a plasma analyzer with Time-of-Flight (TOF) mass spectrometry, will enable the first mass-resolved plasma composition observations in the vicinity of Io. Furthermore, the Jupiter Energetic-Particle Detector Instrument (JEDI) onboard Juno could determine the extent to which there are energetic particle dropouts, which could provide constraints on its extended atmosphere's spatial extent and variability (e.g., Huybrighs et al., 2024). In addition to the close flybys, Juno transits the Io plasma torus multiple times. While the plasma and particle properties in this region are significantly different from those that JADE and JEDI were designed to measure, they could still provide an important set of measurements with which to improve our understanding of the plasma-neutral interactions, plasma chemistry, and mass transport from Io and the Io torus. Future plans for observations of Io are also elaborated on in Keane et al. (2022) and McEwen et al. (2023).

#### 5.1.2 Jupiter Icy Moon Explorer (Juice)

After orbit insertion in 2031, Juice will orbit Jupiter for over three years before going into orbit around Ganymede at the end of 2034. During this Jupiter orbiting phase, observations of Io and its environment will be mostly from a distance of ≥850,000 km. However, there will be several opportunities during this phase, to remotely observe Io at around 400,000 km distance. Several instruments may take observations relevant to the topic of mass loss and we briefly mention such possible studies.

The visible camera JANUS (covering wavelengths between 350 and 1064 nm) aims to study different aspects with remote high-resolution images: (1) Changes in Io's surface identified through repeated coverage; (2) plume detections using high phase angle and eclipse observations; (3) monitoring Io's sodium extended clouds with its sodium filter; and (4) imaging Io's aurora in eclipse as diagnostic for the plasma interaction and gaseous plumes. The Submillimetre Wave Instrument (SWI) has the capabilities to measure sub-mm wave emissions from $SO_2$ as well as other less abundant molecules in Io's atmosphere like KaCl, NaCl, SO and $O_2$. The SWI measurements may allow the extraction of vertical profiles and atmospheric dynamics through Doppler shifts, line shapes and ratios. Juice's Ultraviolet Spectrograph (UVS) will monitor Io's torus and neutral clouds remotely through S and O atom and ion emissions and determine the plasma production rates (Masters et al., Juice WG3 SSR, in review). In addition, it can take remote observations of the Io local aurora and footprint to probe the plasma interaction state. Observations of Io's aurora obtained during eclipse ingress and egress periods, like the JANUS eclipse observations, can inform our understanding of variability in the relative plume to sublimation source contributions over the three year tour period. UV surface reflectance measurements, while only available at hemispherical-scale spatial resolutions, will be monitored as a function of orbital phase, with Lyman-α variations potentially constraining Io's $SO_2$ atmosphere asymmetries. Several stellar occultations are planned, and could provide important new constraints to its nightside atmospheric density especially (not viewable from Earth). At least one Juice-UVS Jupiter transit observation of Io's atmosphere is also planned, possibly informing plume influences on Io's hemisphere-scale atmospheric asymmetries (e.g., Retherford et al., 2019). The Moons And Jupiter Imaging Spectrometer (MAJIS, a visible and near-infrared imaging spectrometer covering wavelengths 0.5 to 5.54 μm) will map Io's surface with spatial resolutions below 100 km at the closest distances with the potential monitor, e.g., $SO_2$ frost abundances and changes. In addition to remote studies, the Jovian Neutrals Analyser (JNA) part of the Particle Environment Package (PEP) onboard Juice could monitor S and O Energetic Neutral Atoms (ENA) of 10 eV-3 KeV from the torus (Futaana et al., 2015). The ratio of S/O obtained from such measurements could reveal that the plasma torus originates from volcanic Io materials ($SO_2$). The Juice-Magnetometer (J-MAG), the Radio and Plasma Wave Instrument (RPWI), and the other sensors on PEP will broadly study Jupiter magnetospheric variability, and potential correlations of Io-based volcanic or atmospheric-escape events with plasma injections and potentially other magnetospheric processes related to Io's plasma interaction.

### 5.1.3 Europa Clipper

The science objectives of the NASA Europa Clipper mission focus exclusively on Europa and its habitability (Pappalardo et al., 2024). Clipper was launched in October 2024 and arrival at Jupiter will be in April 2030 – about one year before Juice. Similar to Juice, the trajectory of Clipper avoids the inner magnetosphere and the spacecraft will not be closer than 250 000 km to Io. The spacecraft instrumentation is partially similar to that of Juice with a near identical Ultraviolet Spectrograph (UVS) instrument, a visible camera (Europa Imaging System – EIS), a near-infrared spectrograph (Mapping Imaging Spectrometer for Europa – MISE), and an ion and neutral mass spectrometer (Mass Spectrometer for Planetary Exploration – MASPEX). Potentially, the instruments provide capabilities to take similar measurements mentioned for Juice above. Europa-UVS will make neutral cloud and torus stare observations obtained ~1-2 days from closest approach. These measurements will point at Europa and its extended, escaping atmosphere but are intended to help assess the state of the plasma environment. Likewise, Clipper's pair of plasma sensors (Plasma Instrument for Magnetic Sounding – PIMS) assess the ion composition and thermal electron densities while its magnetometer (Europa Clipper Magnetometer - ECM) measures fields continually throughout the magnetosphere to provide context for its Europa ionosphere and induced-field measurements near closest approach. In addition, The SUrface Dust Analyzer (SUDA) on Clipper has capabilities to constrain Io-genic dust streams with much higher precision and improved mass resolution compared with previous measurements. If dust ejections are correlated to volcanic activity (Section 3.8) and loss of the bulk gaseous material from Io, SUDA measurements could provide a valuable observatory platform to monitor the activity of Io throughout Europa Clipper's mission. Also, the E-THEMIS experiment has the ability to measure Io's heat flow, much of which occurs at longer wavelengths and cannot be measured by a near-IR instrument.

Although Europa science has driven the development of Europa Clipper, a joint working group with Juice is studying how Europa Clipper can contribute to Jupiter system science, including Io and the plasma torus. Post-launch the Clipper team is expecting to continue discussions of expanded observations of Jupiter system targets for calibrations, operations exercises, and eventually added value science (pending availability of future funds).

### 5.1.4 A dedicated Io mission

A mission with Io as the main target could potentially address many questions. Despite difficulties to realize an Io mission due to the harsh radiation environment, interesting concepts have been put forward in the past. The Io Volcano Observer (IVO) concept completed a Phase A study as a NASA Discovery mission in 2021, but was not selected to proceed (McEwen et al., 2023). The mission could provide much better monitoring of active volcanism and the links between hot spots and plumes. High-resolution visible and thermal observations of vent regions would provide constraints on eruption processes. Magnetometer and plasma instruments could provide monitoring of the atmosphere-plasma interaction, Jupiter's magnetosphere as well as the plasma torus and its variability relative to volcanic activity. Plasma composition measurements with mass spectrometry capability would be critical to improving our understanding of the chemistry and interaction between the atmosphere and plasma environment. Perhaps most important for understanding the atmosphere would be the first neutral mass spectrometer to operate close to Io, to understand what neutral species and abundances are erupting and present in the atmosphere. For these reasons, NASA's New Frontiers program includes an Io mission as one of several predetermined targets allowed for the next proposal opportunity, as recommended through the 2023 Decadal Survey.

## 5.2 Remote Earth-based observations

Observations from the ground or by space telescopes have provided important contributions to understand the Io-Jupiter system, not least because they allow to cover longer timescales of many years or even decades. The observational possibilities and sensitivity of specific observational methods is continuously being improved and remote observations may be key for addressing the issue of Io's mass loss in the future.

### 5.2.1 Role of remote observations and limitations

Almost all parts of the Jupiter-Io system can be observed in some way remotely from Earth: Io's volcanic hot spot emissions are monitored frequently by ground-based telescopes with decent spatial resolution since the availability of Adaptive Optics (Section 2.1). The atmosphere is observed with a variety of methods at various wavelengths (Section 2.2). On the contrary, it is relatively difficult to observe gas or dust plumes remotely, with most notable observations by the Hubble Space Telescope (HST) (Spencer et al., 2000; Section 2.1) or ALMA (Section 2.2). Even Io's interaction with the plasma environment can be indirectly probed from Earth through UV observations of Io's local aurora or the moon's footprint in Jupiter's aurora (Section 2.4). The neutral clouds and plasma torus are observable also primarily in the UV and thus from space-based telescopes, as, for example, monitored regularly in the last decade by the Hisaki satellite (Sections 2.5 and 2.6). Visible observations from the ground (or space) are a tool to monitor not only the trace species (primarily Na) near Io or in the extended nebulae but also sulfur ion torus emissions or even neutral oxygen emissions. And lastly, Jupiter's UV aurora has been regularly imaged with HST for more than three decades now (Section 2.7).

Advances in the capabilities of telescopes, e.g., enhanced spatial resolution capable of resolving Io, enabled new insights as for example the recent detection of SO IR emission at 1.7 μm directly above a volcanic hot spot (de Pater et al., 2023). More frequent observations over longer times similarly provided relevant insights like the apparent stability of the atmosphere or a more complete picture of hot spot variability (Sections 2.1 and 2.2).

A key part of the system is difficult to constrain with remote observations: the loss of material from Io either as neutrals to the neutral clouds or as plasma to the torus. UV observations of the neutral S and O in the region to 10 $R_J$ around Io (beyond the Hill sphere radius ~5.8 $R_{Io}$) provides some idea of atomic neutrals in the process of escaping Io's gravity. In-situ plasma measurements provide means to constrain production of new atomic and molecular ions around Io (Section 4.1). There are, however, no observational results on loss of neutral molecular species, which may constitute a large fraction of the neutral loss and may play a key role for a transient increase of loss, if this loss enhancement is driven by plasma processes (Sections 2.3 and 2.4).

### 5.2.2 Ongoing observing programs and future opportunities

There are currently several ongoing programs that observe Io. Some of them are in support of the close and distant flybys of the Juno spacecraft in 2023 and 2024, in particular for providing constraints on the neutral atmosphere, which cannot be measured with Juno's instrumentation.

A large program with the Hubble Space Telescope and the James Webb Space Telescope (HST GO 17470) targets different observables in the system from surface composition through solar reflection, to hot spot activity, Io's local aurora, and to the neutral clouds and plasma torus out to radial distance of Europa (~10 $R_J$). A tailored program with only JWST (GO 4078) aims to map the gas distribution on Io's dayside through the 7.3 μm $SO_2$ band during the Juno flyby on February 3, 2024. The 7.3 μm band was successfully detected in an earlier JWST program (1373) but the work is not yet published. These mid-IR observations will provide additional information on the hot spot activity, if successful. Another longer-term program to regularly measure Io's dayside $SO_2$ abundance is currently being carried out with the Submillimeter Array (SMA, PI W. Tseng). The observations are similar to those published by Moullet et al. (2010) and the program targeted Io 9 times during observing seasons in 2022-2023.

Efforts in ground-based monitoring of the thermal IR emissions with the Keck telescope (e.g., de Kleer et al., 2016; 2019), and of the sodium cloud, nebula and Io plasma torus with small ground-based telescopes (Yoneda et al., 2013, 2025; Morgenthaler et al., 2019; Kondo et al. 2024) are continuing. The increasing number of observations and thus temporal coverage on the different parts in the systems may enable further tests of correlations and connections.

Two observations could be of particular interest: One is a sensitive observation of the $SO_2$ atmosphere (density and also temperature) just at the onset of an increase in emissions from the neutral clouds. If Io triggers the transient event through an enhancement in the mass loss, the atmosphere should undergo some considerable change at least around the starting time of the enhancement in the neutral cloud. Another valuable observation would be to detect molecular species escaping from Io through e.g., spatially resolving exospheric layers, which is extremely challenging. None of the available telescope facilities and previously applied methods for the bulk $SO_2$ atmosphere (from UV with HST, to IR from ground or now JWST, and sub-mm interferometry) provide the sensitivity to detect the expected $SO_2$ abundances of the escaping population or in the neutral clouds.

Future telescopes - planned or under construction - may provide capabilities to address some aspects like direct measurements of escaping neutral gases. The currently constructed Extremely Large Telescope (ELT) with its ~40-m primary mirror has a nominal spatial resolution of 5 μarcsec, which corresponds to ~20 km on Io or ~200 pixels across Io's diameter. With state-of-the-art high-resolution spectrographs it may provide high sensitivity for accurate $SO_2$ measurements and thermal emissions (and mapping) at infrared wavelengths.

LAPYUTA (Life-environmentology, Astronomy, and PlanetarY Ultraviolet Telescope Assembly) is a future UV space telescope, which was selected as a candidate for JAXA's 6th M-class mission in 2023 (Tsuchiya et al., 2024). Launch is planned for the early 2030s. LAPYUTA would perform spectroscopic and imaging observations in the far ultraviolet spectral range (110-190 nm) with a large effective area (>300 $cm^2$) and a high spatial resolution (0.1 arcsec). LAPYUTA will have capabilities to monitor mass loss from Io's $SO_2$ atmosphere to Io' neutral cloud and plasma torus as well as their effects on the magnetospheric dynamics, similar to but enhancing the successful observations of Hisaki (Sections 3.6 and 3.7).

Powerful space telescopes in planning include the concept of the Habitable Worlds Observatory (HWO) for observations from UV to infrared wavelengths as part of the Great

Observatory Maturation Program (GOMAP) recommendation of the Pathways to Discovery in Astronomy and Astrophysics for the 2020s (Astro2020) Astrophysics Decadal Survey (see also Large Ultraviolet Optical Infrared Surveyor final report, 2019). With currently discussed mirror diameters of 8 m or 15 m and being located at Lagrange Point L2 (continuous view and unaffected by the geocorona), such a space telescope would increase the sensitivity and spatial resolution in the UV as compared to HST by more than an order of magnitude and a factor of 3-6, respectively. Other telescopes built or planned by different agencies and organizations like the Giant Magellan Telescope (GMT), or the Thirty Meter Telescope (TMT) may also allow useful observational advances.

## 5.3 Modeling efforts

To understand the physics of how Io provides mass to the magnetosphere, measurements must ultimately be explained by models. Modeling depends on the applicability and correct implementation of the relevant included physics and choices about boundary and initial conditions. Primary processes sustaining the atmosphere are sublimation/condensation and volcanic outgassing. In the modeling of Io's atmosphere, one must account for the fact that the atmospheric escape to populate the magnetosphere is a secondary physical process; escape is not the major contributor to mass, momentum or energy input to/from the plumes or atmosphere and thus does not play a major role.

### 5.3.1 Spatial scales, time scales and undetermined feedback

Mass-loading of the Jovian magnetosphere presumably results from a long chain of processes that happen on different temporal and spatial scales: The spatial scales range from (i) outgassing from the volcanic plumes and sublimation and the (ii) formation of the bound atmosphere (10-400 km altitude), to the exosphere (~$R_{Io}$), continuing with (iii) the supply to the neutral clouds by plasma-atmosphere interaction (several $R_{Io}$), (iv) the formation of Io's plasma torus (~2 $R_J$), and finally (v) the radial transport of the plasma from the torus through the whole Jovian magnetosphere (several 10s of $R_J$). These physical processes are probably linked via feedback mechanisms and some relevant processes may not yet be recognized. Clearly, these processes cannot all be accommodated in a single model or simulation. Various models for the different parts in the system have been already developed; the easiest next step is to iterate between and/or patch together multiple sub-models.

### 5.3.2 Current sub-models and their limitations

Currently, separate sub-models focus on describing a few aspects of this chain of processes and parameterize (or assume constant) the features not addressed in the model. The parameterization is then constrained by observations. Examples of such sub-models include:

***Atmosphere and plume models***. Sophisticated atmospheric models have been developed, which include the contribution of major plumes and sublimation of the $SO_2$ surface frost. Various escape processes should be considered for a combined sublimation and plume atmosphere. DSMC atmospheric models have used imposed streams of incoming plasma and static electric and magnetic fields (e.g., Moore et al., 2012), which themselves should depend non-linearly on the atmospheric distribution and density. Global-scale winds driven by sublimation/condensation, plumes and plasma impact may provide an additional velocity at the top of the atmosphere which, combined with thermal processes, could yield significant escape. Simple thermal escape rates are exponentially sensitive to the exobase temperature, so it is reasonable to expect possibly *locally enhanced escape* due to winds/plumes, chemical recombination or plasma interactions. Models of planetary escape which establish whether two or more driving processes contribute to high-speed winds and thereby enhanced thermal escape remain to be developed.

***Plasma-interaction models.*** Fluid or kinetic models of the plasma-atmosphere interaction focus on the electromagnetic interaction and properties of the plasma. They include some physical chemistry (e.g., ionization, charge exchange, collisions) but the simulations to date generally rely on a prescribed static atmospheric distribution and composition. Some models also prescribe a static description of plumes (Blöcker et al., 2018). The comparison of the model results with the plasma properties and fields observed along a probe trajectory or the remotely observed auroral emissions constrain static atmosphere and overall electromagnetic interaction state. Interaction models do not include the atmospheric response. This is considered in some atmosphere models (Walker et al. 2010), however, with simplified, static plasma conditions. The details on the atmosphere escape processes are also not the focus of common interaction models (where the focus is on the plasma effects), but have been considered in studies of the plasma/neutral physical chemistry: some reactions lead to neutral escape (e.g., electron-impact dissociation, charge exchange, dissociative recombination) providing neutrals with specific velocity and direction distributions that can escape the gravity of Io (e.g., Dols et al. 2008).

***Neutral cloud models***. Neutral Cloud Models simulate the distribution of neutrals (e.g., Na, O, S, $SO_2$) along the orbit of Io under the gravitational fields of Jupiter and Io (e.g., Smith et al. 2022). These models include some physical chemistry (ionization, charge exchange, etc.) that shape the neutral clouds. These loss processes have been calculated with a prescribed static plasma torus density, composition, and temperature. More importantly, in such models, the source of these neutral clouds is based on a simplified description of the neutral fluxes from Io's exobase. These models prescribe a velocity distribution for the escaping neutrals that is typical of atmospheric sputtering and prescribe the lateral (longitudinal and latitudinal) distribution of these neutral fluxes assuming a purely radial ejection velocity. Comparison of the simulated neutral cloud with the observations of neutrals along Io's orbit constrains the velocities, lateral location and composition of the neutral ejection from Io's exobase. However, sub-models (earlier in the modeling chain) that simulate the plasma-atmosphere interaction conclude that the neutral loss comes from not only sputtering but also from other processes (e.g., charge exchange, molecular dissociation, and photo-processes). These processes, which provide neutrals with velocities sometimes much larger than a sputtering velocity distribution and ejection directions that are not radial, are not currently considered in torus modeling.

***Plasma torus models.*** Sub-models of the plasma torus include a detailed description of the physical chemistry that calculates the ion composition and energy to simulate the time-averaged plasma properties of the torus (Section 3.6). The simplifications involved in this modeling include the parameterization of the neutral supply rate, of the neutral S/O ratio and of the radial plasma transport. Comparison of the simulated plasma properties with *in situ* measurements constrains those values. All of the parameters are generally assumed constant with time for each model.

#### 5.3.3. Future progress in modeling

Considering the open questions (Section 4.4.4) it is clear that we have not yet identified some dominant processes or quantitatively estimated some significant feedback mechanisms. Future modeling should iterate between two or more physically distinct sub-models of the Io environment. Such iterations are already in progress (e.g., atmosphere/torus or atmosphere/neutral cloud). A more complex approach is to physically couple two subsequent sub-models in a single simulation. DSMC simulations of the atmosphere are already moving in this direction, and substantial progress is in sight (Klaiber 2024) but require large computing resources. With current computing power it should be possible to simulate a full 3D coupled model of Io's torus, plumes and atmosphere with radiative transport and solid body heat transfer through an entire Io orbit, including eclipse. We roughly estimate that on ~$10^4$ processors such simulation may require run times of only a few days. The resulting highly resolved global circulation model could serve as a community baseline dataset upon which to examine different escape mechanisms. But the parameter space to be explored (e.g., plume, atmosphere, local plasma interaction variations, torus, magnetosphere) is still extremely large. Both partial differential equation solvers and stochastic solvers are required in different regimes and may need to be coupled: Navier-Stokes/DSMC, DSMC/PIC (Particle-In-Cell) or PIC/MHD hybrid methods, applied in the appropriate physical regime(s), could help reduce the simulation computing time and allow a more efficient exploration of the large parameter space.

# Appendix

## List of terms relevant to the mass supply from Io (alphabetical order within each category)

### A) Surface and Volcanism

**Active volcano**
On Earth an active volcano is a structure that is either erupting or is likely to erupt in the future. A terrestrial active volcano which is not currently erupting is known as a dormant volcano (while extinct volcanoes are not expected to become active again). On Io, every volcano is probably either active or dormant.
We note also that from the perspective of *hot spot* detections or other measurements of volcanic activity at Io, it has not been possible to identify different states or levels of the global volcanic activity. This means there are no indications of general "volcanically active" or "volcanic quiet" periods (yet).

**Hot spot**
This term has different definitions depending on context. It is best to specify **mantle** hot spot, **volcanic** hot spot, or **IR** hot spot. A mantle hot spot may manifest itself as seismic anomalies, elevated topography, gravity anomaly, concentration of volcanoes, etc. A volcanic hot spot should provide evidence for current or recent volcanic activity, such as eruptions observed and recorded by humans, gas venting, or volcanic deposits with very young radiometric dates. Io astronomers call a remotely sensed IR emission enhancement that is clearly above the expected background emission a *(IR) hot spot*.

**Plume (gas vs dust)**
A plume consists of gas and particulates rising from a volcanic source often creating a large umbrella-shaped structure capped by a gasdynamic shockwave. Plume constituents may move differently and form umbrellas or jets within umbrellas, for example with large grains forming a more compact shape within a much larger gas canopy. Ionian plumes tend to be dense enough to be collisional (intermolecular collisions matter to the dynamics and molecules/grains do not simply move ballistically).

**Plume type (Pele vs Prometheus)**
The size of a plume depends on the energy given to the rising gas/particle flow at the source. Giant Pele-class plumes appear to be sourced directly from hot lavas, perhaps a bubbling lava lake, and rise 200-400 km and have a high gas-to-dust mass ratio. Smaller (50-70 km high) Prometheus-class plumes contain more dust and appear to arise from a region where lava is encroaching upon pre-existing sulfur dioxide ice.

**Stealth plume**
A predominantly gaseous plume that has an insufficient dust component to make it detectable in visible light. Most images of lo's plumes, and especially the best known ones, show them through Mie-like scattering from dust grains of solar light in the visible, often at high phase angles (forward scattering). If there is very little dust in plumes and only gas, they are not seen in visible images.

**Volcanic eruption**
Generally, a volcanic eruption is an event in which lava and/or gases are expelled from a volcanic site through a vent or fissure. At Io, the term *eruption* is primarily used for detection of a hot spot, i.e., thermal emission from hot lava. This is the only type of observations taken with sufficient cadence and coverage to infer temporal changes and thus observe and define an eruption event. The cadence of detections of (large) dust (or gas) plumes determined by the availability and timelines for plume activities is hardly constrained yet.

---

B) Mass supply from Io to the magnetosphere

**Atmospheric mass loss**
Neutral gas lost from Io's gravitationally bound atmosphere and corona to space (not to the surface). Material is lost as neutrals through acceleration (above the exobase) to velocities higher than the escape speed (through various processes including collisions with plasma, heating of the neutrals, or recombination), or as ions through local ionization and pick-up by the magnetic field. The fraction lost as neutrals at velocities below Jupiter's escape velocity at Io's distance (25 km/s in the reference frame of Jupiter) feeds into the neutral clouds. Neutrals with effective velocity exceeding the Jupiter escape speed populate the extended nebulae and leave the system. Locally ionized material is supplied to the plasma torus directly.

**Canonical number / Neutral source rate**
Production of fresh torus plasma (ions) through ionization of neutrals (kg/s or particles/s) from the neutral clouds or Io's neutral atmosphere. Ionization and ion-neutral collisions supply slow ions (and electrons) to the torus, which are then accelerated to corotate contributing to the supply of energy powering the torus UV emissions (the other significant energy contribution is hot electrons ~40-400 eV). To sustain the balance, the supply of new slow ions equals the rate at which neutrals must be resupplied, and therefore it is often called the *neutral source rate* in Neutral Cloud Theory modeling. The rate of this generation of fresh ions (and corresponding destruction of neutrals) was estimated based on the emitted energy at ~ 1 ton/s or $(1\text{-}3)\times10^{28}$ particles/s of S and O neutrals. This value is the canonical number of mass transfer frequently cited in the literature. However, it is often used inaccurately for related but not identical rates (or processes) like the mass loss from Io's atmosphere (which is larger) or the net torus mass-loading rate, i.e., the production of a new plasma particle without losing a plasma particle (which is lower and only from ionization and not ion-neutral collisions).

**Local torus ion supply at Io**
The number of ions coming directly from Io has been estimated from the J0 pass of Galileo through the Io wake (Bagenal et al., 1997) at 18-58% of the canonical 1 tons/s. This ion loss rate was further refined to ~300 kg/s (Saur et al., 2003; Dols et al., 2008), which amounts to roughly 20-30% of the rate of neutrals being supplied to the neutral clouds from Io (see *canonical number*). As the plasma is slowed down in the vicinity of Io, the energization from pick-up of the ions generated in this region decreases significantly and likely becomes negligible compared to the energy from ionization in the neutral clouds at full corotation.

**Mass loss from Io**
The total *mass loss from Io* is the *atmospheric mass loss* plus direct mass loss (thermal or sputtered) from the solid surface or direct escape from volcanic plumes. Both loss to space from the surface and volcanic plume escape are likely much smaller than the atmospheric mass loss and thus the *mass loss from Io* is nearly identical to the *atmospheric mass loss*. Loss from the surface is likely small because the shielding atmosphere is even sustained in absence of sunlight by volcanoes as recently shown. The velocity of ejected plume gas (or dust particulates) is well below Io's escape (less than half for the largest plumes) velocity and the plumes interact with the sublimated atmosphere.
The *mass loss from Io* is larger than the *neutral source rate* for the torus (assuming Io is the only viable source). This is because some processes eject neutrals at a velocity larger than Jupiter's escape velocity at the orbit of Io (25 km/s), which are then lost to the Io/Jupiter system and do not contribute to the supply of the torus.

**Mass loss from the Jupiter system**
Combined loss of ions and neutrals from Jupiter's magnetosphere and gravity field. This loss is expected to be similar to the *mass loss from Io*, since no other substantial loss pathways are known.

**Momentum transfer**
Newly generated torus plasma has little to zero momentum. The plasma is then accelerated by the corotating local magnetic field effectively transferring momentum to the plasma flow from the angular momentum of Jupiter's ionosphere, which is collisionally coupled to the planet's upper atmosphere.

**Source for the bound atmosphere**
The most recent observational studies suggest that both sublimation of surface $SO_2$ frost and direct outgassing at volcanic sites sustain Io's bulk dayside atmosphere. There are mutual effects between the two sources and atmospheric abundances. Sublimation equilibrium is maintained above frost patches on the dayside but most of the $SO_2$ should condense at night. A source rate cannot be defined in this dynamic atmosphere.

**Torus mass-loading**
Net production of plasma (ions) in the torus due to ionization of the extended neutral clouds and Io's atmosphere. This is smaller than the neutral source rate, because charge exchange and momentum transfer collisions do not change the net number of ions in the plasma torus. Instead, a co-rotating ion is lost in the same process as a new slow ion is generated as well as a fast neutral, and the latter is lost from the system. This net source of torus plasma is balanced primarily by losses from effective radially outward transport.

C) Atmospheric and plasma-atmosphere processes

**Atmospheric sputtering**
A process in which high-energy particles, either ions or atoms, collide with the atmospheric neutrals, causing the ejection or removal of atoms or molecules from the atmosphere. The process itself is similar to *surface* (knock-out) *sputtering*, but the cascade/recoil processes are localized in few scale-heights instead of few nm. The yield (Y) for this process is defined as the average number of target neutrals released per incident projectile. For each sputtered particle, a cascade of multiple collisions is usually necessary. Atmospheric loss by sputtering has been assumed in a series of publications to model the formation of the Na extended cloud (banana cloud) and O and S extended clouds. The *surface sputtering* rate at Io is negligible on the dayside hemisphere but may contribute to some neutral losses at night or in eclipse.

**Charge exchange**
A process that occurs when a molecule or atom (neutral or not) collides with a charged ion and one or more electrons are transferred from one to the other. In the case of Io, the most common reactions involve atoms of O and S and molecules of $SO_2$, charged or not. Charge exchange rates/cross sections depend on the relative velocity of the colliding particles, which is ~60 km/s for the corotating plasma at Io's radial distance. If a (slow, gravitationally bound) neutral atom is produced, it will have roughly the corotation speed of the ion, which is large enough to escape the Jupiter system. This is the main generation mechanism for ENAs. Charge exchange processes are thought to dominate the production of slow oxygen and sulfur ions at Io's orbit, although they do not significantly change the net ionization level of the plasma.

**Io-genic material**
Neutral gas, plasma or dust in the Jovian system that ultimately originates from the interior or surface of Io. Most sulfur (S) and sodium (Na) material is likely to come from Io, while there are possible other viable sources for O (icy moons) and H (Jupiter, icy moons).

**Joule heating**
Joule heating refers to the process when electromagnetic energy is converted into heat through $\mathbf{j}\cdot\mathbf{E}$ within Poynting's theorem. In case of an ionosphere, the dissipation occurs via Ohmic heating controlled by the anisotropic conductivity of the moon's ionosphere. The total dissipated power at Io depends on the conductances of Io's ionosphere and the Alfvén conductance of the surrounding plasma (see Eq. A19 of Saur et al. 1999) and maximizes in the case when the corotational electric field $E_0$ is reduced to half inside Io's ionosphere ($E_i = 0.5\ E_0$). Heating of the atmosphere through the currents driven in the ionosphere may be the dominant heating mechanism at Io above a certain altitude (at pressures roughly below 1 nbar, Strobel et al. 1994). Numerical simulations by Saur et al. (1999) suggest a total heating rate of ~$4\times10^2$ GW. Strong Joule heating in the upper atmosphere might increase the escape, particularly of lighter species.

**Pick-up process**
Process of entrainment of freshly ionized neutrals initially at rest in Io's reference frame (in Io's atmosphere or in Io's extended neutral clouds) in the torus plasma flow. Fresh ions resulting from electron-impact ionization, photo-ionization or charge exchange in Io's atmosphere (or in neutral clouds) experience an $\mathbf{E}\times\mathbf{B}$ drift in the frame of Io. They are entrained in the local bulk flow of the plasma (corresponding to the velocity of their gyrocenter) and also start a gyromotion at the local flow velocity. In the frame of Io, the trajectories of pick-up ions (and electrons) are cycloids perpendicular to the local magnetic field. An O or S ion picked-up in the neutral cloud at the corotation velocity ~ 60 km/s results in a supply of energy to the torus of 270 eV or 540 eV respectively, larger than the average ion energy of the torus ~100 eV. The pick-up of ions in the atmosphere of Io results either in a local heating or cooling of the plasma depending on the velocity of the local plasma flow where the pick-up takes place (slower than corotation close to Io, and faster than corotation on the flanks of Io).

---

D) Specific regions and components in the systems

**Auroral footprint**
Electron or ion impact excited emission from Jupiter's atmosphere triggered by Io. Charged particle acceleration is powered by Io's electrodynamic interaction with Jupiter's magnetosphere. Particles are accelerated along the Alfvén wings connecting Io with Jupiter. Among the three auroral moon footprints that are clearly recognizable, Io's is considerably brighter, and always visible both in UV and IR. Especially in IR, Io's auroral footprint reveals detailed structures, like secondary spots or a tail, at preceding and posterior longitudes (Section 2.7). Some of these structures are caused by the reflections of the Alfvén waves on density gradients, and they depend on the latitudinal location of the moon within the plasma sheet, while part of the fine morphology of the footprint still requires a clear explanation.

**Banana, jet, and stream**
The *banana*, *jet,* and *stream* terminology for extended neutral cloud structures derives from sodium studies (e.g., Smyth and Combi, 1988b; Wilson et al., 2002), but has been broadly applied to other species in existing literature (see Section 2.5 for more details).

**Exobase**
The general definition is given in Section 2.3. As Io's atmospheric density and temperature vertical distributions are still unknown, the exobase altitude is still undetermined. A wide range of estimates of the exobase altitudes have been given in the literature, ranging from several thousand kilometers to a few tens of kilometers on the dayside atmosphere, down to Io's surface in eclipse. Our ISSI group consensus places the exobase at a few hundred kilometers, based on numerical simulations of the plasma properties along the Galileo flybys of Io, atmospheric modeling that includes the plasma/atmosphere interaction, and OI (630 nm) emissions, which can be collisionally quenched within the 110s radiative lifetime yet glow closely to Io's limb (Geissler et al., 1999; 2004), precluding higher exobase altitudes.

**Io corona (exosphere)**
Neutral gas bound to Io beyond the exobase of Io. In this region, the gases are non-collisional but still bound gravitationally to Io. The corona extends to the Hill sphere (~5.8 $R_{Io}$ with $R_{Io}$ ~ 1821 km) and smoothly merges with the neutral clouds. Atomic O and S neutral coronae have been detected by HST in the UV, extending radially to ~10 $R_{Io}$ in all directions (Wolven et al., 2001). Energetic ion absorption features detected by Galileo far from Io are also consistent with an extended corona but the neutral species cannot be determined (Huybrighs et al., 2024). Electro-Magnetic Ion Cyclotron waves (EMIC) at the $SO^+$ and $SO_2^+$ gyro frequencies have been observed by Galileo extending as far as ~7-20 $R_{io}$ from Io in the downstream direction (Russell et al., 2003), which suggests a pickup process in extended $SO_2$ and SO coronae.

**Io local aurora**
Electron or ion impact excited emission from Io's atmosphere. This choice of nomenclature is non-unique as no universal definition of aurora exists and many authors require in their definition active electron acceleration as part of the auroral processes. The latter is not the case at Io, i.e., electrons are not actively accelerated near Io (in contrast to Ganymede) and the aurora is excited by the thermal and/or suprathermal electrons in the plasma torus.

**Neutral cloud(s)**
Structures of neutral gas extending along Io's orbit, subjected to the gravitational field of Jupiter. These neutral clouds are fed by the plasma-atmosphere interaction at Io and are shaped by magnetospheric loss processes. Depending mostly on the ionization energy of the neutrals, these clouds have a limited extension along Io's orbit (Na banana cloud, S cloud etc.). The use of plural in *clouds* alludes to the different species and different shapes of the neutral structures (Figure 15, Section 2.5). They can also encompass the whole Io orbit as for atomic O (Koga et al., 2018a; Smith et al., 2022; Section 2.5) and can in that sense also be named *neutral torus* or *tori*. However, in stark contrast to the azimuthally rather homogeneous *plasma torus*, the neutral density is significantly larger close to Io for all species and we propose *neutral clouds* as the general term. The S and O neutral clouds are (likely) the main source of plasma for the torus (~80%).

**Neutral nebulae / sodium nebula**
Neutral gas abundance that extends more than 100 (1000?) $R_J$ around Jupiter and is not subject to the gravitational field of Jupiter. The only nebula observed so far is the Na nebula fed by neutral sodium ejection from Io's atmosphere or Io's torus at velocity larger than the escape velocity at the orbit of Io (25 km/s in the Jovian reference frame). The existence of nebulae in the major species O and S (and maybe SO and $SO_2$) is suspected to exist but has not yet been detected.
In the analysis of images, the emissions are analyzed in differently extended regions, like up to a radial distance of 25 $R_J$ (with much contribution from the neutral cloud in the signal), or up to 100 $R_J$ (with relatively more contributions from the nebulae).

**Plasma torus / Io plasma torus**
The *Io plasma torus* is a structure that encompasses the orbit of Io and is mainly composed of S and O multiply-charged ions. It comprises three main regions: (1) the *warm torus* (peak density near 5.9 $R_J$), (2) the *ribbon* (peak ~5.6 $R_J$), and (3) the *cold torus* (peak ~5.2 $R_J$).

**Plasma sheet**
Structure of plasma in the magnetosphere, radially beyond the *warm torus* (beyond ~7-10 R_J) and out to roughly 30 RJ where the density drops below 1/$cm^3$ . It is sourced from the effective radially outward transport of torus Io-genic material. The plasma sheet is mostly corotating, but the corotation breaks down between 20-30 $R_J$. Ion and electron temperatures increase with radial distance, thus are higher in the plasma sheet than in the warm torus.

**Transient torus event / brightening / enhancement**
Identified by an intensification of UV or optical ion emissions from the *plasma torus* for a limited period of usually 1-3 months w.r.t. a common stable background level. Such intensification is commonly explained by an enhanced *mass loading* of the torus, which leads to a higher density and possibly temperature in the torus causing the brighter emissions.

# Acknowledgements

This review is based on discussions within International Team project #515 ("Mass Loss from Io's Unique Atmosphere: Do Volcanoes Really Control Jupiter's Magnetosphere?") funded by the International Space Science Institute (ISSI) in Bern, Switzerland. LR is supported by the Swedish National Space Agency through grant 2021-00153. Further travel support from the Swedish Foundation for International Cooperation in Research and Higher Education (STINT) is acknowledged. We thank Márton Galbács very much for preparing the sketch in Figure 10. HH is supported by a DIAS Research Fellowship in Astrophysics. VD is supported by the NASA System Workings 2020-0139 Grant 80NSSC22K0109 and as a part of NASA's Juno mission supported by NASA through contract 699050X with the Southwest Research Institute. HH gratefully acknowledges financial support from Khalifa University's Space and Planetary Science Center (Abu Dhabi, UAE) under Grant no. KU-SPSC-8474000336. CS acknowledges the support of NASA through programs 80NSSC21K1138 and 80NSSC22K0954, and the NSF through program AST-2108416. KdK acknowledges funding from the National Science Foundation grant 2238344 through the Faculty Early Career Development Program. SVB was supported by STFC Fellowship and grant ST/M005534/1 and ST/V000748/1. CT, RK, MK, and FT are supported by JSPS KAKENHI under Grant Number 20KK0074. AB is supported by the Volkswagen Foundation Grant Az 97742. JS has received funding from the European Research Council (ERC) under the European Union's Horizon 2020 research and innovation programme (grant agreement No. 884711).